\pdfoutput=1
\newcommand*{\ATLASLATEXPATH}{}
\documentclass[UKenglish,texlive=2013,cernpreprint,biber]{\ATLASLATEXPATH atlasdoc}
\usepackage{\ATLASLATEXPATH atlaspackage}
\usepackage{rotating}
\usepackage{soul}
\usepackage{\ATLASLATEXPATH atlasbiblatex}
\usepackage{\ATLASLATEXPATH atlascontribute}
\usepackage{subfig}
\usepackage{\ATLASLATEXPATH atlasphysics}

\usepackage{siunitx}

\AtlasTitle{Modelling radiation damage to \\ pixel sensors in the ATLAS detector}
\author{The ATLAS Collaboration}
\AtlasJournalRef{JINST 14 (2019) P06012}
\AtlasDOI{10.1088/1748-0221/14/06/P06012}
 \PreprintIdNumber{CERN-EP-2019-061}
 
\AtlasAbstract{Silicon pixel detectors are at the core of the current and planned upgrade of the ATLAS  experiment at the LHC.  Given their close proximity to the interaction point, these detectors will be exposed to an unprecedented amount of radiation over their lifetime.  The current pixel detector will receive damage from non-ionizing radiation in excess of $10^{15}$ 1~MeV~$\text{n}_\text{eq}/\text{cm}^2$, while the pixel detector designed for the high-luminosity LHC must cope with an order of magnitude larger fluence. This paper presents a digitization model incorporating effects of radiation damage to the pixel sensors.  The model is described in detail and predictions for the charge collection efficiency and Lorentz angle are compared with collision data collected between 2015 and 2017 ($\leq 10^{15}$ 1~MeV~$\text{n}_\text{eq}/\text{cm}^2$).}

\hypersetup{pdftitle={ATLAS document},pdfauthor={The ATLAS Collaboration}}

\begin{document}

\maketitle

\tableofcontents

\vspace{-3mm}

\section{Introduction}
\label{sec:intro}

\vspace{-1mm}

As the subdetector in closest proximity to the interaction point, the ATLAS pixel detector will be exposed to an unprecedented amount of radiation over its lifetime. The modules comprising the detector are designed to be radiation tolerant, but their performance will still degrade over time.  It is therefore of crucial importance to model the impact of radiation damage for an accurate simulation of charged-particle interactions with the detector and the reconstruction of their trajectories (\textit{tracks}).  Modelling radiation damage effects is especially relevant for the high-luminosity (HL) upgrade of the Large Hadron Collider (LHC); the instantaneous and integrated luminosities will exceed current values by factors of 5 and 10, respectively.  The simulations for the present (Run 1: 2010-12, Run 2: 2015-18) and future ATLAS detectors currently do not model the effect of silicon sensor radiation damage~\cite{Aad:2010ah,ATL-PHYS-PUB-2016-025}.  
 
This article documents the physics and validation of the pixel radiation damage models that will be incorporated into the ATLAS simulation.  Section~\ref{sec:detector} briefly introduces the specifications of the ATLAS pixel detector and provides an overview of the impact of radiation damage effects.  Measurements of the fluence and depletion voltage are presented in Section~\ref{sec:conditions}.  A model of charge deposition and measurement that includes radiation damage effects is documented in Section~\ref{sec:fullmodel}.  Comparisons and validation of the simulation with data are presented in Section~\ref{sec:validation} and conclusions are given in Section~\ref{sec:conclusions}.
 
 \section{The ATLAS pixel detector and radiation damage effects}
 \label{sec:detector}
 
 The ATLAS pixel detector~\cite{Aad:2008zz,Capeans:1291633,Abbott:2018ikt} consists  of four barrel layers and a total of six disc layers, three at each end of the barrel region. The four barrel layers are composed of $n^+$-in-$n$ planar oxygenated~\cite{Lindstrom:421210,Lindstrom:2000gd} silicon sensors at radii of 33.5, 50.5, 88.5, and 122.5 mm from the geometric centre of the ATLAS detector~\cite{Detector}.  The sensors on the innermost barrel layer (the insertable $B$-layer or IBL~\cite{Capeans:1291633,Abbott:2018ikt}, installed between Runs 1 and 2) are $200$ \textmu m  thick, while the sensors in the other layers are $250$ \textmu m  thick.  At high $|z|$\footnote{ATLAS uses a right-handed coordinate system with its origin at the nominal interaction 
point (IP) in the centre of the detector and the $z$-axis coinciding with the axis of the beam pipe.  The 
$x$-axis points from the IP towards the centre of the LHC ring, and the $y$-axis points upward. 
Cylindrical coordinates ($r$,$\phi$) are used in the transverse plane, $\phi$ being the azimuthal angle 
around the $z$-axis. The pseudorapidity is defined in terms of the polar angle $\theta$ as $\eta = - \ln 
\tan(\theta/2)$. %
} on the innermost barrel layer, there are $n^+$-in-$p$ 3D sensors~\cite{PARKER1997328} that are $230$ \textmu m  thick.   The innermost barrel layer
 pixel pitch is $50\times 250$ \textmu m$^2$; everywhere else the pixel pitch is  $50\times 400$ \textmu m$^2$.   Charged particles traversing the sensors deposit energy by ionizing the silicon bulk; for typical LHC energies, such particles are nearly minimum-ionizing particles (MIP).  The deposited charge drifts through the sensor and the analogue signal recorded by the electrode is digitized, buffered, and read out using an FEI4~\cite{GarciaSciveres:2011zz} (IBL) or FEI3~\cite{Aad:2008zz} (all other layers) chip.  Non-ionizing interactions from heavy particles and nuclei lead to radiation damage, which modifies the sensor bulk and can therefore alter the detection of MIPs. %
 
 Radiation damage in the sensor bulk is caused primarily by displacing a silicon
 atom out of its lattice site resulting in a silicon interstitial site and a leftover vacancy (Frenkel pair)~\cite{Moll:1999kv,niel}.  These primary defects build, depending on the recoil energy, cluster defects and point defects in the silicon lattice that cause energy levels in the band gap.  When activated and occupied, these states lead to a change in the effective doping concentration, a reduced signal collection efficiency due to charge trapping, and an increase in the sensor leakage current that is proportional to the fluence received. The change in effective doping concentration has consequences for the depletion voltage and electric field profile. For the pixel planar sensors before irradiation, the depletion region grows from the back side of the sensor towards the pixel $n^+$ implant. After irradiation, the effective doping concentration decreases with increasing fluence until the sensor bulk undergoes space-charge sign inversion (often called \textit{type inversion}) from $n$-type to $p$-type. After this type inversion, the depletion region grows from the $n^+$ implant towards the back side of the sensor and the depletion voltage gradually increases with further irradiation (more details are given in Section~\ref{sec:annealing}). The effective doping concentration is further complicated by annealing in which new defects are formed or existing defects dissociate due to their thermal motion in the silicon lattice~\cite{Moll:1999kv}.  Consequently, radiation damage effects depend on both the irradiation and temperature history.  The silicon bulk of the IBL planar sensors underwent type inversion after about 3 fb$^{-1}$ of data collected in 2015 and the second innermost layer ($B$-layer) inverted in the 2012 run after about 5 fb$^{-1}$.  The outer two layers inverted between Runs 1 and 2.

In ATLAS, complex radiation fields are simulated by propagating inelastic proton--proton interactions, generated by Pythia 8~\cite{Sjostrand:2006za,Sjostrand:2014zea} using the MSTW2008LO parton distribution functions~\cite{Martin:2009iq} and the A2 set of tuned parameters~\cite{ATLAS:2012uec}, through the ATLAS detector material using the particle transport code FLUKA~\cite{Battistoni:2007zzb,Ferrari:898301}. Particles are transported down to an energy of 100~keV, except for photons (30~keV) and neutrons (thermal). It is important to model as accurately as possible all the inner detector and calorimeter geometry details because high-energy hadron cascades in the material lead to increased particle fluences in the inner detector, especially neutrons. A description of the ATLAS FLUKA simulation framework can be found in Ref.~\cite{Baranov:814823}.

\begin{figure}[h!]
\centering
\hspace*{-0.5in}
\centerline{\subfloat[]{\includegraphics[width=0.54\textwidth]{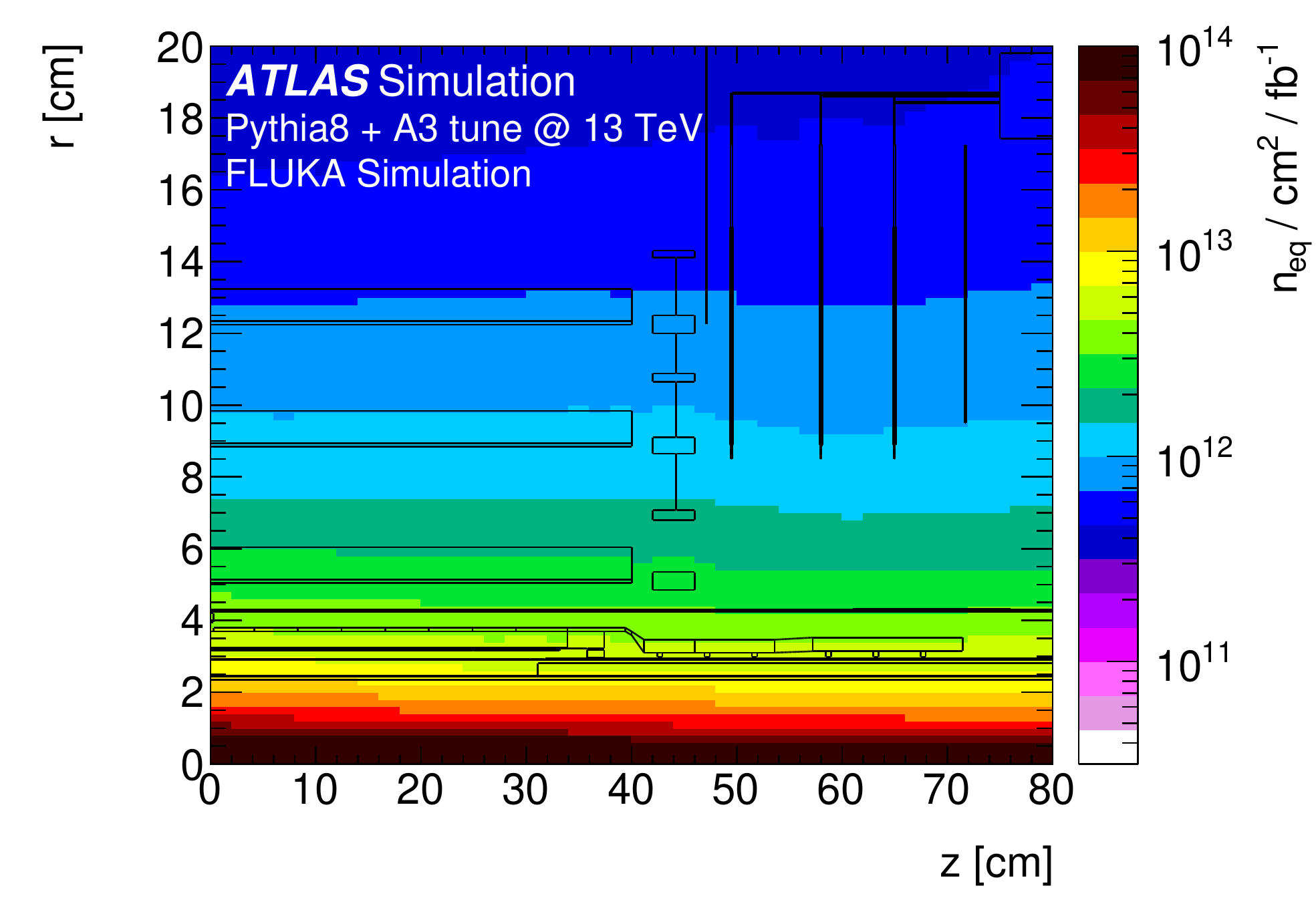}}\hspace{3mm}
\subfloat[]{\includegraphics[width=0.41\textwidth]{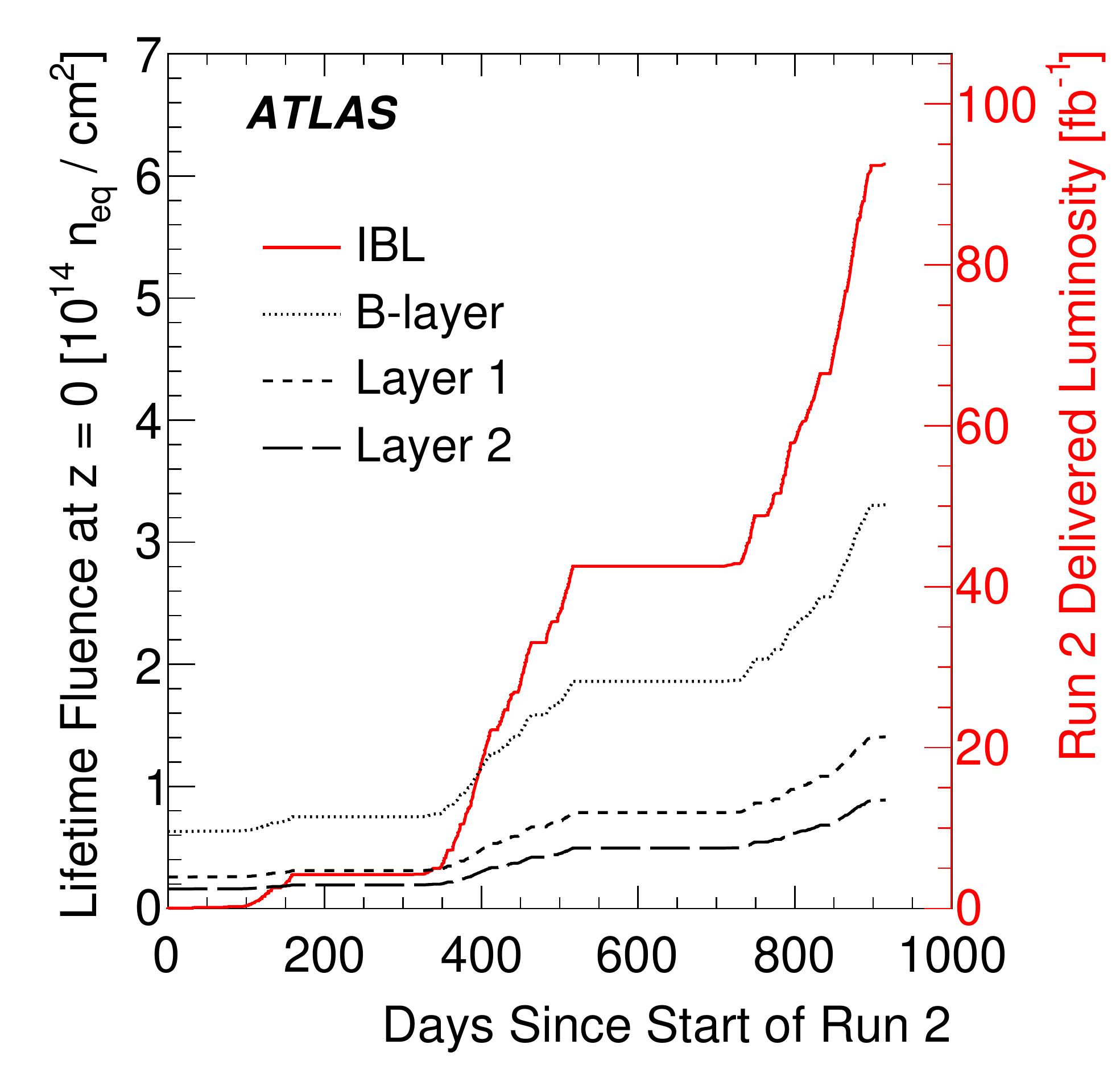}}}
  \caption{(a) Simulated 1~MeV~$\text{n}_\text{eq}$ fluence predictions shown as a function of the radial and longitudinal distance from the geometric centre of the detector for a one-quarter slice ($z>0$ and above the beam) through the ATLAS FLUKA geometry.   (b) Predictions for the lifetime fluence experienced by the four layers of the current ATLAS pixel detector as a function of time since the start of Run 2 (June 3, 2015) at $z\approx 0$ up to the end of 2017.  For the IBL, the lifetime fluence is only due to Run 2 and for the other layers, the fluence includes all of Run 1.  The IBL curve represents both the fluence on the IBL (left axis) as well as the delivered integrated luminosity in Run 2 (right axis).}
\label{fig:fluenceoverview2}
\end{figure}

Predictions of the 1~MeV neutron-equivalent fluences\footnote{For silicon sensors
the relevant measure of the radiation damage is the non-ionizing energy loss (NIEL), normally expressed as the equivalent
damage of a fluence of 1~MeV neutrons ($\text{n}_\text{eq}/\text{cm}^2$).} per fb$^{-1}$ for silicon in the ATLAS FLUKA inner detector geometry are shown in Figure~\ref{fig:fluenceoverview2}(a). The dominant contribution is from charged pions originating directly from the proton--proton collisions. The fluence values averaged over all barrel modules for the four pixel layers starting from the innermost one are $6.1\times 10^{12}$, $2.9\times 10^{12}$, $1.2\times 10^{12}$ and $7.8\times 10^{11}$ $\text{n}_\text{eq}/\text{cm}^2/\text{fb}^{-1}$, respectively.  The fluence depends on the $z$ position as the material and particle composition are $\eta$-dependent.
For example, in the IBL the maximum predicted value of $6.6\times 10^{12}$ $\text{n}_\text{eq}/\text{cm}^2/\text{fb}^{-1}$ in the central location is about $10\%$ higher than in the end regions (studied further in Section~\ref{sec:fluence}). Figure~\ref{fig:fluenceoverview2}(b) shows the 1~MeV neutron-equivalent fluence as a function of time, based on the FLUKA simulation. The luminosity is determined by a set of dedicated luminosity detectors~\cite{Aaboud:2016hhf} that are calibrated using the van der Meer beam-separation method~\cite{vanderMeer:296752}. By the end of the proton--proton collision runs in 2017, the IBL and $B$-layer had received integrated fluences of approximately $\Phi=6\times 10^{14}$ and $3\times 10^{14}$~$\text{n}_\text{eq}/\text{cm}^{2}$, respectively. The two outer layers have been exposed to less than half the fluence of the inner layers.

The goal of this paper is to present a model for radiation damage to silicon sensors that is fast enough to be incorporated directly into the \textit{digitization} step of the ATLAS Monte Carlo (MC) simulation, i.e.\ the conversion from energy depositions from charged particles to digital signals sent from module front ends to the detector read-out system.  In the context of the full ATLAS simulation chain~\cite{Aad:2010ah}, digitization occurs after the generation of outgoing particles from the hard-scatter collision and the simulation of their interactions with the detector and before event reconstruction, which is the same for data and simulation. The CMS Collaboration has developed a model of radiation damage~\cite{Contardo:2020886,Swartz2002,Chiochia:2004qh,Swartz:2005vp,Contardo:2020886},\footnote{This model is used in some HL-LHC projection studies, but there is currently no public documentation with a detailed description of the implementation in the CMS software.} validated with test-beam data, but it is used to apply template corrections to the total deposited charge in simulation from a model without inherent radiation damage effects and so is not directly comparable to the methods described here.

There are two types of microscopically motivated effective radiation damage models used for the studies presented here:  Hamburg\footnote{See Ref.~\cite{Moll:1999kv} and references therein.  This model is a phenomenological approach that includes some physically well-motivated components and other aspects that are not directly based on the microphysics of defect states.} and models developed in the framework of Technology Computer Aided Design (TCAD) simulations.  In reality, the microphysics is complex, involving many defect states, but each model includes a small number of effective components to capture the main effects.  The Hamburg model describes annealing and is only used to validate conditions data (Section~\ref{sec:conditions}).  Stand-alone implementations of this model simulate the time-dependent leakage current (Section~\ref{sec:fluence}) and doping concentration (Section~\ref{sec:annealing}) for checking the fluence and depletion voltage.  The second type of model (TCAD) is used directly in the digitizer (software that performs digitization) described in Section~\ref{sec:fullmodel}.  In contrast to the Hamburg model, radiation damage implemented in TCAD predicts a non-uniform spatial distribution of space-charge density and thus a more realistic electric field profile (Section~\ref{sec:EField}) for computing charge propagation inside the sensor bulk.  

Multiple radiation damage models are required since no model accounts for both annealing and a non-uniform space-charge density distribution.  Therefore, each model is used where it is most appropriate.  An approximate combination of model predictions is described in Section~\ref{sec:TCADAnneal}.  However, for the present levels of annealing, the combination yields variations in electric field profiles that are smaller than the uncertainty in the TCAD radiation damage model parameters (Section~\ref{sec:Efieldmodelcomparisons}) and so is not used for the final results (Section~\ref{sec:validation}) -- only TCAD input without annealing is currently used for the digitizer.  In the future, when there is more annealing and the radiation damage model parameters are further constrained from data, it will become a crucial and challenging project to combine the power of both types of models.  

\section{Validating sensor conditions}
\label{sec:conditions}

\subsection{Luminosity to fluence}
\label{sec:fluence}

 The most important input to the radiation damage digitization model is the estimated fluence.  Section~\ref{sec:detector} introduced the baseline FLUKA simulation that is used to determine the conversion factor ($\Phi/L_\text{int}$) between integrated luminosity and fluence.  In order to estimate systematic uncertainties in these predictions, the fluence is converted into a prediction for the leakage current. The leakage 
current can be precisely measured and therefore provides a solid validation for the FLUKA simulation.  For $n$ time intervals, the predicted leakage current is given by Ref.~\cite{Moll:1999kv}:

\begin{align}
\label{eq:leakage_modified}
I_\text{leak}=(\Phi/L_\text{int})\cdot\sum_{i=1}^n V_i\cdot L_\text{int,$i$}\cdot\left[\alpha_I\exp\left(-\sum_{j=i}^n\frac{t_j}{\tau(T_j)}\right)+\alpha_0^*-\beta\log\left(\sum_{j=i}^n\frac{\Theta(T_j)\cdot t_j }{ t_0}\right)\right],
\end{align}

where $L_\text{int,$i$}$ is the integrated luminosity, $t_i$ is the duration, and $T_i$ is the temperature in time interval $i$.  The first sum is over all time periods and the two sums inside the exponential and logarithm functions are over the time between the irradiation in time period $i$ and the present time.   The other symbols in Eq.~(\ref{eq:leakage_modified}) are $t_0=1\;$min, $V_i$ is the depleted volume (in cm$^3$), $\alpha_I=(1.23\pm0.06)\times 10^{-17}\text{ A}/\text{cm}$, $\tau$ follows an Arrhenius equation $\tau^{-1}=(1.2^{+5.3}_{-1.0})\times 10^{13}\text{ s}^{-1}\times \text{e}^{(-1.11\pm 0.05)\text{ eV}/k_\text{B}T}$, where $k_\text{B}$ is the Boltzmann constant, $\alpha_0^*=7.07\cdot 10^{-17}\;$A/cm, and\footnote{A small temperature dependence has been observed in the value of $\beta$~\cite{Moll:1999kv}.  For this analysis, the reported value at 21$^\circ$C is taken as it is closest to the operational temperature range of the detector.} $\beta=(3.29\pm 0.18)\times 10^{-18}$~A/cm.  The time scaling function $\Theta(T)$ is defined by\footnote{This is not the only way to incorporate time-dependence in the thermal history.  Another proposal is to sum the inverse temperatures~\cite{Barney:1709387}.  Such a method has been compared with Eq.~(\ref{eq:tempscaling}) and results in similar predictions for the leakage current with the current fluence levels and annealing times.}

\begin{align}
\label{eq:tempscaling}
\Theta(T) = \exp \left[ - \frac{E_I^*}{k_\text{B} } \left( \frac{1}{T} - \frac{1}{T_\text{ref} } \right) \right],
\end{align}

where $E_I^*=(1.30\pm 0.14)$~eV and $T_{\text{ref}}$ is a reference temperature, typically 20$^\circ$C.

\begin{figure}[h!]
\centering
\includegraphics[width=0.85\textwidth]{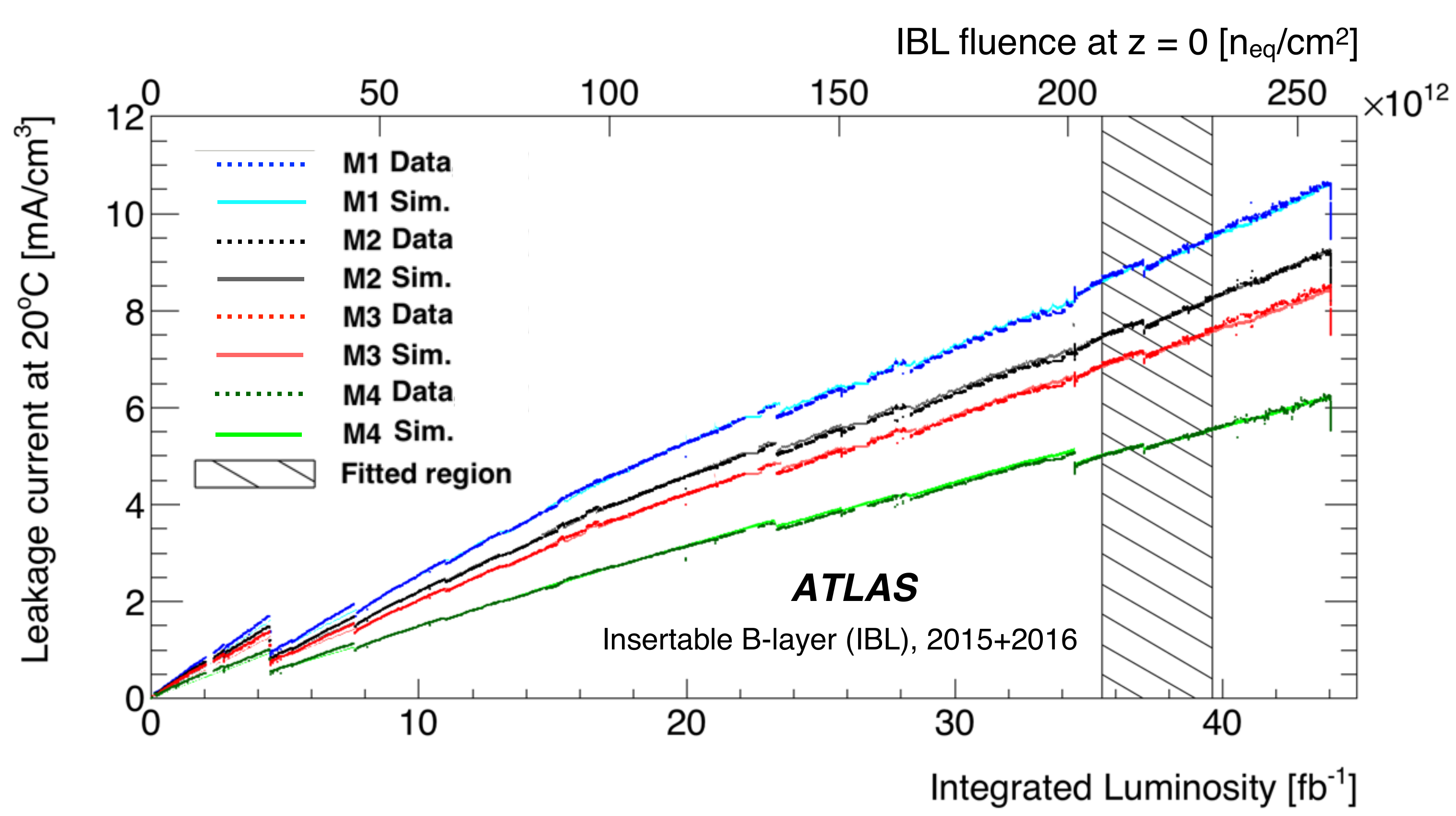}
  \caption{The measured (``Data'') and predicted (``Sim.'', fitting for $\Phi/L_\text{int}$) leakage current for the four module groups of the IBL as a function of integrated luminosity since the start of the Run 2.  The predicted leakage current is obtained from Eq.~(\ref{eq:leakage_modified}) and by fitting the data in the dashed region to determine the luminosity-to-fluence factor.   The IBL pixel module groups M1, M2, M3 approximately span the ranges $z\in[-8,8]$~cm, $|z|\in[8,16]$~cm, and $|z|\in [16,24]$~cm, respectively.  The M4 modules use 3D sensors and span the range $|z|\in [24,32]$~cm.  Sharp drops correspond to periods without collisions.}
\label{fig:leackagecurrent:beta}
\end{figure}

Using the measured module temperature as a function of time, Eq.~(\ref{eq:leakage_modified}) is used to predict the leakage current as shown in Figure~\ref{fig:leackagecurrent:beta}. The leakage current is scaled to correspond to a temperature of 20$^\circ$C using the factor (see e.g., Ref.~\cite{book}) $(T_\text{R}/T)^2\exp\left(-E_\text{eff}(T_\text{R}^{-1}-T^{-1})/2k_\text{B}\right)$, where $T_\text{R}=20^\circ$C and $E_\text{eff}=1.12$~eV. The value of $E_\text{eff}$ is lower than the one measured in Ref.~\cite{Chilingarov_tscale}, but was found to agree better with the data.  Measurements of the properties describing the modules were updated every ten minutes.  Since the IBL was newly inserted before the 2015 run, the initial leakage current level is compatible with zero.  A constant $\Phi/L_\text{int}$ conversion factor is fit to the data per module group.  Module groups differ by their distance along the beam direction from the geometric centre of the detector.  Each module group is $8$~cm long on both sides of the detector along the beam direction.  The groups M1, M2, M3 approximately span the ranges $z\in[-8,8]$~cm, $|z|\in[8,16]$~cm, and $|z|\in [16,24]$~cm, respectively.  The M4 modules use 3D sensors; M4 spans the range $|z|\in [24,32]$~cm.  Only the time interval indicated by a dashed region in Figure~\ref{fig:leackagecurrent:beta} is used in the fluence rate extraction.  Prior to this time, the IBL was under-depleted and after this time, the frequency of measurements decreased.  A sensor volume correction is applied to the under-depleted data.  After this correction, the adjusted simulation reproduces the trends observed in the data both inside and outside of the fit region.  

\begin{figure}[h!]
\centering
\includegraphics[width=0.65\textwidth]{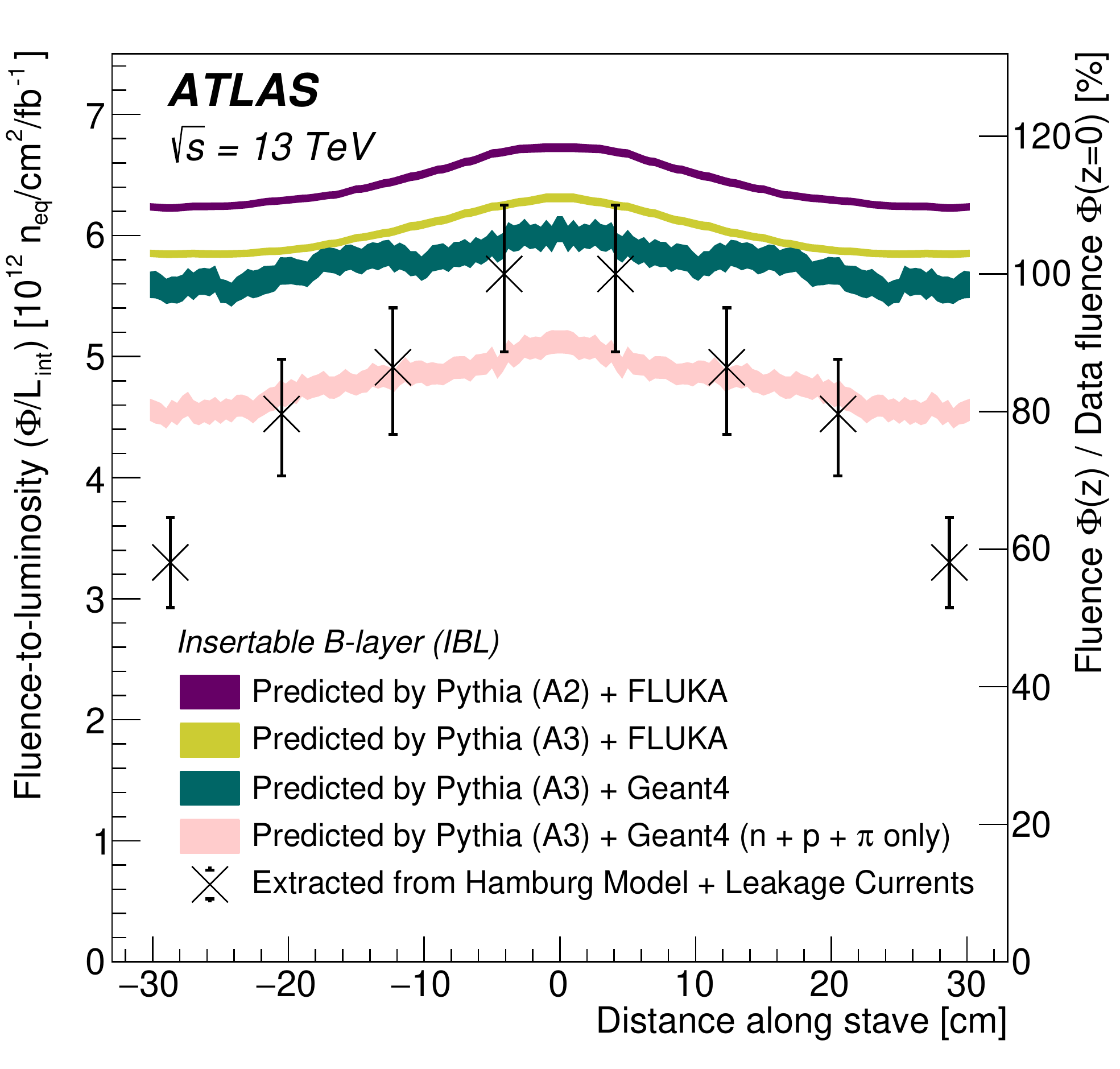}
\caption{The fluence-to-luminosity conversion factors (extracted from leakage current fits) as a function of $z$, compared with the Pythia+FLUKA and Pythia+Geant4 predictions.}
\label{fig:leackagecurrent:eta}
\end{figure}

Figure~\ref{fig:leackagecurrent:eta} shows that there is a stronger $z$-dependence in the measured fluence compared with the FLUKA predictions described in Section~\ref{sec:detector}.  The error bars on $\Phi/L_\text{int}$ predicted by the Hamburg model fitted to data are dominated by a conservative $10\%$ uncertainty, accounting for the possible difference between the leakage current at the operational bias voltage and the current at the full depletion voltage (see Section~\ref{sec:annealing}).  After irradiation, the leakage current increases with increasing bias voltage also after full depletion, while the Hamburg model predicts a constant leakage current above the full depletion voltage. Therefore, the choice of voltage for the leakage current measurement is crucial for comparison with the Hamburg model prediction~\cite{ATL-INDET-PUB-2014-006}. Uncertainties due to the annealing model ($0.1\%$) and data fit ($0.5\%$) are subdominant. 
The predictions in Figure~\ref{fig:leackagecurrent:eta} deviate from the measured values by about $1.5\sigma$ of the uncertainty at $z=0$, with larger deviations at higher $|z|$.  In addition to the Pythia+FLUKA prediction described in Section~\ref{sec:detector}, Figure~\ref{fig:leackagecurrent:eta} also shows predictions with an updated Pythia set of tuned parameters (A3~\cite{ATL-PHYS-PUB-2016-017}) as well as an alternative geometry and transport model using Geant4~\cite{Agostinelli:2002hh}.  Neither of these variations can account for the $z$-dependence, but this does illustrate part of the uncertainty due to the transport model and particle generator.  There is also a significant source of uncertainty from the silicon hardness factors~\cite{niel} (common to both the Geant4 and FLUKA models).  The hardness factors used here are from the RD50 database~\cite{rd50,osti_10115441,KONOBEYEV1992117,HUHTINEN1993580,273529}, but all of these values are without uncertainty and many are based only on simulation.   The uncertainty in the hardness factors affects both the prediction and the Hamburg model (through the $\alpha$ parameters).  Future collision data may be able to constrain these hardness factors.  As shown in Fig.~\ref{fig:leackagecurrent:eta}, most of the damage is due to charged pions, protons, and neutrons, so the larger uncertainties on other particle species is a subdominant source of total uncertainty for the hardness factors.

The remainder of this paper focuses on central $|z|\approx 0$, using the FLUKA simulations without modification for the central value, but with a $15\%$ uncertainty in the fluence taken from this leakage current study.  The ATLAS tracking acceptance is $|\eta|<2.5$, which corresponds to $|z|< 20$~cm in the IBL.

\clearpage

\subsection{Annealing and depletion voltage}
\label{sec:annealing}

 As already introduced in Section~\ref{sec:detector}, the irradiation and thermal history are accounted for in the prediction of the effective doping concentration with the Hamburg model.  In this model, the effective doping concentration ($N_\text{eff}(t)$) has the following form:

\begin{align}
\label{eq:hamburg0}
N_\text{eff}(t)=N_\text{D}^\text{non-removable}(0)+N_\text{D}^\text{removable}(t)-N_\text{A}^\text{stable}(t)-N_\text{A}^\text{beneficial}(t)-N_\text{A}^\text{reverse}(t),
\end{align}

where $N_\text{D}^\text{(non)-removable}(0)$ is the initial concentration of (non)-removable donors\footnote{Where $N_\text{eff}(0)=N_\text{D}^\text{non-removable}(0)+N_\text{D}^\text{removable}(0)$.} and the other terms are described below.  The fraction of removable donors at the doping concentrations used for silicon sensors is predicted to be 100\% of the initial doping concentration for charged-particle irradiation, which dominates the inner pixel layers in the ATLAS detector.  The time-dependence of the terms on the right-hand side of Eq.~(\ref{eq:hamburg0}) are described by the following differential equations:

\begin{align}
\label{eq:hamburg1}
&\frac{\text{d} }{\text{d} t}N_\text{D}^\text{removable}(t) &&=-c\phi(t) N_\text{D}^\text{removable}(t)&&\text{removal of donors for $n$-type during irradiation,}\\\label{eq:hamburg2}
&\frac{\text{d} }{\text{d} t}N_\text{A}^\text{stable}(t) &&= g_\text{C}\phi(t)&&\text{addition of stable acceptors during irradiation,}\\\label{eq:hamburg3}
&\frac{\text{d} }{\text{d} t}N_\text{A}^\text{beneficial}(t)  &&=g_\text{A}\phi(t)-k_\text{A}(T)N_\text{A}^\text{beneficial}(t) &&\text{beneficial annealing,}\\\label{eq:hamburg4}
&\frac{\text{d} }{\text{d} t}N_\text{N}^\text{reverse}(t)  &&= g_\text{Y}\phi(t)-k_\text{Y}(T)N_\text{N}^\text{reverse}(t) &&\text{reverse annealing -- neutrals,}\\\label{eq:hamburg5}
&\frac{\text{d} }{\text{d} t}N_\text{A}^\text{reverse}(t)  &&= k_\text{Y}(T)N_\text{N}^\text{reverse}(t) &&\text{reverse annealing -- acceptors,} 
\end{align}

\noindent where $\phi(t)$ is the irradiation rate in $\text{n}_\text{eq}/\text{cm}^2/\text{s}$.  Equation~(\ref{eq:hamburg1}) represents the effective removal of the initial donors by mobile defects.  The removal constant is $c=6.4\times 10^{-14}\text{ cm}^2$~\cite{Moll:1999kv}.  The second equation, Eq.~(\ref{eq:hamburg2}), represents the constant addition of stable (non-annealable) defects which act electrically as acceptors.  Two additional defects are introduced in Eqs.~(\ref{eq:hamburg3}) and~(\ref{eq:hamburg4}).  These defects, introduced during irradiation with introduction rates $g_\text{A}$ and $g_\text{Y}$, are short-lived at sufficiently high temperatures ($\gtrsim 10^\circ$C).  The temperature-dependence of the decay rates is modelled with an Arrhenius equation, $k_i(T)=k_{i,0}\,\text{e}^{-E_i/k_\text{B}T}$, where $k_\text{A,0}=2.4^{+1.2}_{-0.8}\times 10^{13}/\text{s}, k_\text{Y,0}=1.5^{+3.4}_{-1.1}\times 10^{15}/\text{s}$, $E_\text{A}=(1.09\pm0.03)$~eV and $E_\text{Y}=(1.33\pm0.3)$~eV~\cite{Moll:1999kv}.  For the beneficial annealing (Eq.~(\ref{eq:hamburg3})), the acceptor-like defects introduced during irradiation decay into neutral states with a time constant that is $\mathcal{O}(\text{days})$ at $20^{\circ}$C and $\mathcal{O}(\text{years})$ at $-15^{\circ}$C .  In contrast, for reverse annealing, neutral defects are introduced during irradiation (Eq.~(\ref{eq:hamburg4})). The neutral defects can decay into acceptor-like states (Eq.~(\ref{eq:hamburg5})), decreasing (increasing) the effective doping concentration before (after) space-charge sign inversion.  The timescale for reverse annealing is $\mathcal{O}(\text{weeks})$ at $20^{\circ}$C.  

\begin{figure}[h!]
\centering
  \subfloat[]{\includegraphics[width=0.49\textwidth]{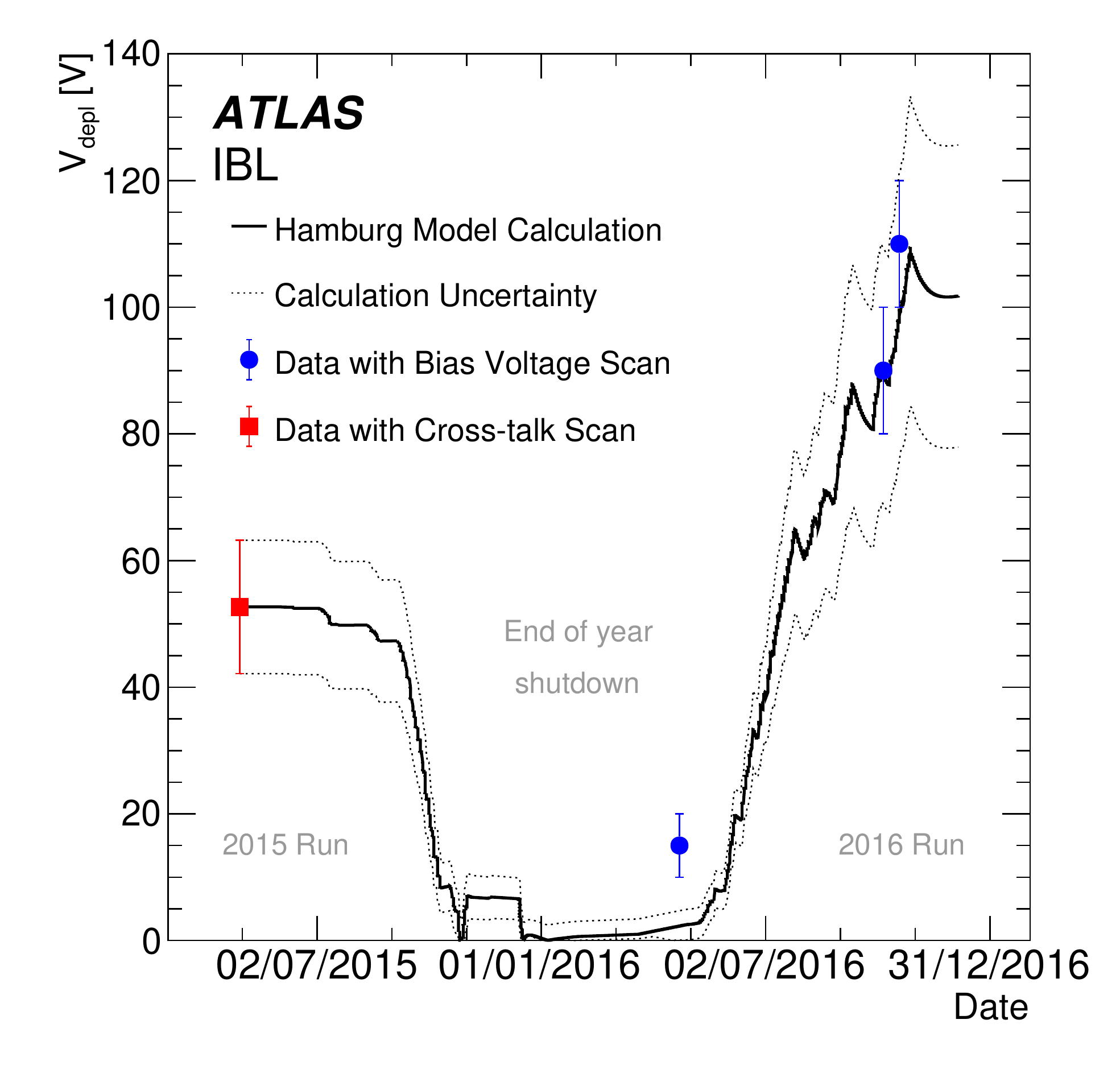}}
  \subfloat[]{\includegraphics[width=0.49\textwidth]{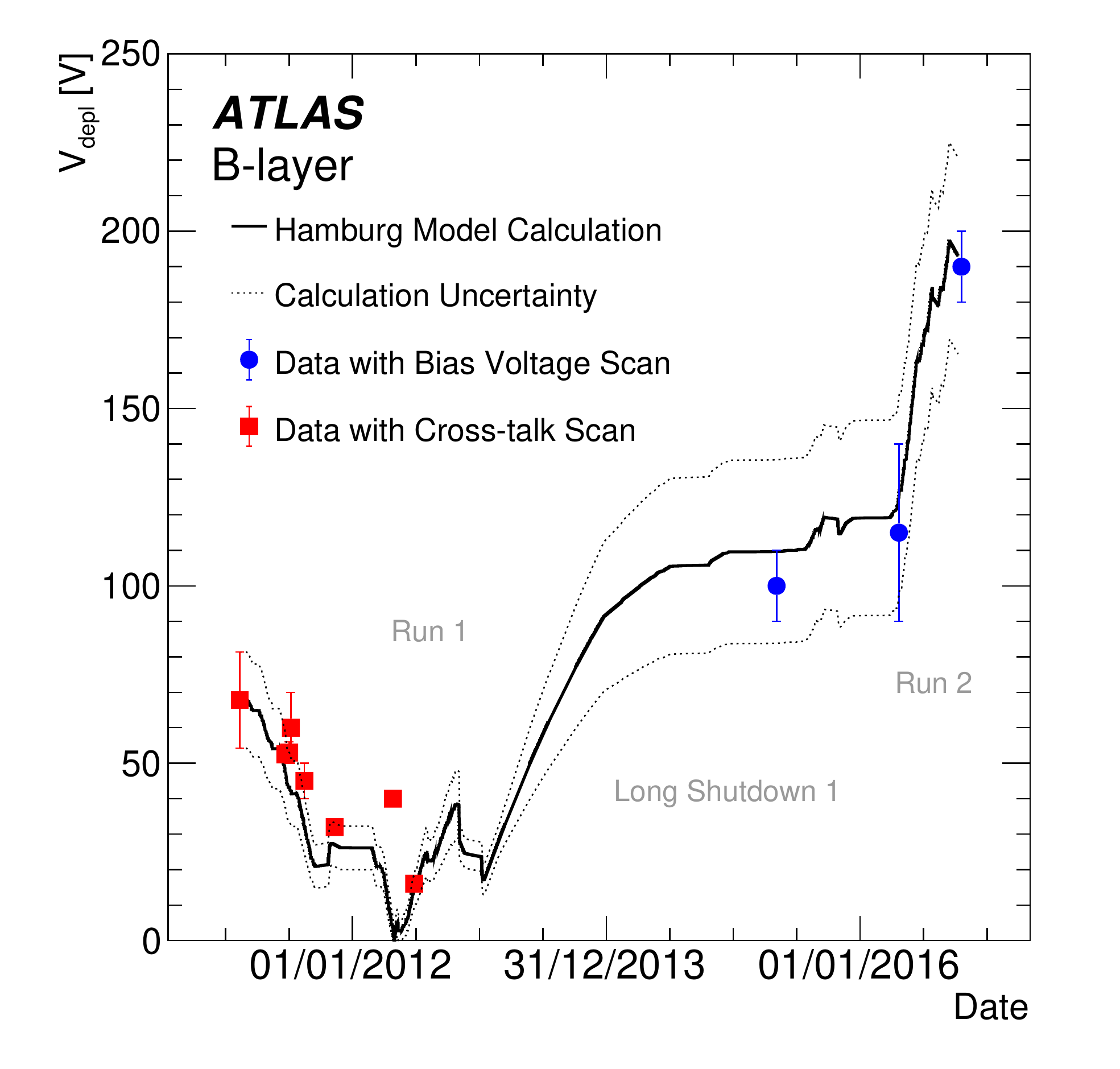}}
\caption{Calculated depletion voltage of (a) IBL and (b) $B$-layer according to the Hamburg model as a function of time from the date of their installation until the end of 2016. The calculations shown use the central values of the fitted introduction rates listed in Table~\ref{tab:depletionvoltage:introductionrates}. Circular points indicate measurements of the depletion voltage using the bias voltage scan method while square points display earlier measurements using cross-talk scans.}
\label{fig:electricfield:depletionvoltage}
\end{figure}

While the introduction rates $g_\text{C},g_\text{A}$, and $g_\text{Y}$ have been measured elsewhere (e.g.\ Ref.~\cite{Lindstrom:421210}), the reported values vary significantly amongst different materials and irradiation types, and so are fit with depletion voltage data from the ATLAS pixel detector.  The notion of full depletion is not well-defined for highly irradiated sensors where the regions inside the sensor bulk can have a very low field (see Section~\ref{sec:Efieldprofile}).  However, at moderate fluences, the depletion region is well-defined and is important for calibrating the parameters of the Hamburg model specifically for the ATLAS pixel sensors, as the full depletion voltage ($V_\text{depl}$) is calculated in terms of the effective doping concentration:

\begin{align*}
\label{eq:hamburg13}
V_\text{depl} = |N_\text{eff}| \cdot \frac{ed^2}{2\epsilon\epsilon_0},
\end{align*}

where $d$ is the sensor thickness, $e$ is the charge of the electron, $\epsilon$ is the dielectric constant, and $\epsilon_0$ is the vacuum permittivity. Figure~\ref{fig:electricfield:depletionvoltage} shows the calculated $V_\text{depl}$ using the Hamburg model as a function of time for (a) the IBL and (b) the $B$-layer.  In situ measurements of $V_\text{depl}$ of the sensors were performed with two different methods using the ATLAS pixel detector: the cross-talk scan and the bias voltage scan. The first method uses the cross-talk between adjacent pixels (square points in Figure~\ref{fig:electricfield:depletionvoltage}). Since the pixels are isolated (i.e., no cross-talk) only after full depletion, this is a powerful measurement tool. However, this method is only applicable before space-charge sign inversion since afterwards the pixels are already isolated at low bias voltages, much before full depletion. The bias voltage scan uses the mean time over threshold (ToT)~\cite{603658} of clusters of hits on reconstructed particle trajectories, measured in units of bunch crossings (25 ns).  The depletion voltage is extracted by fitting two linear functions to the rising and plateau regions of the measured data. The intersection of the two lines is defined to be the depletion voltage (circular points in Figure~\ref{fig:electricfield:depletionvoltage}).  The initial calculated $V_\text{depl}$ is chosen to match the value measured during quality assurance of the IBL planar pixel sensors.   The total uncertainty band for the calculations is due to varying the input parameters within their uncertainties in addition to a 20\% uncertainty in the initial doping concentration (see e.g.\ Ref.~\cite{Wittig}).

Due to the huge parameter space given by the defect introduction rates, the time necessary for the simulation, and a focus on physically rather than mathematically correct parameter combinations, the adjustment of the introduction rates\footnote{The initial effective doping concentration also has some uncertainty, but the fitted parameters are mostly set by the measurements following space-charge sign inversion and therefore are largely insensitive to this uncertainty.} was performed using particular periods of time or available data, described in the following. The derived introduction rates are summarized in Table~\ref{tab:depletionvoltage:introductionrates}.  The central value of $g_\text{A}$ is from the literature~\cite{Lindstrom:421210} since the measurements reported here were performed at times where there was no sensitivity to beneficial annealing.  The uncertainty in $g_\text{A}$ reported in Table~\ref{tab:depletionvoltage:introductionrates} is determined by adjusting $g_\text{A}$ according to the same prescription as for $g_\text{Y}$ and $g_\text{C}$ -- one parameter is varied at a time until there is a large deviation (more details provided below). The value of $g_\text{Y}$ was extracted from the reverse annealing during the long shutdown 1 (LS1), which was an extended period when the detector was maintained at room temperature without further irradiation.  Since the IBL was installed after LS1 and has not undergone significant reverse annealing, the $g_\text{Y}$ value of the $B$-layer is used also for the IBL. %
In contrast to $g_Y$ and $g_A$, $g_C$ can be well-constrained during any data-taking period when constant damage is accumulated.  
The extracted values for $g_\text{C}$ are different for the IBL and the $B$-layer because the particle compositions are different (relatively more neutrons for the $B$-layer and more charged pions for the IBL). 
The uncertainties arise from the procedure, from the luminosity-to-fluence conversion, and from the uncertainty in the temperature in the actual sensor.   

For comparison, the values obtained by the ROSE Collaboration~\cite{Lindstrom:421210} for oxygen-enriched silicon are also reported.  The values for $g_\text{C}$ and $g_\text{Y}$ are within the range given by the ROSE Collaboration when neutron and proton irradiations are considered.  The predictive power of the simulation would benefit from more precise measurements of $g_\text{A}$, $g_\text{Y}$ and $g_\text{C}$, which may be possible with future ATLAS data, but are beyond the scope of the present study.

\begin{table}[h!]
\centering
\begin{tabular}{|c|c|c|c|}
  \hline
  Parameter & IBL [$\times 10^{-2} \text{cm}^{-1}$] & $B$-layer [$\times 10^{-2} \text{cm}^{-1}$] & ROSE Coll. [$\times 10^{-2} \text{cm}^{-1}$]	\\
   \hline	
   $g_\text{A}$ & $1.4\pm0.5$ & $1.4\pm0.5$ & 1.4 ($n$)  \\
   $g_\text{Y}$ & $6.0 \pm 1.6$& $6.0\pm 1.6$ & 2.3 ($p$), 4.8 ($n$) \\
   $g_\text{C}$ & $1.1\pm 0.3$ & $0.45\pm 0.1$ & 0.53 ($p$), 2.0 ($n$) \\
  \hline  
\end{tabular}
\caption{Introduction rates of the Hamburg model as obtained by adjusting the simulated depletion voltage to the available measurements.  For comparison, in the last column the 
values reported by the ROSE Collaboration~\cite{Lindstrom:421210} are listed for oxygen-enriched silicon, separately for protons ($p$) and neutrons ($n$).}
\label{tab:depletionvoltage:introductionrates}
\end{table}

Table~\ref{tab:NeffHamburg} collects predictions from the Hamburg model for the effective doping concentration $N_\text{eff}$ in the IBL for two points in time based on the parameter values discussed above, corresponding to lifetime fluence values of $1\times 10^{14}$ and $2\times 10^{14}$~$\text{n}_\text{eq}/\text{cm}^2$, respectively. The thermal history 
of the IBL modules was taken into account.  The uncertainty column includes all contributions shown in Figure~\ref{fig:electricfield:depletionvoltage}.  For the uncertainty in the depletion voltage fit to the introduction rates, one parameter is varied at a time until there is a large deviation; for the luminosity-to-fluence conversion uncertainty, the fluence is varied by $\pm15\%$ (see Section~\ref{sec:fluence}), and for the temperature uncertainty, the input temperature in all phases is varied by $\pm 5^{\circ}$C.  All three sources are added in quadrature to determine the total uncertainty.

\begin{table}[!htbp]
\centering
\begin{tabular}{|c|c|c|c|c|c|c|}
  \hline
     $\Phi$ [n$_\text{eq}/\text{cm}^{2}$] &Approx. date & $N_\text{eff}$ [cm$^{-3}$] & $N_\text{eff}$ uncert. [\%]
   & $N_\text{D}$ [cm$^{-3}$]	& $N_\text{A}$ [cm$^{-3}$] & $V_\text{depl}$  [V]	\\
   \hline	
 1$\times10^{14}$ &  9/7/2016 & $-1.62\times10^{12}$ & 9 &0.02$\times10^{12}$ & $-1.64\times10^{12}$ & 50 \\
 2$\times10^{14}$ &  8/9/2016 & $-2.72\times10^{12}$ & 21 & 0.02$\times10^{12}$ & $-2.74\times10^{12}$ & 85\\
  \hline  
\end{tabular}
  \caption{\label{tab:NeffHamburg}Nominal predictions from the Hamburg model for the effective doping concentration $N_\text{eff}$ and for donor (acceptor) concentration $N_\text{D(A)}$ for two points in time during Run 2.  The value of $N_\text{D}$ was chosen to be numerically small (for technical reasons, it cannot be exactly zero) and the actual value has little impact on the result.  The fluence  2$\times10^{14}$ n$_\text{eq}/\text{cm}^{2}$ was reached near a time of annealing where the effective doping concentration changed by about 4\% over a short period in fluence.  The reported doping concentration and corresponding bias voltage correspond to approximately the midpoint of the concentration during this brief period.}
\end{table}

The operational conditions of the sensor bulk studied in this section are crucial inputs to the simulation of digitization to be presented in Section~\ref{sec:fullmodel}.  Overall, the Hamburg model provides an excellent description of the shape of the leakage current dependence on time; FLUKA + Pythia 8 predict the fluence at $|\eta|\approx 0$ within 15\%, but deviate much more at higher $|z|$.  Even though the Hamburg model does not incorporate a non-uniform electric field, it accurately describes the depletion voltage dependence on time at the current fluence levels.  This may need to be revisited in Run 3 when there will be significant distortions in the space-charge density spatial distribution due to radiation damage.

\section{Digitizer model}
\label{sec:fullmodel}

\subsection{Overview}

\begin{figure}[h!]
\centering
  \subfloat[]{\includegraphics[width=0.7\textwidth]{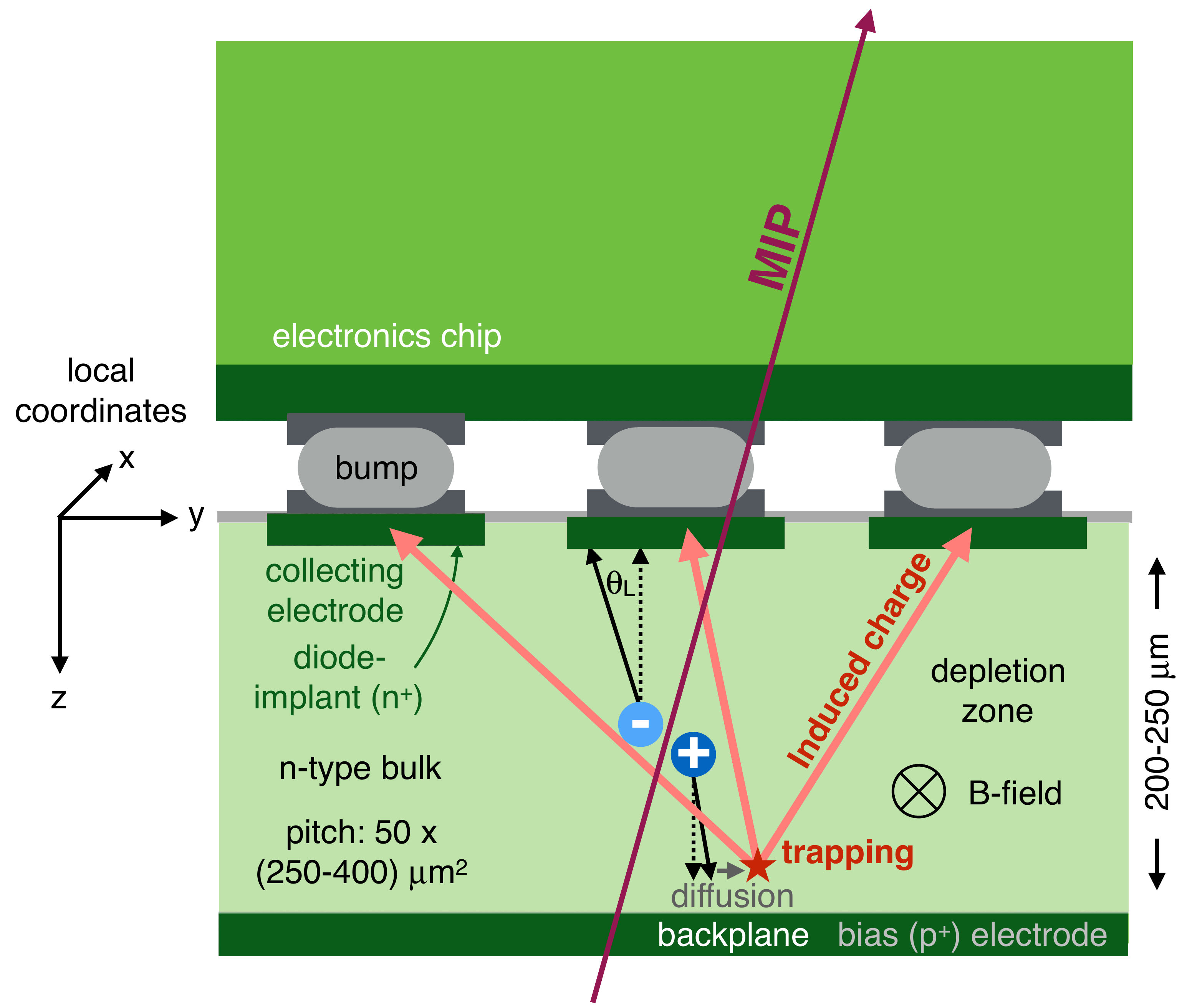}}\\
  \subfloat[]{\includegraphics[width=0.7\textwidth]{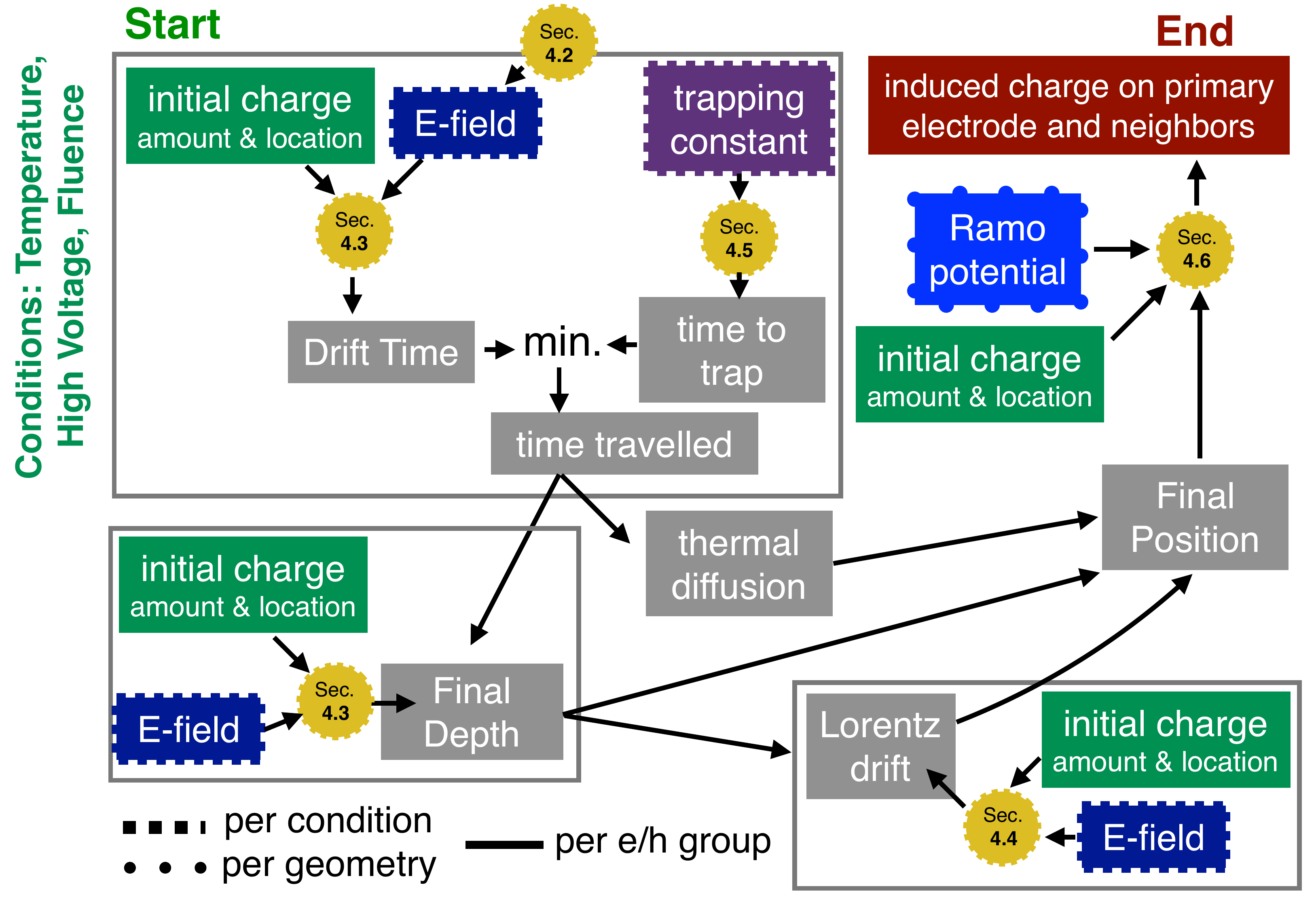}}

  \caption{ (a) A schematic diagram of the digitizer physics.  As a MIP traverses the sensor, electrons and holes are created and transported to the electrodes under the influence of electric and magnetic fields.  Electrons and holes may be trapped before reaching the electrodes, but still induce a charge on the primary and neighbour electrodes. (b) A flowchart illustrating the components of the digitizer model described in this article.  The digitizer takes advantage of pre-computation to re-use as many calculations as possible.  For example, many inputs are the same for a given condition (temperature, bias voltage, fluence).  The Ramo potential~\cite{Shockley,Ramo} only depends on the sensor geometry and the quantities in dashed boxes further depend only on the condition information (see also Section~\ref{sec:ramo}).  The output of the algorithm described in this paper is an induced charge on the primary electrode and the neighbours, which is then converted into a ToT by the existing software.}
\label{fig:app:raddamge1}
\end{figure}

Figure~\ref{fig:app:raddamge1} presents a schematic overview and flowchart of the physics models included in the digitization model.   Upon initialization, the digitizer receives global information about the detector geometry (pixel size and type, tilt angle) and conditions, including the sensor bias voltage, operating temperature, and fluence.  For the calculation of individual charge deposits within the pixel sensor, the digitizer takes as input the magnitude and location of energy deposited by a charged particle, and outputs a digitized encoding of the measured charge.  The input is produced by Geant4 with possible corrections for straggling in thin silicon~\cite{Bichsel:1988if}.  A TCAD tool is used to model the electric field, including radiation damage effects (Section~\ref{sec:ElectricField:SimulationDetails}). 

The TCAD simulations consider a limited number (2--3) of effective deep defects that capture the modification of macroscopic variables such as leakage current, operational voltage and charge collection efficiency.  Ionization energy is converted into electron--hole pairs ($\sim 3.65$~eV/pair) which experience thermal diffusion and drift in electric and magnetic fields.  In order to speed up the simulation, groups of $\mathcal{O}(10)$ charge carriers drift and diffuse toward the collecting electrode (electrons) or back plane (holes), with a field- and temperature-dependent mobility.  Charge groupings are chosen by dividing the deposited energy into a fixed number of pieces.  The number of fundamental charges per charge grouping is set to be small enough so that the overestimation of fluctuations is negligible.\footnote{Suppose a fraction $(1-p)$ of electrons are trapped while drifting toward the electrode, assuming $p$ is constant for illustration and ignoring holes.  If $n$ electrons are deposited, the number of electrons that reach the electrode is $np$ on average with a variance of $p(1-p)n$.  If instead $m<n$ charge groupings with $n/m$ electrons per group are propagated and have trapping probability $p$, the average number of electrons that reach the electrode is still $mp\times (n/m)=np$ but the variance is $p(1-p)n^2/m\geq p(1-p)n$.}  For each charge group, a fluence-dependent time-to-trap (Section~\ref{sec:chargetrapping}) is randomly generated and compared with the drift time (Section~\ref{sec:maps}). If the drift time is longer than the time-to-trap, the charge group is declared trapped, and its trapping position is calculated. Since moving charges induce a current in the collecting electrode, a signal is induced on the electrodes also from trapped charges during their drift.  This induced charge also applies to neighbouring pixels, which contributes to charge sharing.  The induced charge is calculated from the initial and trapped positions using a weighting (`Ramo') potential (Section~\ref{sec:ramo}).  The total induced charge is then converted into a ToT that is used by cluster and track reconstruction tools.

The schematic diagram in Figure~\ref{fig:app:raddamge1}(a) shows a planar sensor, but the digitization model also applies for 3D sensors.  In the simulation, the only differences between planar and 3D pixels 
are that different TCAD models are used (Section~\ref{sec:ElectricField:SimulationDetails}) and charge carrier propagation occurs in two dimensions (transverse to the implants) instead of one (perpendicular to the electrodes).
  The digitization model description (Section~\ref{sec:fullmodel}) and validation (Section~\ref{sec:validation}) focus on planar sensors, in part because they constitute most of the current ATLAS pixel detector and the 3D sensors are formally outside of the tracking acceptance ($|\eta|<2.5$).  Some 3D sensor simulation results are nonetheless described in Section~\ref{sec:3D} in order to highlight the main differences relative to planar sensors.

\clearpage

\subsection{Electric field}
\label{sec:EField}

The radiation-induced states in the silicon band gap affect the electric field in the pixel cells by altering the electric field distribution in the bulk.\footnote{There are also changes at the surface, but the focus here is on the deformations of the electric field within the sensor.}  Since the signal formation in silicon sensors depends directly and indirectly on the electric field shape (Sections~\ref{sec:maps},~\ref{sec:mapsLorentz}), a careful parameterization of the field profile is required.  Section~\ref{sec:ElectricField:SimulationDetails} introduces the default two-trap TCAD model used for subsequent studies. The resulting field profiles are shown in Section~\ref{sec:Efieldprofile}.  In Section~\ref{sec:Efieldmodelcomparisons}, systematic 
uncertainties in electric field profiles determined by TCAD simulations
are discussed.   This section ends with the presentation of a method to incorporate annealing in Section~\ref{sec:TCADAnneal}.

\subsubsection{Simulation details}
\label{sec:ElectricField:SimulationDetails}

Since the charge collection is significantly different in planar and 3D sensors due to the different electrode geometries, two different set-ups are used to implement the radiation damage in the simulation model.  The simulation is set up for both sensor types, but the focus is on planar sensors as the 3D sensors are outside of the standard $|\eta| < 2.5$ tracking acceptance.  Validation studies in Section~\ref{sec:validation} are therefore only presented for planar sensors.  In TCAD simulations, impurities are only added and not removed;\footnote{This is because structure simulation and device simulation are two separate processes in TCAD.  See e.g.\ Section~3.3.1 in Ref.~\cite{silvacomanual}.} therefore one must balance initial shallow defects with radiation-induced defects.  As a result, different TCAD models of effective defect states are used for each bulk type.  Since the planar sensors are $n$-type and the 3D sensors are $p$-type, different TCAD models are used for the two sensors.  Details for the 3D sensor simulation can be found in Section~\ref{sec:3D}.

Investigations of the electric field profile in the bulk of irradiated silicon sensors have shown that the electric field is no longer linear with the bulk depth after irradiation (see, for example, Refs.~\cite{bib:DP,Chiochia2}). Irradiated planar sensors with non-linear profiles are simulated using the Chiochia model~\cite{Chiochia2}, implemented in the Silvaco TCAD package~\cite{silvaco,silvacomanual}.  The Petasecca~\cite{1710302} $n$-type model was also investigated, but was found to not predict space-charge sign inversion below $\Phi=1\times 10^{14}$~$\text{n}_\text{eq}/\text{cm}^2$ and was therefore not considered further.

The simulation is performed over an area that corresponds to a quarter of an ATLAS IBL pixel sensor cell, to take advantage of symmetry.  The electric field is computed at $T=-10^{\circ}$C using an effective doping concentration of $1.6\times 10^{12}/\text{cm}^3$ (corresponding to about 50~V full depletion voltage for unirradiated sensors~\cite{Wittig}) with a discretization resolution of $1$~$\text{\textmu m}^2$.   During Run 2, the operational temperature of the pixels was adjusted multiple times.  For example, the IBL temperature was set to $-4^\circ$C in 2015, +20$^\circ$C for the first part of 2016, +10$^\circ$C for the rest of 2016, and was $-15^\circ$C in 2017.  The TCAD simulations are all performed at $-10^\circ$C since this is where the models were developed.  A naive temperature variation from scaling the trap occupation probability according to $\exp(-E_\text{t}/k_\text{B}T)$ ($E_\text{t}$ is the trap energy) predicts variations in the leakage current that are about 20\% larger than the observations.\footnote{TCAD simulations with the Chiochia model were performed at $\Phi=10^{14}$~$\text{n}_\text{eq}/\text{cm}^2$ at 150~V and between $-20^\circ$C and 20$^\circ$C.  The rescaling factor between $-10^\circ$C and the standard 20$^\circ$C is 20\% lower with the Chiochia model compared with other studies in the literature~\cite{Chilingarov_tscale}.}  The reason is that the TCAD models only include a small number of effective states, and in reality the temperature dependence is reduced when a more complex (but computationally intractable) combination of states is present.  The trap energy level $E_\text{t}$ varies by 10\% of thermal energy $k_\text{B}T$ (see Section~\ref{sec:Efieldmodelcomparisons}) are found to be consistent with naive temperature variations that bracket all Run~2 operational temperatures ($-15^{\circ}$C to $+20^{\circ}$C) and therefore provide a conservative bound on the predictions presented in Section~\ref{sec:validation}.  In the future, high-statistics collision data may be used to tune models in situ and avoid this complication.

The Chiochia model is a double-trap model with one acceptor and one donor trap with activation energies set to $E_\text{c}-0.525$~eV and $E_\text{v}+0.48$~eV~\cite{bib:DP} for the conduction band energy level $E_\text{c}$ and the valence band energy level $E_\text{v}$, respectively. This model was developed using CMS diffusion oxygenated float zone $n^+$-in-$n$ pixel module prototypes and was chosen since the bulk material type is the same as in the ATLAS IBL and pixel layers and the initial effective doping concentration is similar: 50~V depletion voltage for the ATLAS IBL and 75~V for CMS with slightly thicker sensors.  Sensor annealing in Ref.~\cite{Chiochia2} is different than for the operational ATLAS detector, but the partially unaccounted annealing situation is incorporated in the model variations discussed in Section~\ref{sec:Efieldprofile}.  Table~\ref{tab:Chiochia2006} documents the radiation damage model parameter values used for the planar sensor TCAD simulations.  Note that the concentrations are not comparable to the ones from the model presented in Table~\ref{tab:NeffHamburg} because the defects presented here are deep traps while the ones for the depletion voltage Hamburg model are shallow and thus have a higher occupation probability. 

\begin{table}[!htpb]
\centering
  \begin{tabular}{|c|c|c|c|c|c|c|c|}   
    \hline
  \begin{tabular}{@{}c@{}}$\Phi$ \\ $[{\textrm n_\text{eq}/\text{cm}^{2}}]$ \end{tabular}     & \begin{tabular}{@{}c@{}} $N_\text{A}\times10^{-15}$ \\ $[{ \text{cm}^{-3}}]$ \end{tabular}&  \begin{tabular}{@{}c@{}} $N_\text{D}\times10^{-15}$ \\ $[{ \text{cm}^{-3}}] $\end{tabular} & \begin{tabular}{@{}c@{}}  $\sigma_{e}^\text{A/D}\times10^{15}$ \\ $[{ \text{cm}^{2}}]  $\end{tabular} &  \begin{tabular}{@{}c@{}} $\sigma_{h}^\text{A}\times10^{15}$ \\ $[{ \text{cm}^{2}}]$\end{tabular} &  \begin{tabular}{@{}c@{}} $\sigma_{h}^\text{D}\times10^{15}$ \\ $[{ \text{cm}^{2}}]$\end{tabular} & \begin{tabular}{@{}c@{}}$g^\text{A}_\text{int}$ \\ $[{\text{cm}^{-1}}]$\end{tabular}&\begin{tabular}{@{}c@{}} $g^\text{D}_\text{int}$\\ $[{\text{cm}^{-1}}]$\end{tabular} \\
    \hline
    $1\times10^{14}$ & $0.36$ & $0.5$ & $6.60$ & $1.65$  & $6.60$ &3.6 &5  \\
    $2\times10^{14}$ & $0.68$ & $1$ & $6.60$ & $1.65$  & $6.60$ &3.4 &5 \\
    $5\times10^{14}$ & $1.4$ & $3.4$ & $6.60$ & $1.65$  & $1.65$ &2.8  &6.8\\
    \hline
  \end{tabular}
  \caption{\label{tab:Chiochia2006}Values used in TCAD simulations for 
  deep acceptor (donor) defect concentrations $N_\text{A}$ ($N_\text{D}$) and for their 
  electron (hole) capture cross sections ($\sigma_{e,h}^\text{A,D}$) for three different fluences. Values are derived from the Chiochia model~\cite{Chiochia2} for temperature $T=-10^{\circ}$C.  Reference~\cite{Chiochia2} gives values for $\Phi=0.5\times 10^{14}$, $2\times 10^{14}$, and $5.9\times 10^{14}~\text{n}_\text{eq}/\text{cm}^{2}$.  In between the reported values, the interpolated value is given by the average of the neighbouring low and high fluence points scaled to the target fluence: $N_\text{A/D}(\Phi)=\frac{1}{2}(N_\text{A/D}(\Phi_\text{low})/\Phi_\text{low}+N_\text{A/D}(\Phi_\text{high})/\Phi_\text{high})\Phi\equiv g_\text{int}\Phi$, where $g_\text{int}$ is the effective introduction rate.  For fluences below $0.5\times 10^{14}$ or above $5\times 10^{14}$~$\text{n}_\text{eq}/\text{cm}^{2}$, the value is scaled, based on the nearest reported value: $g_\text{int}=N_\text{A/D}(\Phi_\text{bench})/\Phi_\text{bench}$, where $\Phi_\text{bench}$ is the nearest reported fluence.}
\end{table}

\subsubsection{Electric field profiles}
\label{sec:Efieldprofile}

\begin{figure}[h!]
\centering
  \subfloat[]{\includegraphics[width=0.5\textwidth]{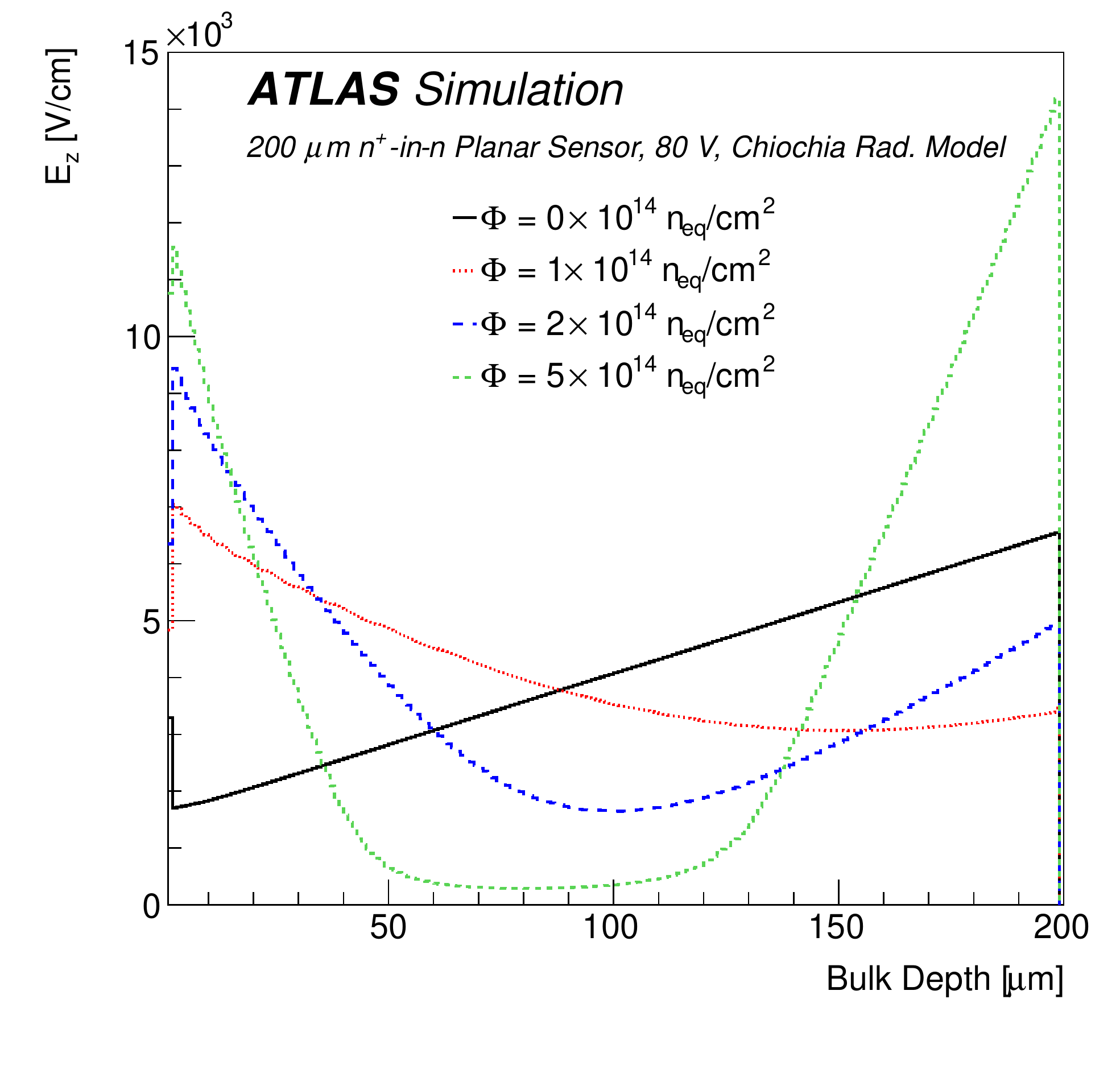}}  \subfloat[]{\includegraphics[width=0.5\textwidth]{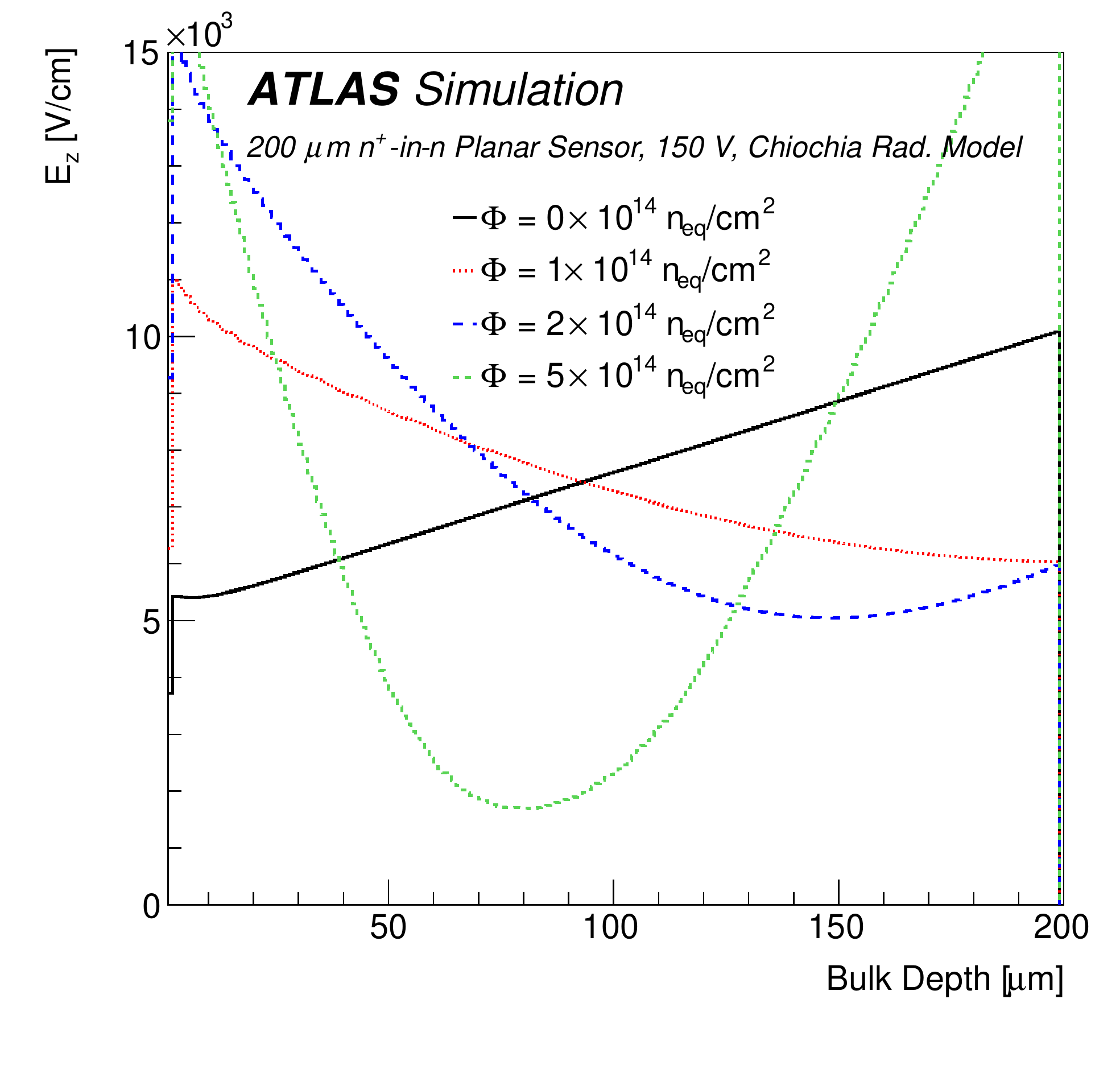}}
\caption{The simulated electric field magnitude in the $z$ direction along the bulk depth, averaged over $x$ and $y$ for an ATLAS IBL sensor biased at: (a) 80~V and (b) 150~V for various fluences.}
\label{fig:electricfield:profilesRun23}
\end{figure}

For planar sensors, the field is largely independent of $x$ and $y$, and perpendicular to the sensor surface. Figure~\ref{fig:electricfield:profilesRun23} shows the $z$-dependence of the electric field, averaged over $x$ and $y$, for an ATLAS IBL planar sensor for various fluences and bias voltages of 80~V and 150~V.  Before irradiation the field is approximately linear as a function of depth.  Just after type inversion (at about $2\times 10^{13}$~$n_\text{eq}/\text{cm}^2$), the field maximum is on the opposite side of the sensor.  With increasing fluence, there is a minimum in the electric field in the centre of the sensor.  For a fluence of $\Phi=5\times 10^{14}$~$\text{n}_\text{eq}/\text{cm}^2$ and a bias voltage of 80~V (for which the sensors are not fully depleted, as shown in Section~\ref{sec:annealing}), this minimum is broad and occupies nearly a third of the sensor.
The rest of the section considers bias voltages of 80~V and 150~V as they were the operational voltages of IBL planar sensors 
in 2015 and 2016, respectively.

\subsubsection{Electric field profile uncertainties}
\label{sec:Efieldmodelcomparisons}
In this section, systematic uncertainties in the electric field profiles evaluated 
using TCAD simulations are discussed. This includes studying other radiation
damage models for TCAD simulations as well as the effect of varying the Chiochia model 
parameter values.

Extensive model comparisons are beyond the scope of this work, but the data presented in Section~\ref{sec:validation} can be used to constrain various simulations as well as tuning parameters and to derive systematic uncertainties for predictions for higher-luminosity data.  In addition to the Chiochia model for the planar sensors, the Petasecca model~\cite{1710302} was also briefly investigated.  While the model itself is supported by test-beam data, it is found to disagree qualitatively on the fluence for type-inversion with the Chiochia model\footnote{Around $3\times 10^{14}$~$\text{n}_\text{eq}/\text{cm}^2$ (Petasecca) versus $5\times 10^{13}$~$\text{n}_\text{eq}/\text{cm}^2$ (Chiochia); the IBL inverted around $2\times 10^{13}$~$\text{n}_\text{eq}/\text{cm}^2$ and the $B$-layer inverted around $2$--$3\times 10^{13}$~$\text{n}_\text{eq}/\text{cm}^2$ (based on the measurements presented in Figure~\ref{fig:electricfield:depletionvoltage}).} and does not reproduce the observed trend of the Lorentz angle data as described later in Section~\ref{sec:lorentzangle}.  Therefore, this alternative model was not studied in further detail.

Next, the Chiochia model parameters are varied.
Each parameter (capture cross sections and introduction rate) is varied by $\pm$10\% of its value except the trap energy level  $E_\text{t}$, which is varied by $\pm$10\% of the thermal energy 
$V_\text{th}=k_\text{B}T$. The energy of the trap $E_\text{t}$ is defined as the energy difference between the trap 
and the relevant band (conduction for the acceptor-like trap and valence 
for the donor-like trap).  The value 10\% was chosen for illustration in the absence of experimental input; ideally future models or model tunings will provide quantitative uncertainty estimates.

\begin{figure}[h!]
\centering
  \subfloat[]{\includegraphics[width=0.49\textwidth]{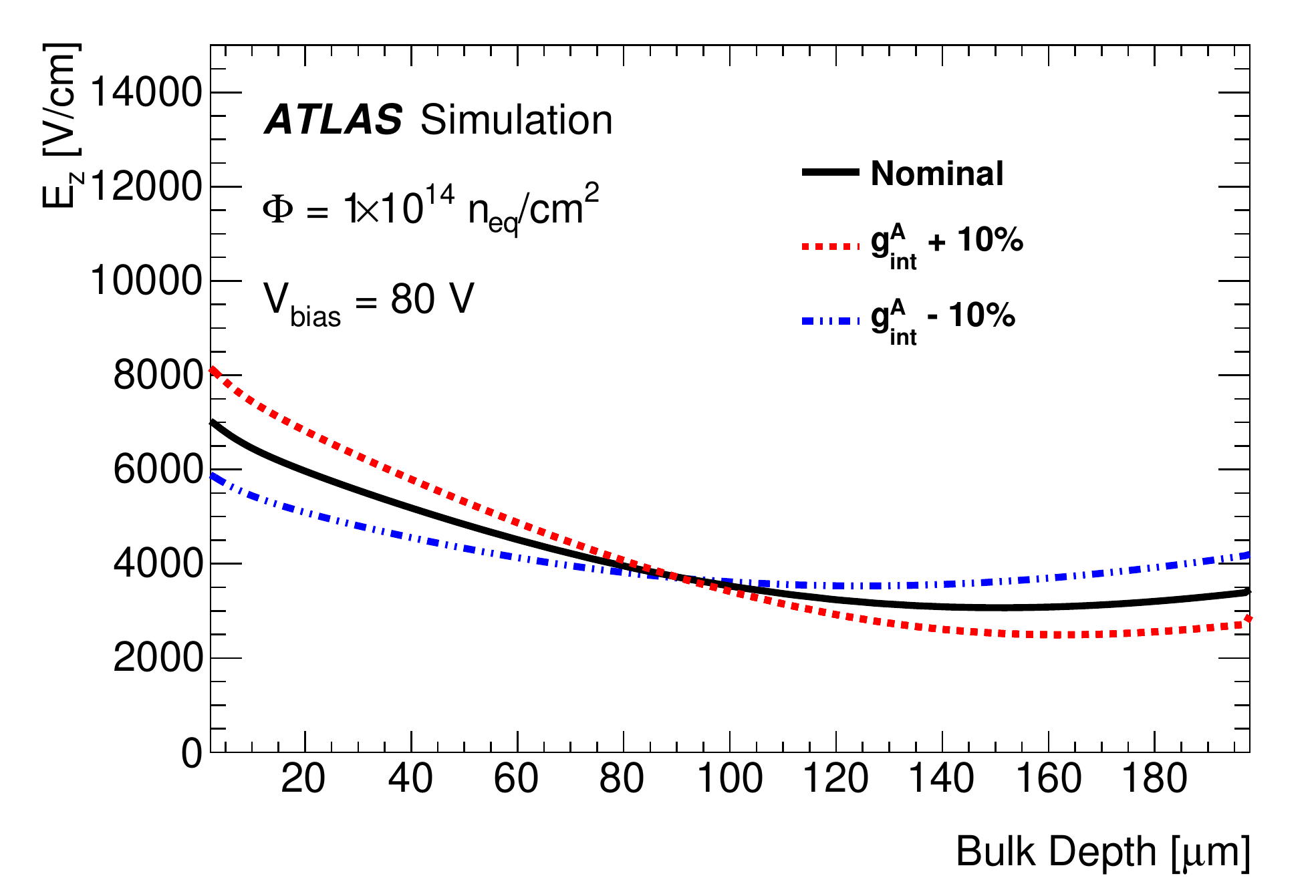}}
  \subfloat[]{\includegraphics[width=0.49\textwidth]{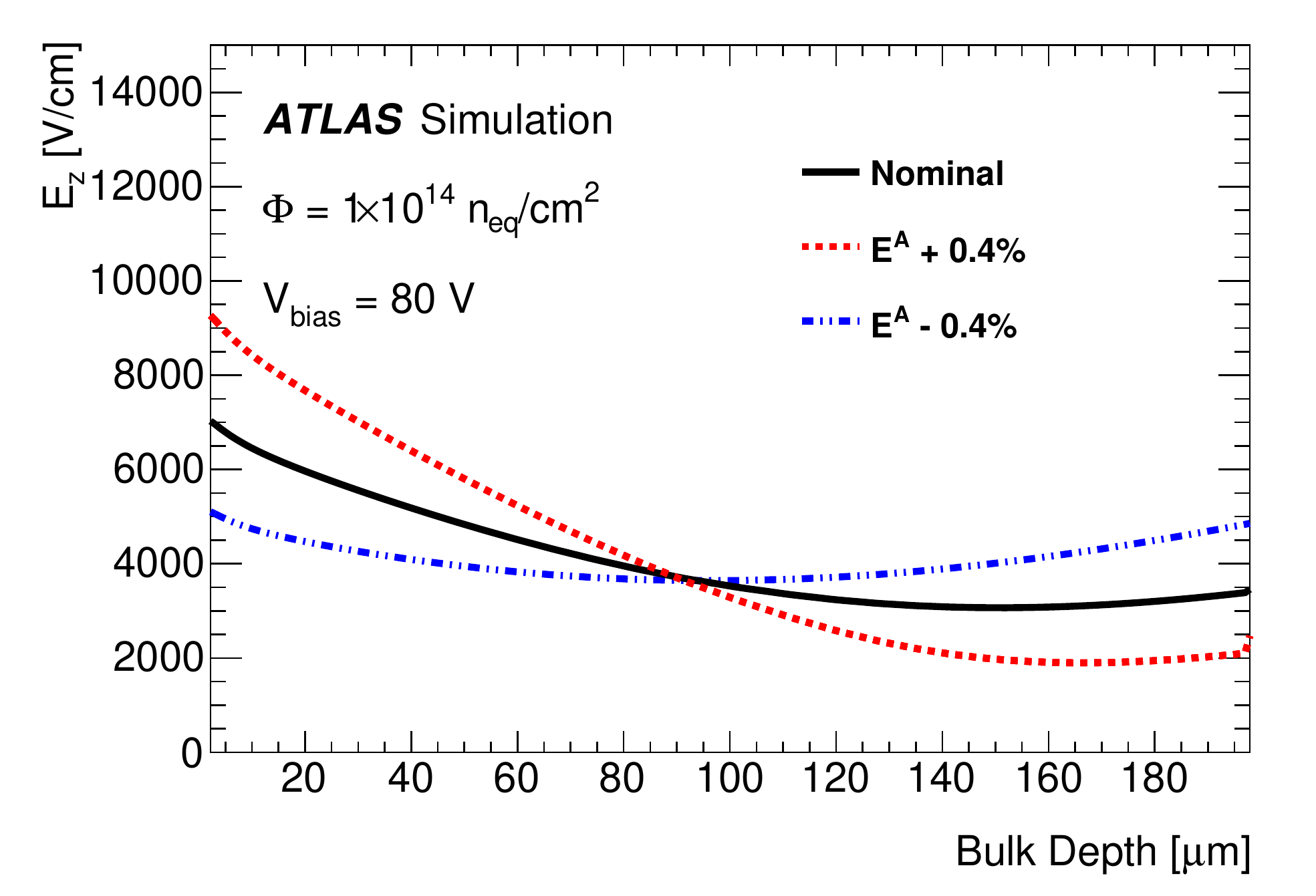}}
\\
  \subfloat[]{\includegraphics[width=0.49\textwidth]{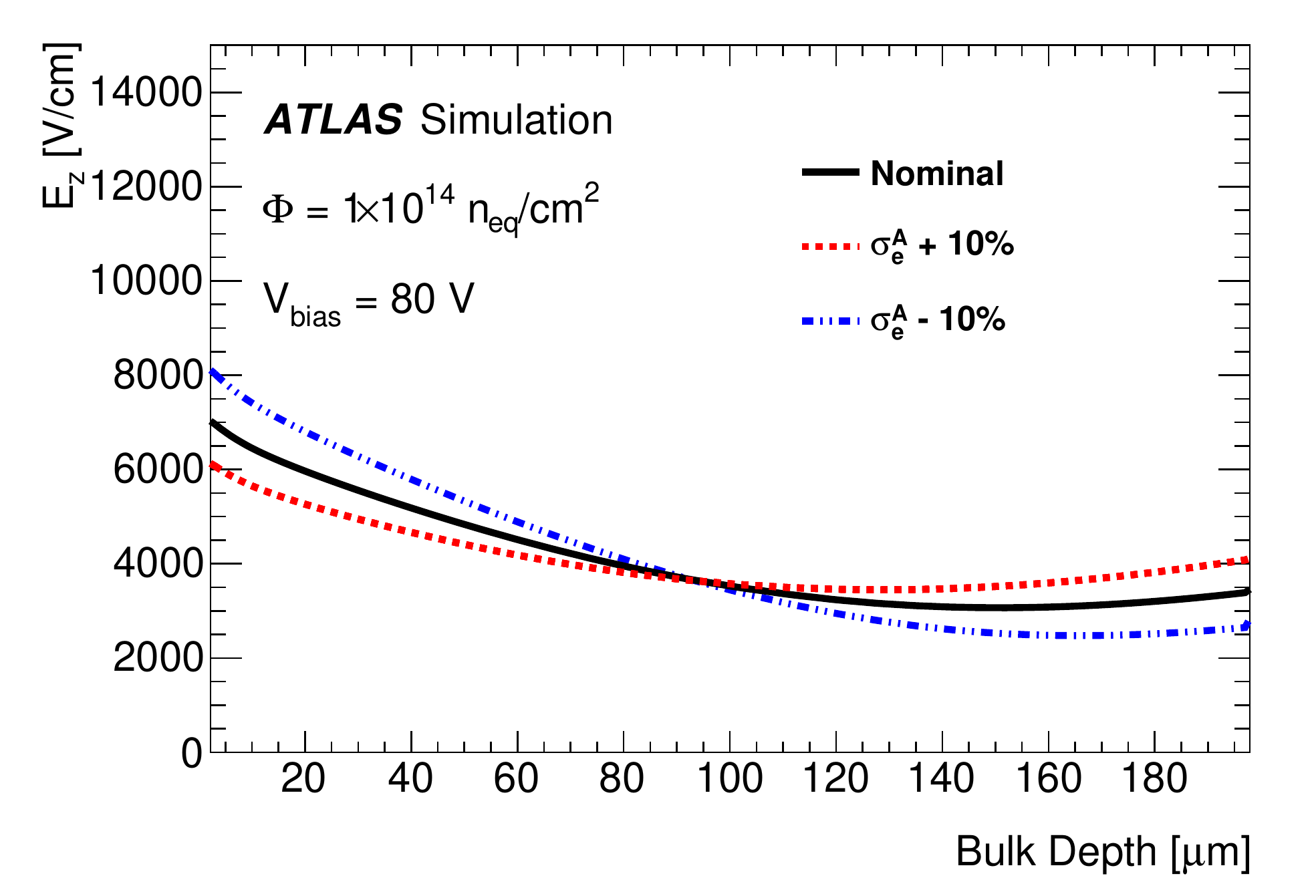}}
  \subfloat[]{\includegraphics[width=0.49\textwidth]{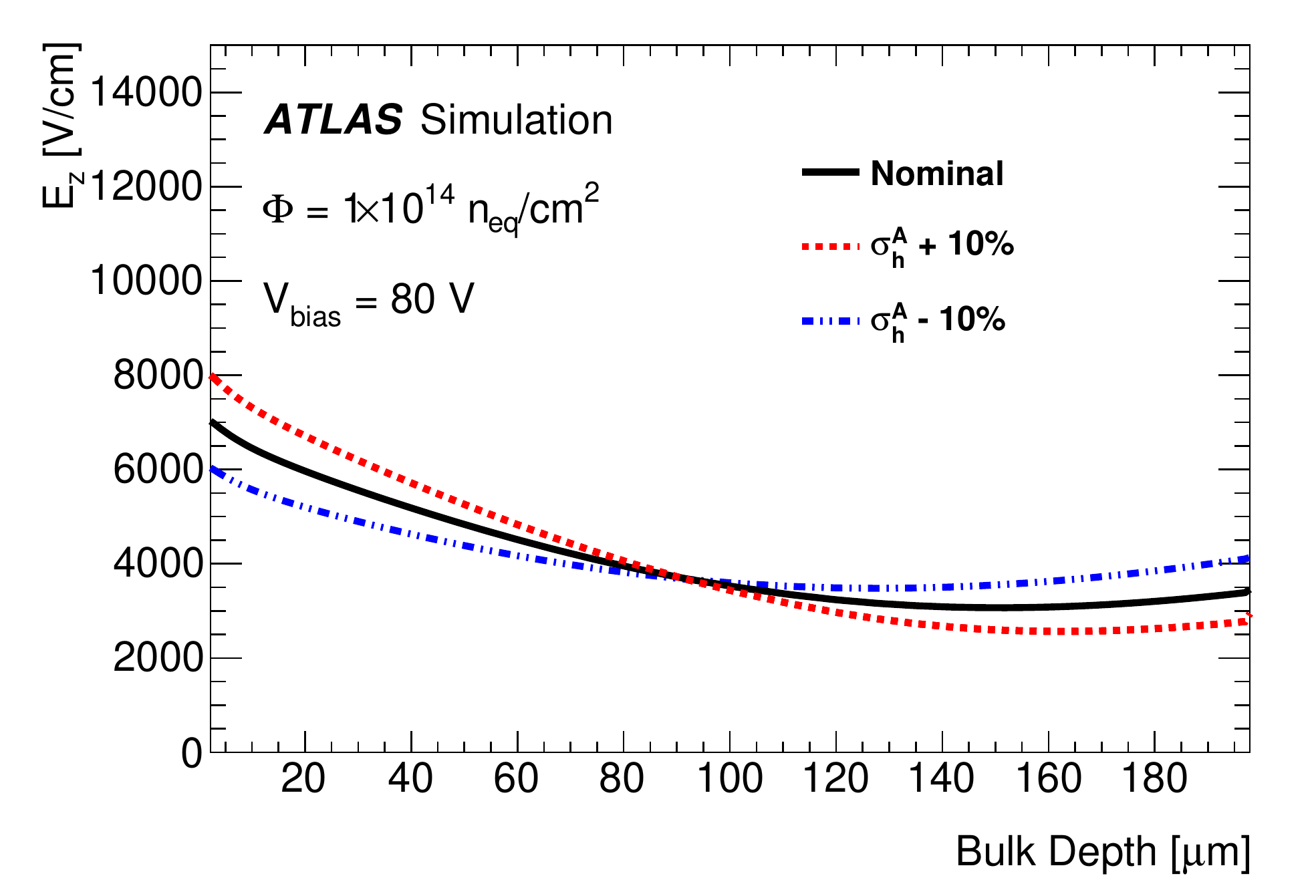}}
\caption{The $z$ dependence of the electric field in an ATLAS IBL planar sensor, averaged over $x$ and $y$, for a simulated
  fluence of $\Phi=1\times10^{14}$~n$_\text{eq}/\text{cm}^{2}$, after varying parameters of the acceptor trap in the Chiochia model. (a) $\pm 10\%$ variation in the fluence dependence ($g^\text{A}_\text{int}$) of the acceptor trap concentrations; (b) variation in the acceptor trap energy level by $0.4\%$ ($0.525 \pm 0.002$~eV from the conduction band level); 
  (c) $\pm 10\%$ variation in the electron capture cross section; (d) $\pm 10\%$ variation in the hole capture cross section.
The bias voltage was set to $80$~V in all cases.}
\label{fig:acc_electricfieldvariations}
\end{figure}

Figure~\ref{fig:acc_electricfieldvariations} shows the electric field for variations in the acceptor trap parameters for a fluence of $10^{14}$~$\text{n}_\text{eq}/\text{cm}^2$ and a bias voltage of 80~V.  The normalization of all the curves is fixed by the bias voltage and therefore all the curves cross at a point. 
Variations in the capture cross sections and introduction rate ($g_\text{int}$) introduce a change in the peak electric field that is between 15\% and 30\%.  Similar variations are observed when the trap concentrations are varied by $\pm10\%$ and the energy levels are varied by $\pm10\%$ of the thermal energy $V_\text{th}$, which corresponds roughly to $0.4\%$ of the energy level.  The latter number is chosen as a benchmark because the occupancy probability scales exponentially with the energy as $\sim \text{e}^{-E_\text{t}/k_\text{B}T}$~\cite{LUTZ1996}.
For example, when the acceptor energy is moved closer to the conduction band by $0.4\%$, the electric field looks 
 symmetric around the mid-plane; moving the acceptor even closer to the conduction band would likely result in depletion 
 starting from the back side.

 As expected, the results for donor traps (not shown) show a behaviour that is opposite to the acceptor case when concentrations are changed. 
All the observed changes in the electric field are consistent with expectations (see for example Ref.~\cite{LUTZ1996}).

\subsubsection{Effective modelling of annealing effects in TCAD simulations}
\label{sec:TCADAnneal} 
\begin{figure}[h!]
\centering
\includegraphics[width=0.54\textwidth]{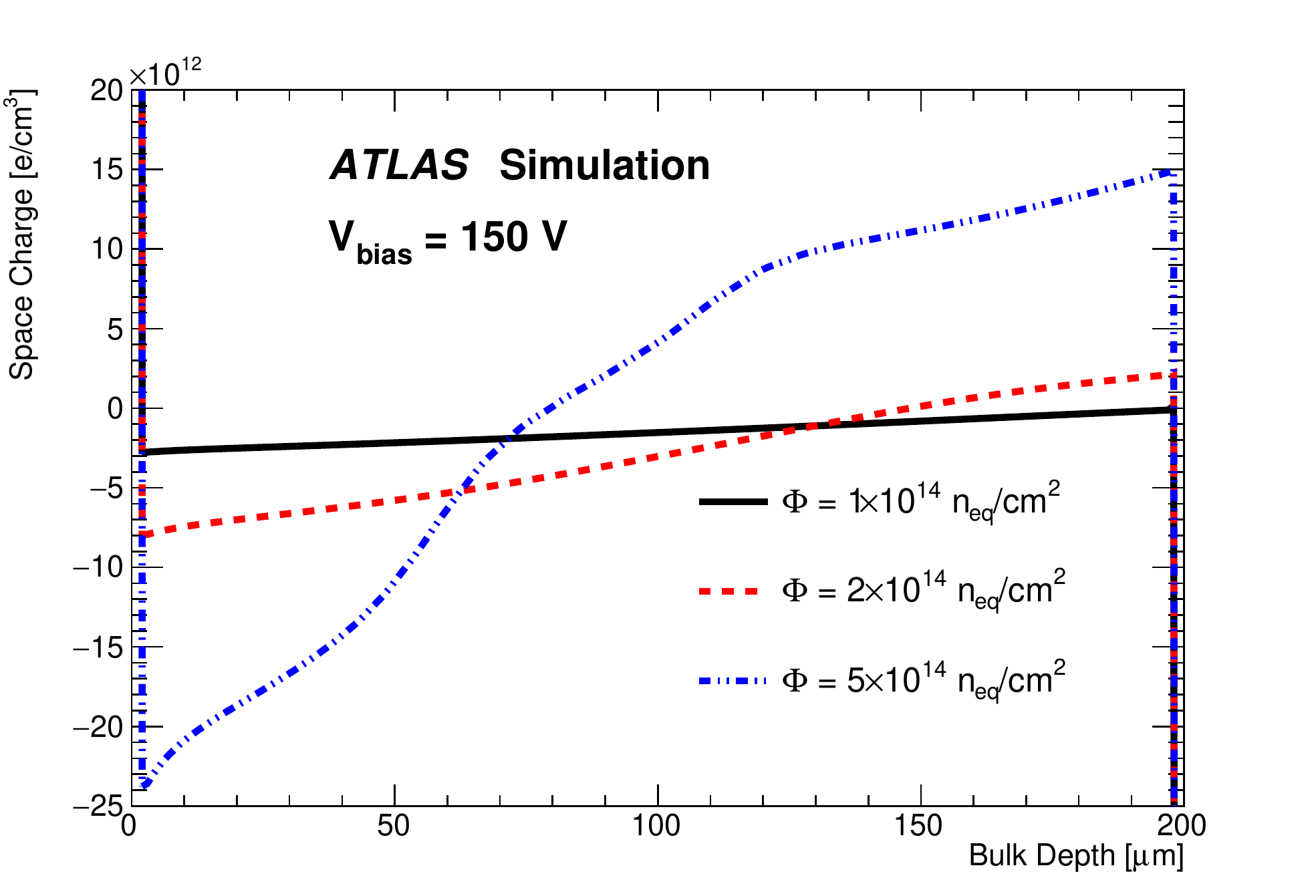}
\caption{\label{fig:TCADSpaceCharge}The $z$ dependence of the 
space-charge density $\rho$ in a simulated 
 ATLAS IBL planar sensor, averaged over $x$ and $y$, for  simulated
  fluences: $1\times10^{14}$, $2\times10^{14}$ and $5\times10^{14}$~n$_\text{eq}/\text{cm}^{2}$. The bias voltage was set to $150$~V in all cases. These are predictions based on the Chiochia model at temperature $T=-10^{\circ}$C.}
\end{figure}

There is no known recipe to include the annealing effects presented in Section~\ref{sec:annealing} in 
TCAD-based predictions.  One challenge for incorporating annealing effects is that both the Hamburg and TCAD models are motivated by multiple effective traps~\cite{Moll:1999kv,bib:DP,Chiochia2} and the effective states are not in one-to-one correspondence (in particular, no cluster defects are directly reproduced by TCAD simulations). In addition to this, the relative abundance of the measured acceptor-like traps changes with annealing.  Lastly, the Hamburg model does not make a prediction for the dependence of the space-charge density on depth while the TCAD model predicts a non-trivial dependence, resulting in the complicated electric field profile discussed in Section~\ref{sec:Efieldprofile}.  The non-constant space-charge density from the TCAD model is shown in Figure~\ref{fig:TCADSpaceCharge} for an ATLAS IBL planar sensor after radiation damage.   For $\Phi=1\times10^{14}$~n$_\text{eq}/\text{cm}^{2}$ the space-charge density is negative and shows an almost linear dependence on the bulk depth, 
whereas for higher fluences the 
functional form is more complicated, exhibiting sizeable regions where the space-charge density is positive, 
in agreement with the model first proposed in Ref.~\cite{bib:DP}. This results in the non-trivial electric field profiles shown in Figure~\ref{fig:electricfield:profilesRun23}. 

 The non-constant space-charge density, despite the simulated traps being uniformly distributed across the sensor bulk, 
 is due to the thermally generated electrons and holes which, drifting in opposite directions and getting trapped along their trajectory, give rise to a more negative (positive) region 
 close to the electrode collecting the electrons (holes)~\cite{bib:DP}. 
  Deviations from linearity in the space-charge density region distribution with respect to the position in the bulk are 
  predicted by the TCAD simulations when the voltage is fixed and the fluence gets larger, as can be seen in Figure~\ref{fig:TCADSpaceCharge}. These deviations can be understood in the following way: as the depletion regions develop from both sides, for fixed voltage and larger fluences, the mid part of the sensors is not depleted. Hence the space-charge density region profile deviates from linearity there.

 One way to emulate annealing effects from the Hamburg model in the TCAD simulation is to match\footnote{They do not agree exactly because the space-charge density in TCAD is dynamically generated and not known a priori.} the effective doping concentration predictions from the former,\footnote{The physical origin of the effective doping concentration is not exactly the same for the Hamburg and TCAD models.  The approach given here is a first approximation that must be expanded upon in the future when annealing effects are much more prominent.} such as the ones presented in Table~\ref{tab:NeffHamburg}, to the average space-charge density (normalized by the electron charge) of the latter: 

\begin{equation}
\langle \rho/e\rangle_\text{TCAD}=(N_\text{eff})_\text{Hamburg}.
\label{eq:SpaceChargeTCADHamburg}
\end{equation}

Two different scenarios to realize the situation described in Eq.~(\ref{eq:SpaceChargeTCADHamburg}) are studied: the \textit{Hamburg} scenario and one in which the concentration of acceptor traps in TCAD simulations $N_\text{A}^\text{TCAD}$ was changed to satisfy Eq.~(\ref{eq:SpaceChargeTCADHamburg}), referred to in the following as \textit{TCAD with effective annealing}.  For the sake of comparison 
a third one was added, called the \textit{Chiochia} scenario, which is the default set-up described in Section~\ref{sec:ElectricField:SimulationDetails} with no modifications to emulate annealing.

For the Hamburg scenario, the concentration of shallow donors in the structure is set to a very low value and the deep acceptor and donor concentrations are adjusted as a function of depth in order to create a constant space-charge density as predicted by the static Hamburg model everywhere in the bulk.
The Hamburg scenario is qualitatively different than the Chiochia one and would predict an electric field that is linear (more below), which is in contrast to various measurements elsewhere~\cite{Chiochia2}.  For the TCAD with effective annealing scenario, the space-charge density can vary in the bulk and Eq.~(\ref{eq:SpaceChargeTCADHamburg}) was solved by varying the acceptor concentrations $N_\text{A}^\text{TCAD}$.

\begin{figure}[h!]
\centering
  \subfloat[]{\includegraphics[width=0.49\textwidth]{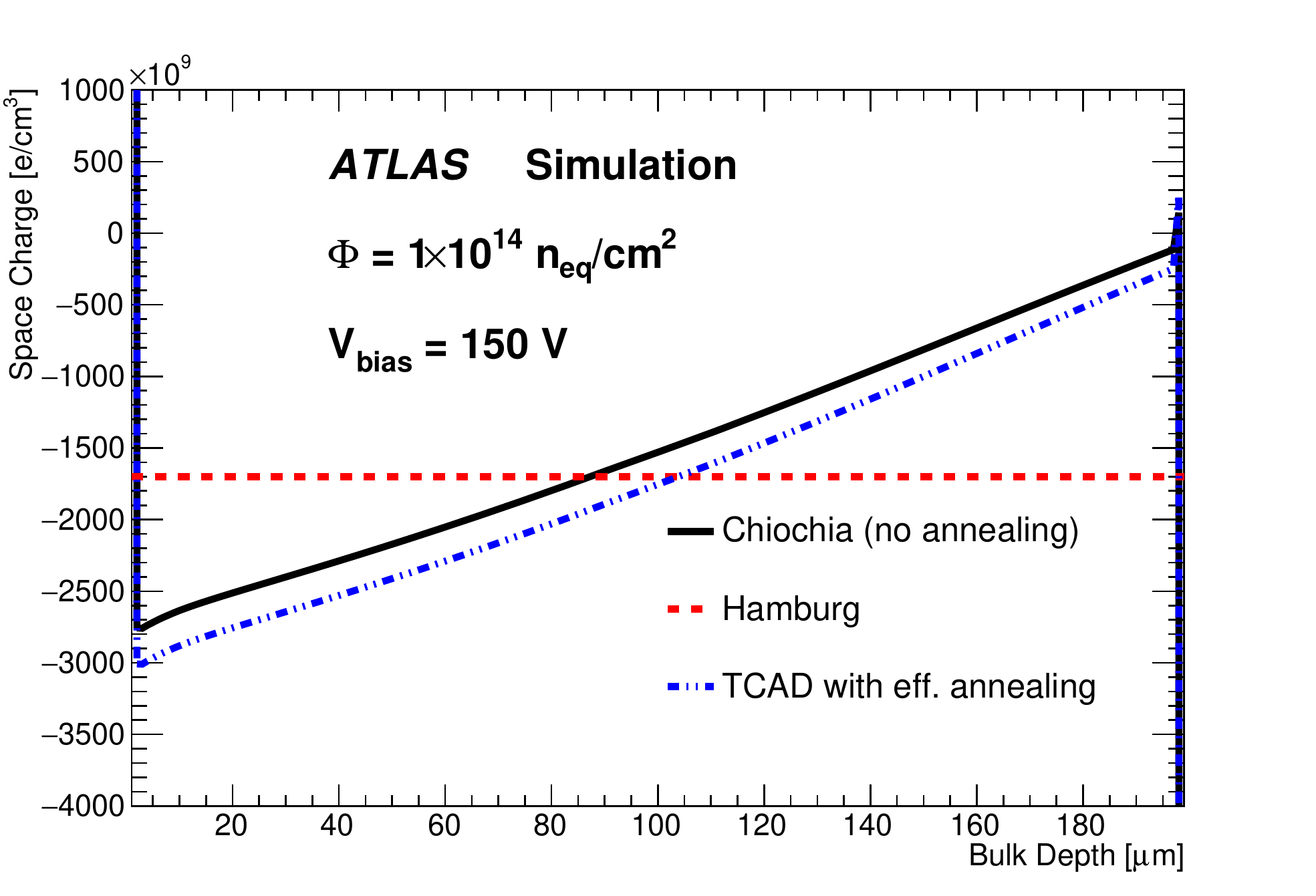}}
\subfloat[]{\includegraphics[width=0.49\textwidth]{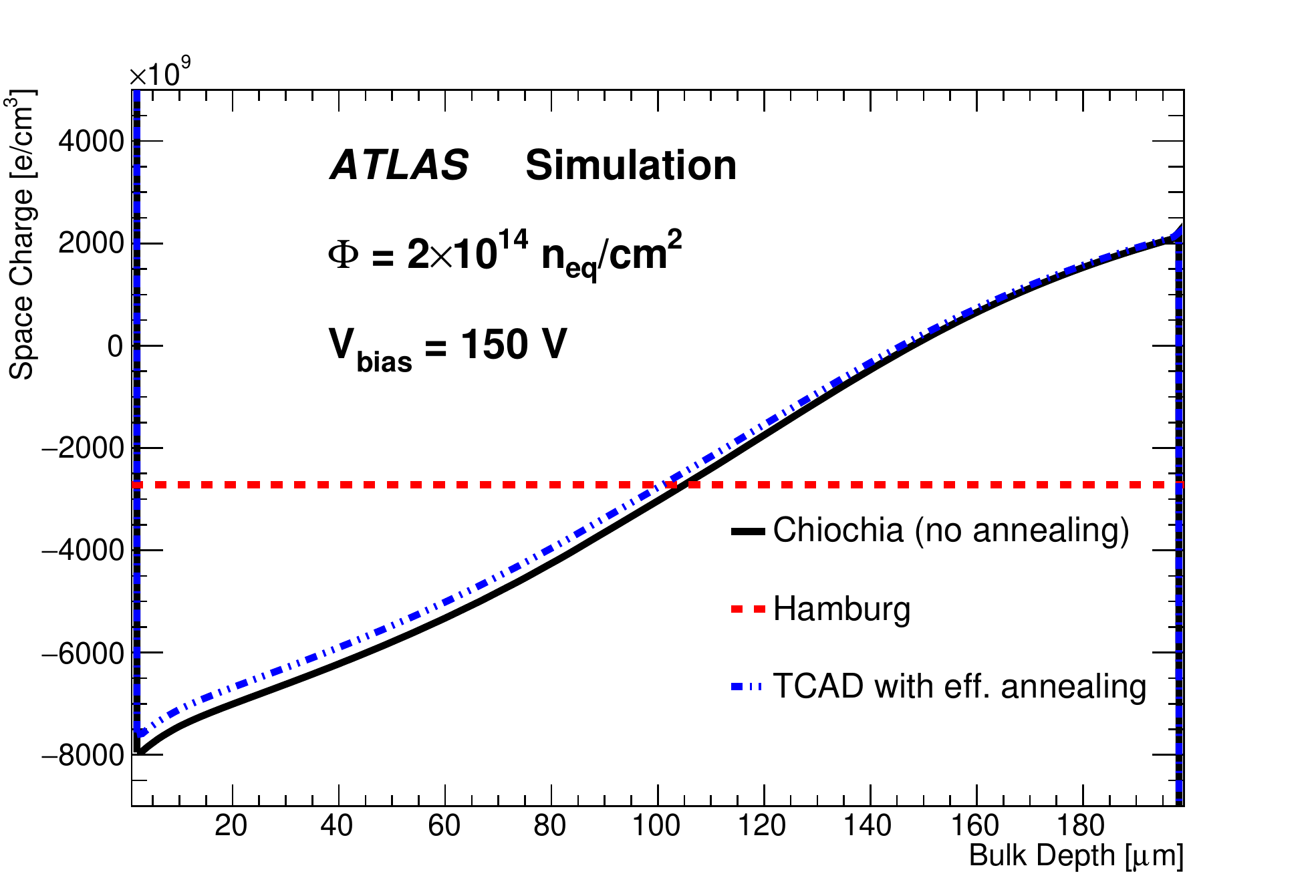}}
\caption{\label{fig:NeffTCADAnnealing}The $z$ dependence of the space-charge density in a simulated ATLAS IBL planar sensor, averaged over $x$ and $y$, for  simulated
  fluences of (a)~$\Phi=1\times10^{14}$~n$_\text{eq}/\text{cm}^{2}$ and (b)~2$\times10^{14}$~n$_\text{eq}/\text{cm}^{2}$. The bias voltage was set to $150$~V in all cases. Three scenarios - Chiochia (no annealing), Hamburg and TCAD with effective annealing - to emulate annealing effects were simulated. }
\end{figure}

\begin{figure}[h!]
\centering
  \subfloat[]{\includegraphics[width=0.49\textwidth]{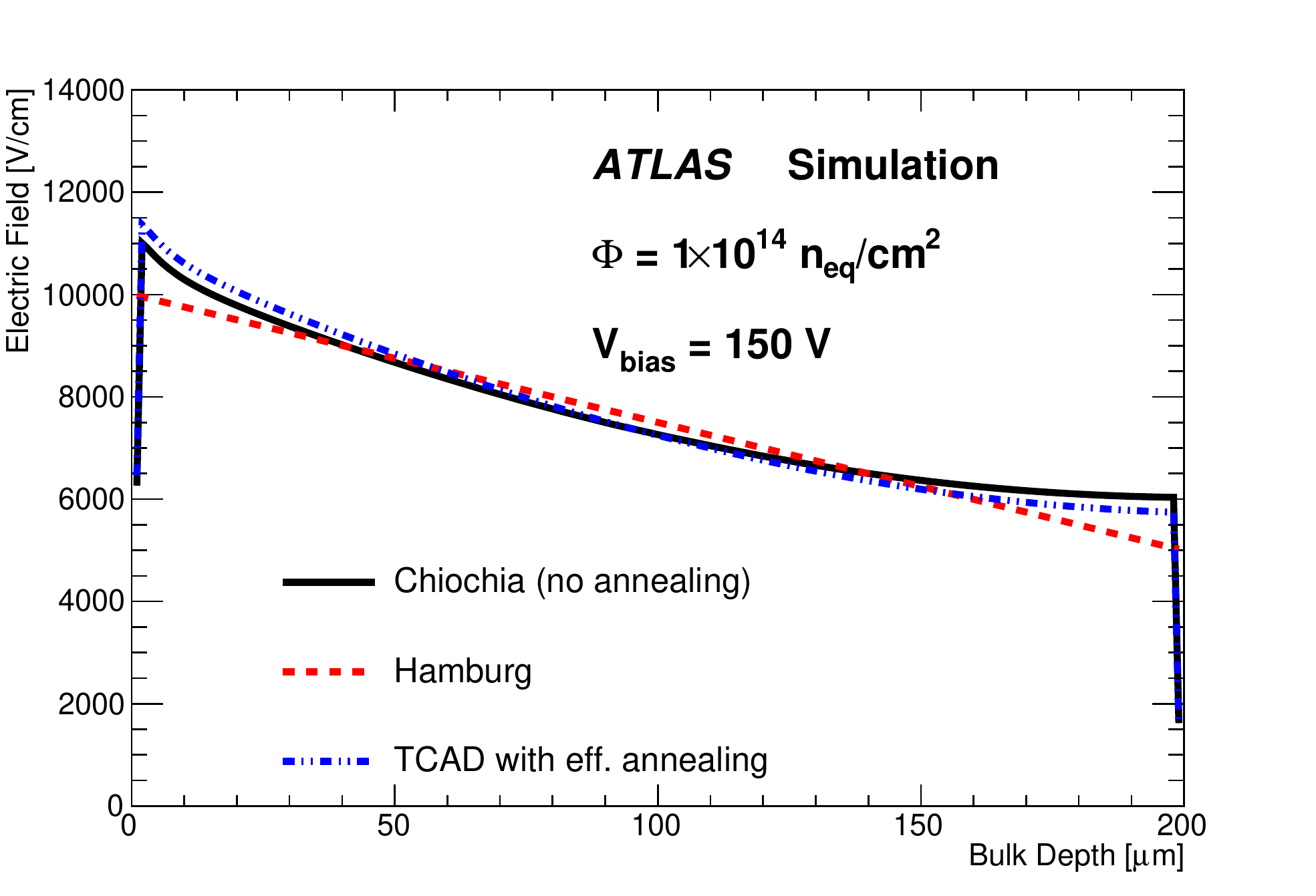}}
  \subfloat[]{\includegraphics[width=0.49\textwidth]{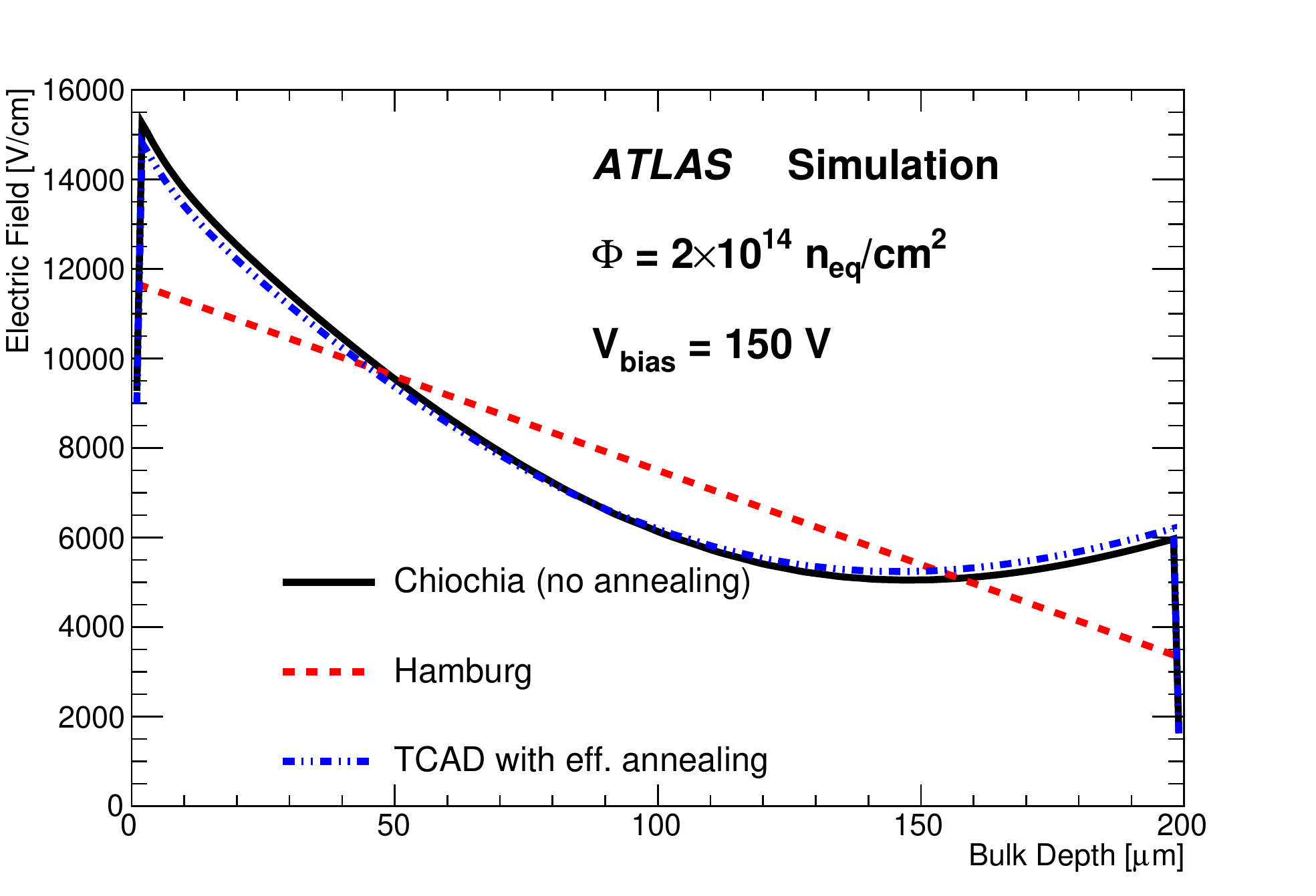}}
\caption{\label{fig:EFieldTCADAnnealing}The $z$ dependence of the electric field in a simulated ATLAS IBL planar sensor, averaged over $x$ and $y$, for  simulated
  fluences of (a)~$\Phi=1\times10^{14}$~n$_\text{eq}/\text{cm}^{2}$~and~(b)~2$\times10^{14}$~n$_\text{eq}/\text{cm}^{2}$. The
bias voltage was set to $150$~V in all cases. Three scenarios - Chiochia (no annealing), Hamburg and TCAD with effective annealing - to emulate annealing effects were simulated.}
\end{figure}

Figure~\ref{fig:NeffTCADAnnealing} shows the space-charge density predicted by TCAD simulations in the three 
scenarios with a bias voltage of 150~V and at the points in the irradiation and temperature history reported in Table~\ref{tab:NeffHamburg}.  The average space-charge density in the various scenarios is summarized in Table~\ref{tab:TCADNeffHamburg1} for the two fluences shown in Figure~\ref{fig:NeffTCADAnnealing}.  The electric field profiles corresponding to the three scenarios shown in Figure~\ref{fig:NeffTCADAnnealing} are presented in Figure~\ref{fig:EFieldTCADAnnealing}. 
  For the Hamburg scenario (shown only for comparison), the profile is linear while this is not the case for the other scenarios, especially at the higher fluence $\Phi=2\times10^{14}$~n$_\text{eq}/\text{cm}^{2}$.

\begin{table}[htbp]
   \centering
   \begin{tabular}{|c|c|c|} %
      \hline
            \multicolumn{3}{|c|}{$\Phi=1\times10^{14}$~n$_\text{eq}/\text{cm}^{2}$  } \\
      \hline %
        Scenario  & $\langle \rho/e\rangle$ [cm$^{-3}$] & RMS [cm$^{-3}$] \\
      \hline
        Chiochia &  $-1.5\times10^{12}$ & 0.7$\times10^{12}$  \\ %
  Hamburg   & $-1.62\times10^{12}$ & 0  \\ %
       $N_\text{A}^\text{TCAD}$+3\% &  $-1.7\times10^{12}$ & 0.8$\times10^{12}$  \\ %
      \hline
         \end{tabular}
   \begin{tabular}{|c|c|c|} %
      \hline
            \multicolumn{3}{|c|}{$\Phi=2\times10^{14}$~n$_\text{eq}/\text{cm}^{2}$  } \\
      \hline %
        Scenario  & $\langle \rho/e\rangle$ [cm$^{-3}$] & RMS [cm$^{-3}$] \\
      \hline
        Chiochia &  $-2.9\times10^{12}$ & 3.0$\times10^{12}$  \\ %
        Hamburg   & $-2.72\times10^{12}$ & 0 \\ %
      $N_\text{A}^\text{TCAD}-1.6$\% &  $-2.7\times10^{12}$ & 3.0$\times10^{12}$  \\ %
      \hline
   \end{tabular}
   \caption{\label{tab:TCADNeffHamburg1}Results for average and RMS of the space-charge density over the sensor bulk from TCAD simulation for different scenarios at fluences $\Phi=1\times10^{14}$ and $2\times10^{14}$~n$_\text{eq}/\text{cm}^{2}$; the last row corresponds to the TCAD with eff. annealing scenario. Bias voltage $V_\text{bias}$ 
was 150~V. Since $\rho$ is constant for the Hamburg scenario, the RMS is zero. }
\end{table}

In summary, the TCAD with effective annealing scenario used an acceptor trap density $N_{A}^\text{TCAD}$ in the TCAD simulations that was increased by 3\% at a fluence of $\Phi=1\times10^{14}$~n$_\text{eq}/\text{cm}^{2}$ and reduced by 1.6\%
at a fluence $\Phi=2\times10^{14}$~n$_\text{eq}/\text{cm}^{2}$ to emulate the effect of annealing predicted by the Hamburg model. These variations are currently within the model variations described in Section~\ref{sec:Efieldmodelcomparisons} that are used to set systematic uncertainties on the radiation damage model parameters.  Therefore, no corrections or uncertainties are applied to the simulation to account for annealing for the current radiation levels.  This must be revisited when the IBL has experienced significant annealing.

\subsection{Time-to-electrode, position-at-trap}
\label{sec:maps}

\begin{figure}[h!]
\centering
\includegraphics[width=0.5\textwidth]{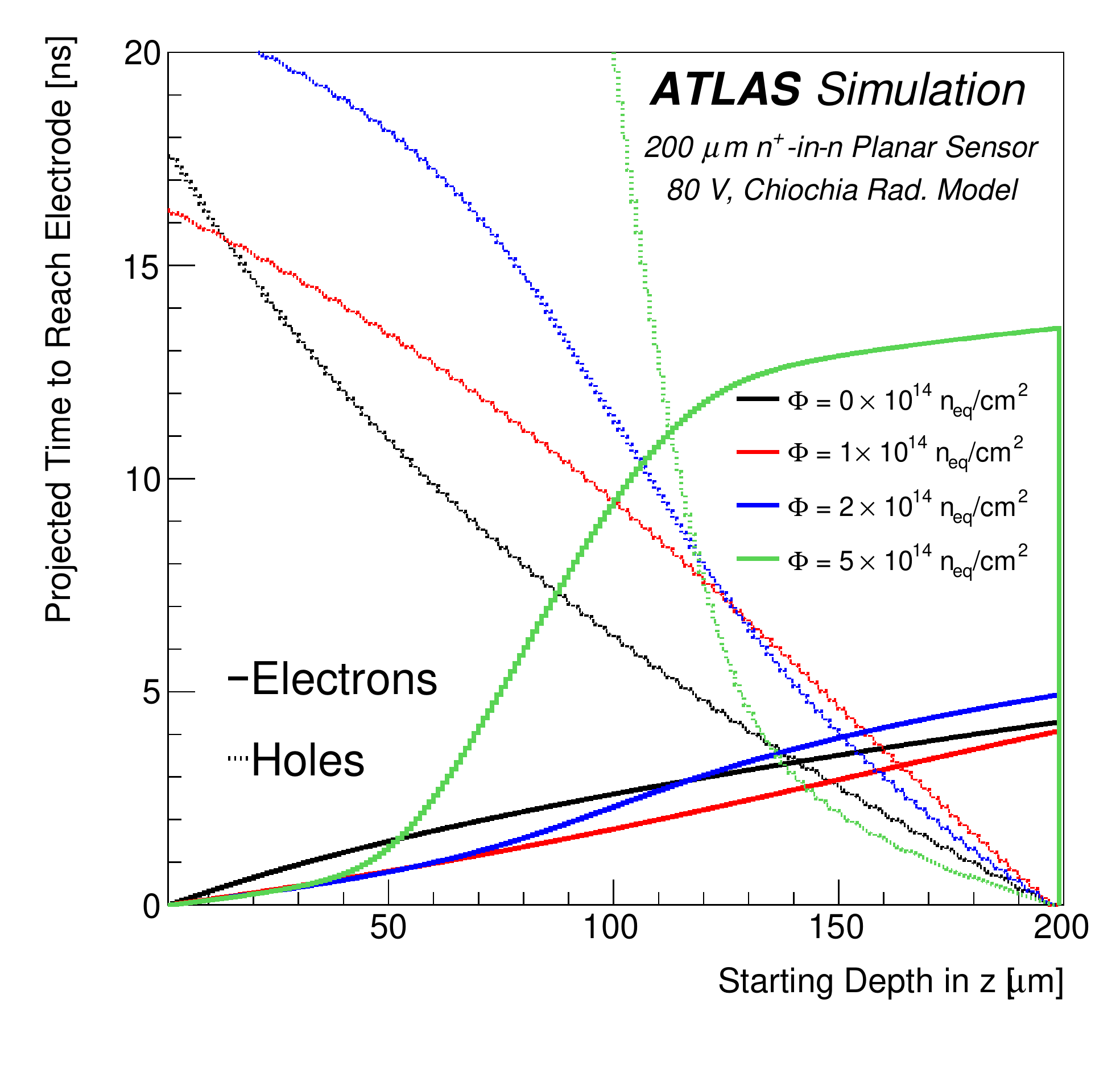}
\caption{The projected time (in the absence of trapping) for an electron or hole to drift all the way from their point of generation to the collecting electrode (electrons) or back plane (holes) in an ATLAS IBL planar sensor biased at 80~V as a function of the depth ($z$) using the averaged $E$ fields predicted by Chiochia model through TCAD simulation (Figure~\ref{fig:electricfield:profilesRun23}).  The $n^+$ electrode is located at the left ($z=0$) and the back plane at $z= 200\,\upmu$m.}
\label{fig:timetoelectrode}
\end{figure}

Numerically, propagating charges through the silicon sensor can be computationally expensive, but fortunately can be computed once per geometry and set of conditions (temperature, bias voltage, and fluence).   Electrons and holes drift with a carrier-dependent mobility ($\mu$) that depends on the electric field ($E$) and temperature~\cite{JACOBONI197777}. The drift velocity is given by $\vec{v}(E)\sim r\mu(E)\vec{E}$ ($r=\text{Hall}$ scattering factor, $r\approx 1$) and the charge collection time is estimated via

\begin{align}
\label{eq:timetoelectrode}
t_\text{collection}(\vec{x}_\text{initial})=\int_C\frac{\text{d}s}{r\mu(E)E},
\end{align}

where $C$ is the path from $\vec{x}_\text{initial}$ to $\vec{x}_\text{final}$ that is determined by the equations of motion $\vec{v}=r\mu(E)\vec{E}$; $\vec{x}_\text{final}$ depends on the type of the charge carrier.  For planar sensors, the field is nearly independent of $x$ and $y$, so the time to the electrode is parameterized in $z$ and the integral in Eq.~(\ref{eq:timetoelectrode}) is one-dimensional. 
Since the mobility of holes is much lower than for electrons, it takes holes much longer (factor of $\gtrsim 3$--$5$) on average to arrive at the electrode. The collection time varies with fluence, bias voltage, and distance to the electrode, but is on average $\mathcal{O}(1)$--$\mathcal{O}(10)$ ns for $\Phi\lesssim 10^{15}$~$\text{n}_\text{eq}/\text{cm}^2$ as shown in Figure~\ref{fig:timetoelectrode}.

\begin{figure}[h!]
\centering
  \subfloat[]{\includegraphics[width=0.45\textwidth]{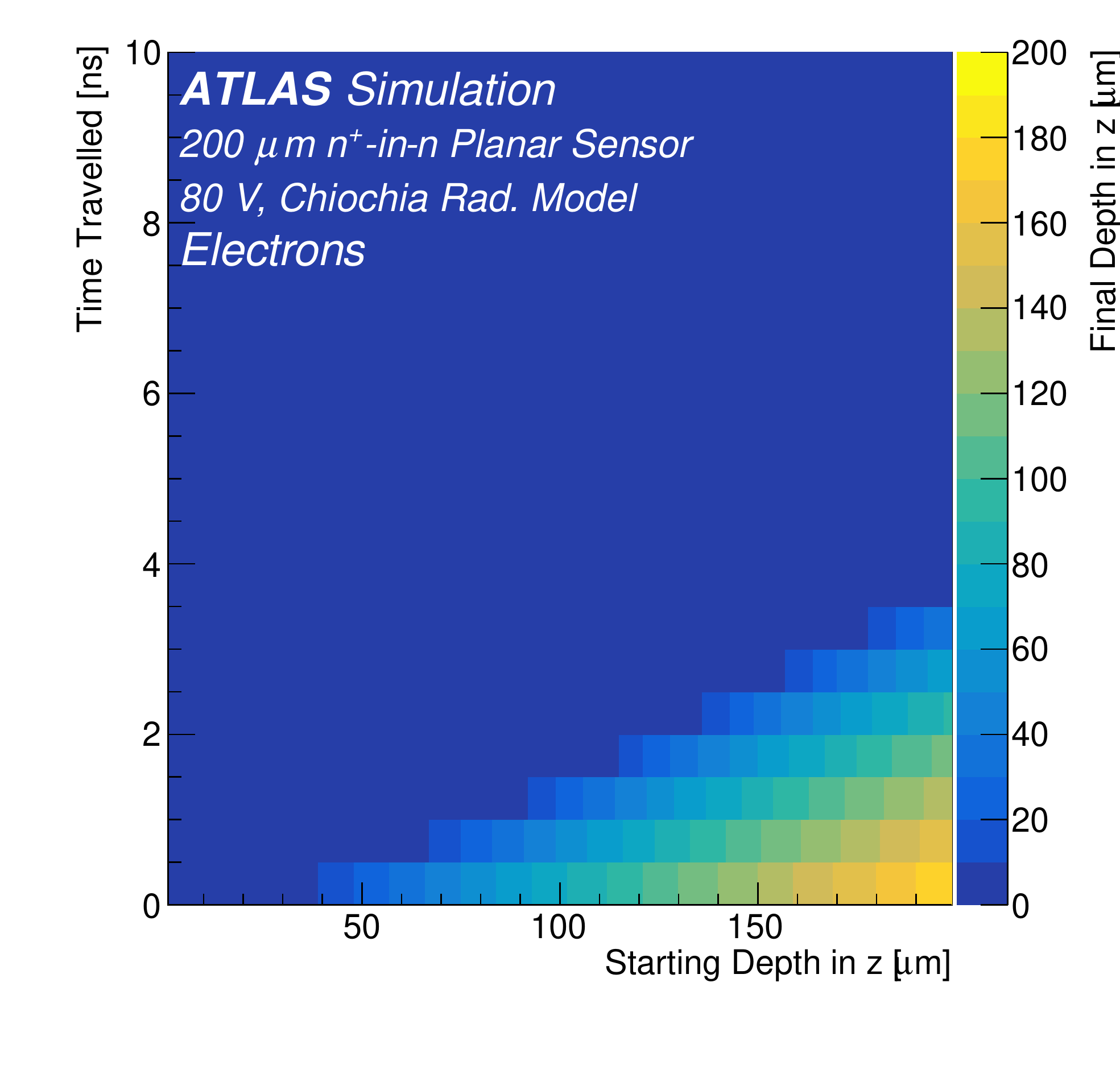}} %
  \subfloat[]{\includegraphics[width=0.45\textwidth]{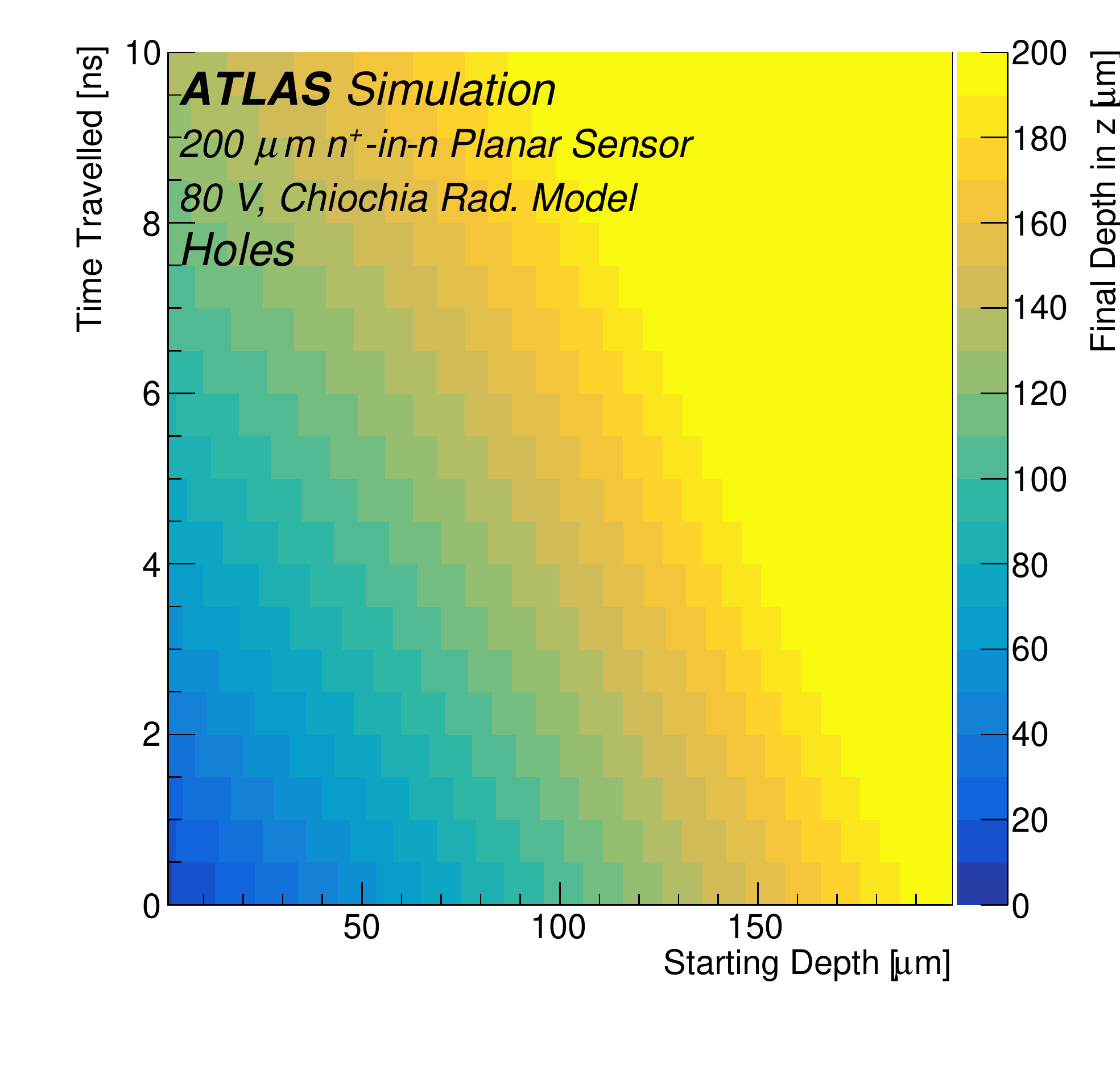}} %
  \caption{The $z$ position of trapped (a) electrons and (b) holes as a function of their starting position and the time travelled for $\Phi=10^{14}$~$\text{n}_\text{eq}/\text{cm}^2$.  The $n^+$ electrode is located at the left ($z=0$) and the back plane at $z= 200\,\upmu$m.}
\label{fig:posattrap}
\end{figure}

For charges that are trapped (see Section~\ref{sec:chargetrapping}), the location of the trapped charge must be known.  The position-at-trap can be calculated in a fashion similar to the time-to-electrode from Eq.~(\ref{eq:timetoelectrode}).  In particular, the location is given by

\begin{align*}
\vec{x}_\text{trap}(t_\text{to trap})=\int_\text{$0$}^\text{$t_\text{to trap}$}r\mu(E)\vec{E}\text{d}t,
\end{align*}

\noindent where $t_\text{to trap}$ is either the drift time (if the charge is not trapped) or a random time set by the trapping constant (Section~\ref{sec:chargetrapping}).  Representative position-at-trap maps are shown in Figure~\ref{fig:posattrap} for planar sensors.  If the time travelled is larger than the drift time, then the electrons reach the collecting electrode ($z=0$) and holes reach the back side ($z=200$ \textmu m).  The corresponding maps for 3D sensors are more difficult to visualize due to their higher dimensionality.

\subsection{Lorentz angle}
\label{sec:mapsLorentz}
\begin{figure}[h!]
\centering
  \subfloat[]{\includegraphics[width=0.49\textwidth]{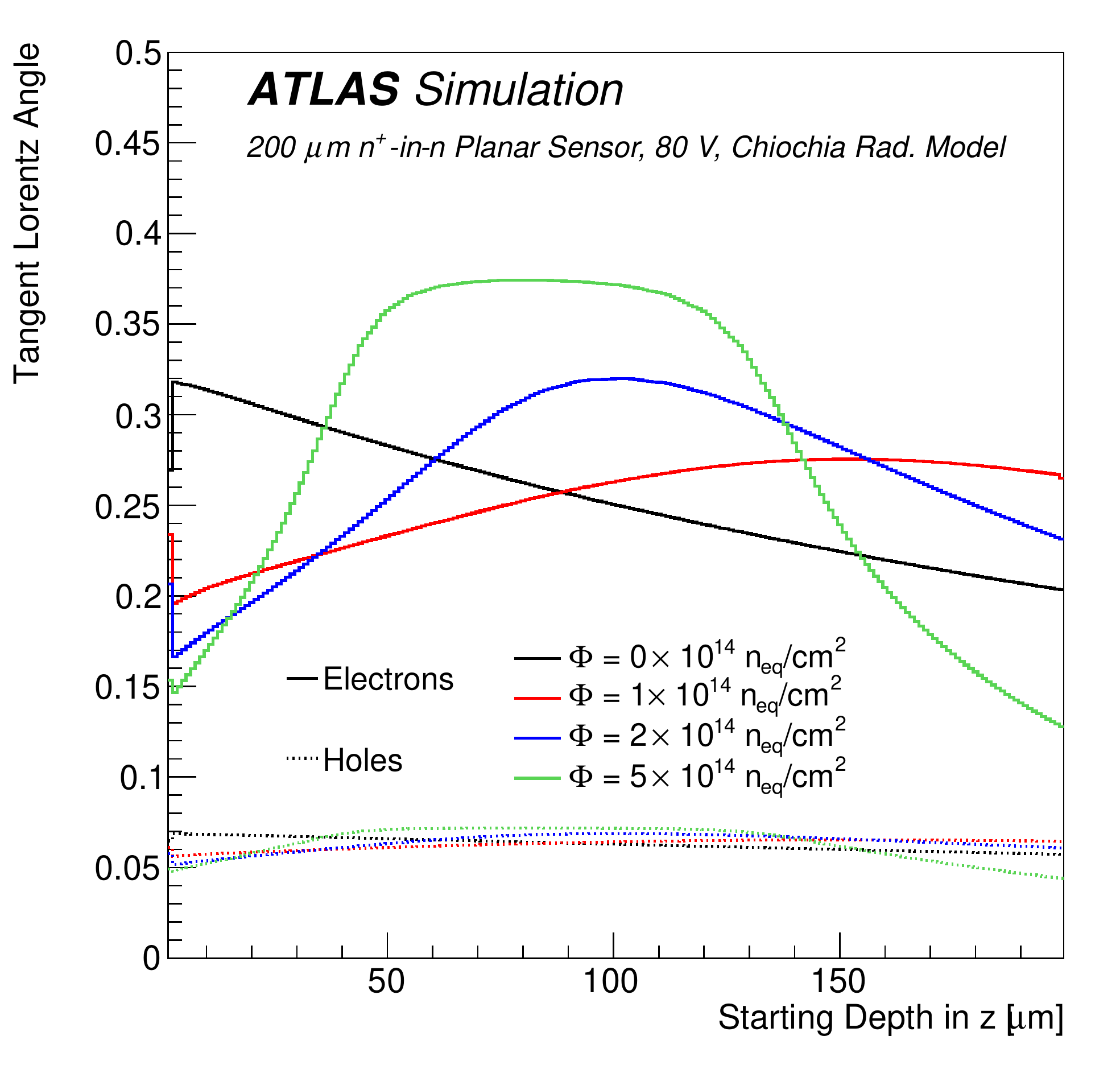}}
  \subfloat[]{\includegraphics[width=0.49\textwidth]{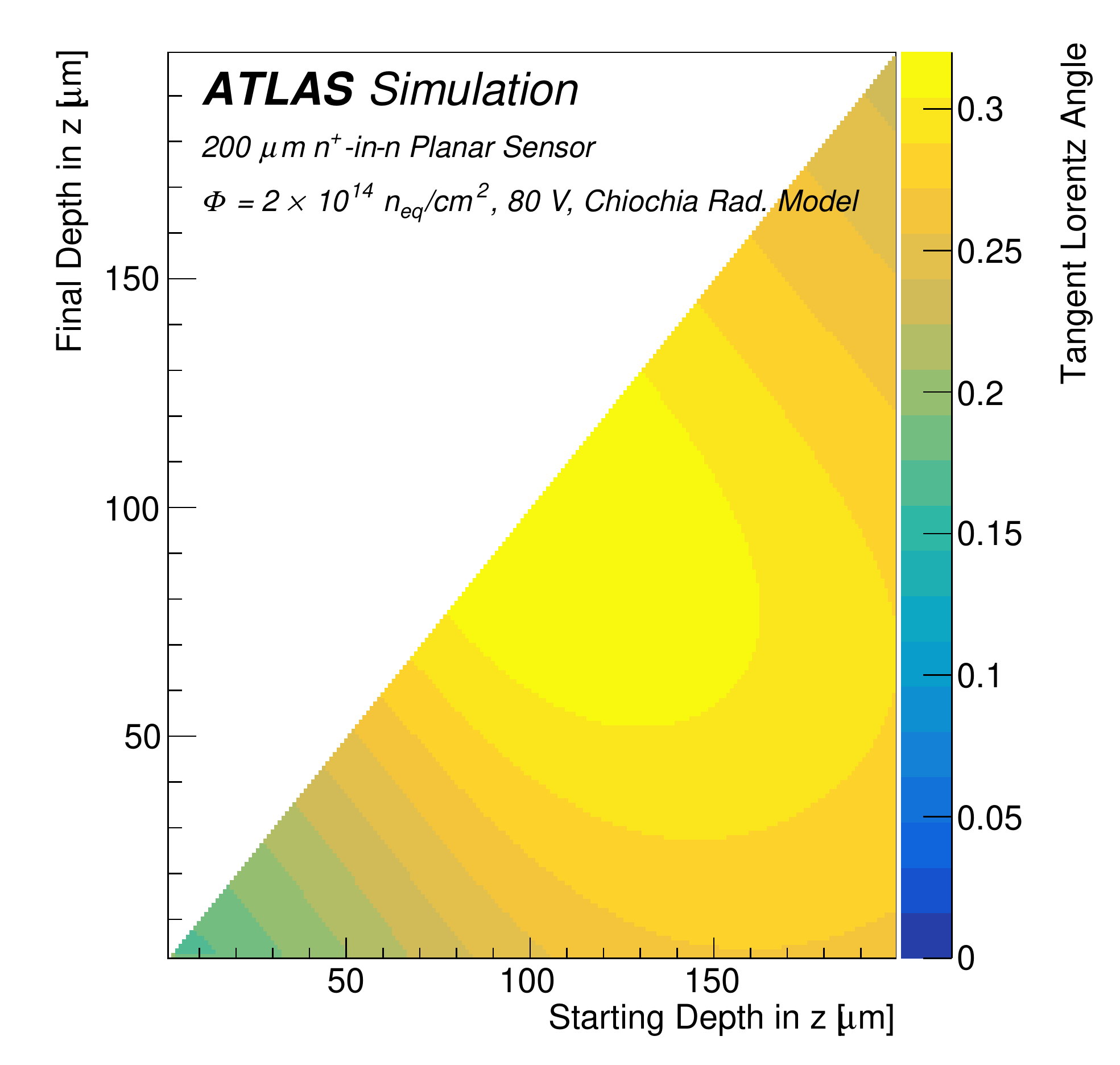}}
  \caption{(a) The depth dependence of the Lorentz angle for electrons and holes for four fluences in an ATLAS IBL planar sensor biased at 80~V.  (b) The integrated Lorentz angle for electrons (see Eq.~(\ref{eq:integratedlorentzangle})) as a function of the starting and ending position for a fluence of $\Phi=2\times10^{14}$~$\text{n}_\text{eq}/\text{cm}^2$.  The collecting electrode is at a $z$ position of $0$.}
\label{fig:lorentzangle:Run2}
\end{figure}

The Lorentz angle ($\theta_\text{L}$) is the result of balancing electric and magnetic forces, and is defined as the incidence angle that produces the smallest cluster size in the transverse direction.  As $\theta_\text{L}$  depends on the shape of the electric field, it is affected by radiation-induced changes of the space-charge density.
Changes in the electric field affect $\theta_\text{L}$ indirectly through the mobility as $\tan\theta_\text{L}=r\mu(E) B$, where $B$ is the magnetic field.  Figure~\ref{fig:lorentzangle:Run2}(a) demonstrates the change in the Lorentz angle along the trajectory of electrons and holes.  As the mobility increases with decreasing electric field strength, the Lorentz angle is largest near the centre of the sensors when irradiated.  The path-dependence of the Lorentz angle is modelled in a manner similar to the position maps from Section~\ref{sec:maps} (Figure~\ref{fig:posattrap}) by averaging the Lorentz angle along the path:

\begin{align}
\label{eq:integratedlorentzangle}
\tan\theta_\text{L}^\text{integrated}(z_\text{initial},z_\text{final})=\frac{rB}{|z_\text{final}-z_\text{initial}|}\int_{z_\text{initial}}^{z_\text{final}}\mu(E(z)) \text{d}z.
\end{align}

The drift along the $\phi$ direction  is then modified as $|z_\text{final}-z_\text{initial}|\tan\theta_\text{L}^\text{integrated}(z_\text{initial},z_\text{final})$, where the direction is the same for both the electrons and holes because both the charge and velocity sign are reversed for holes relative to electrons. The integrated Lorentz angle variations are shown in Figure~\ref{fig:lorentzangle:Run2}; for this fluence and bias voltage, the integrated Lorentz angle can change by as much as a factor of two, depending on the starting and ending position.  

\subsection{Charge trapping}
\label{sec:chargetrapping}

In the simulation, charge carriers are declared trapped if the projected time to reach the electrode, as defined in Section~\ref{sec:maps}, exceeds a randomly set trapping time $t$ that is exponentially distributed with mean value $1/(\beta\Phi)$~\cite{Kramberger:2002zb,Kras04}, where $\Phi$ is the fluence and $\beta$ is the trapping constant.

The linear relation with fluence has been measured and shown to hold with very good precision up to $10^{15}$~$\text{n}_\text{eq}/\text{cm}^2$, but the value of $\beta$ has been found to depend on the type of irradiation, the temperature, the annealing history of the device, and the type of charge carriers (electrons or holes)~\cite{Kramberger:2002zb,Kras04}. The measurements in Refs.~\cite{Kramberger:2002zb,Kras04} were performed with the transient current technique (TCT).  The results of the measurements are reported in Table~\ref{tab:trapping}. Since measurements were performed at different temperatures between $-10^{\circ}$C and 10$^{\circ}$C and a significant decrease of $\beta$ with temperature is found~\cite{Kramberger:2002zb}, the parameterizations of the temperature dependence provided in Ref.~\cite{Kramberger:2002zb} separately before and after annealing have been used to correct the measurements to the value expected at 0$^{\circ}$C (in the middle of the Run 2 temperature range). Test-beam measurements taken with ATLAS pixel sensors~\cite{Aad:2008zz,trapping2} are also reported in the table but they are more indirect and less precise than TCT measurements.

Both TCT references find that annealing results in a decrease of $\beta$ for electrons and an increase for holes, with a plateau reached after
one or two days at 60$^{\circ}$C. The two references agree on the values of $\beta$ for electrons, but Ref.~\cite{Kras04} finds a smaller $\beta$ value for holes
than Ref.~\cite{Kramberger:2002zb}. Irradiation with neutrons is found to result in smaller $\beta$-values than irradiation by charged particles~\cite{Kramberger:2002zb}.
The uncertainties reported in Table~\ref{tab:trapping} do not include an uncertainty of about 10\% in the fluence received by the devices,
and refer to the average found by fitting measurements made on several devices. 

\begin{table}[!htbp]
   \centering
   \begin{tabular}{cccccc} %
     \hline
     Irradiation   & Annealing & $\beta_e$ ($10^{-16} \text{cm}^2$/ns) & $\beta_h$ ($10^{-16} \text{cm}^2$/ns) & Reference & Method  \\
     \hline
     Neutrons      & minimum $V_\text{depl}$ & $4.0 \pm 0.1$ & $5.7 \pm 0.2$ & \cite{Kramberger:2002zb} & TCT \\
     Pions         & minimum $V_\text{depl}$ & $5.5 \pm 0.2$ & $7.3 \pm 0.2$ & \cite{Kramberger:2002zb} & TCT\\
     Protons         & minimum $V_\text{depl}$ & $5.13 \pm 0.16$ & $5.04 \pm 0.18$ & \cite{Kras04} & TCT\\
     \hline
     Neutrons      & $> 50$~hours at 60$^{\circ}$C & $2.6 \pm 0.1$ & $7.0 \pm 0.2$ & \cite{Kramberger:2002zb}& TCT \\
     Protons         & $> 10$~hours at 60$^{\circ}$C & $3.2 \pm 0.1$ & $5.2 \pm 0.3$ & \cite{Kras04}& TCT \\
     \hline
     Protons         & minimum $V_\text{depl}$ & $4.0 \pm 1.4$ &   -    & \cite{Aad:2008zz,trapping2} & Test-beam\\
     Protons         & 25h at 60$^{\circ}$C    & $2.2 \pm 0.4$  &  -    & \cite{Aad:2008zz,trapping2} & Test-beam\\
   \end{tabular}
   \caption{Measurements of the trapping constant $\beta$ are summarized, normalized to a temperature of 0$^{\circ}$C. Some
     measurements are reported after annealing to the minimum in the full depletion voltage $V_\text{depl}$ (reached in about
     80 minutes at 60$^{\circ}$C) while others correspond to the asymptotic values observed after long annealing times.}
   \label{tab:trapping}
\end{table}

For the simulation results reported in this paper, a value of $\beta_e= (4.5 \pm 1.5) \times 10^{-16}$ cm${}^2$/ns was used for
electron trapping and $\beta_h= (6.5 \pm 1.5) \times 10^{-16}$ cm${}^2$/ns for hole trapping. These values were chosen after considering the irradiation conditions of the IBL pixel modules during the LHC Run 2. The 
range of selected $\beta$ values covers the variation of measurements presented in Table~\ref{tab:trapping} including annealing.

\subsection{Ramo potential and induced charge}
\label{sec:ramo}

Charge carrier movement induces a signal on the detector electrodes. The instantaneous current $i$ 
induced in an electrode by a carrier of charge $q(=\pm e)$ moving at velocity $\vec{v}$ can be calculated by means of the Shockley--Ramo theorem~\cite{Shockley,Ramo}:

\begin{equation}
  i=q\vec{E}_\text{w}\cdot\vec{v},
  \label{eq:RamoCurrent}
\end{equation}
where $\vec{E}_\text{w}$ is the Ramo (or \textit{weighting}) field that describes the induction coupling of the moving charge to a specific electrode. The Ramo field can be calculated by applying a unit potential to the electrode under consideration and zero potential to all other electrodes.
Integrating Eq.~(\ref{eq:RamoCurrent}) over a certain drift time, the charge $Q$ induced on the 
electrode can be expressed as:

\begin{align*}
\label{eq:ramo}
Q_\text{induced} = -q[\phi_\text{w}(\vec{x}_\text{end})-\phi_\text{w}(\vec{x}_\text{start})],
\end{align*}

where $\phi_\text{w}$ is the Ramo (or \textit{weighting}) potential  with $\vec{E}_\text{w} = - \nabla \phi_\text{w}$
  and $\vec{x}_\text{start,end}$ are the 
initial (final) positions of the charge carrier under consideration.  The Ramo potential depends only on geometry and therefore can be computed once prior to any event simulation. 

\begin{figure}[h!]
\centering
\includegraphics[width=0.5\textwidth]{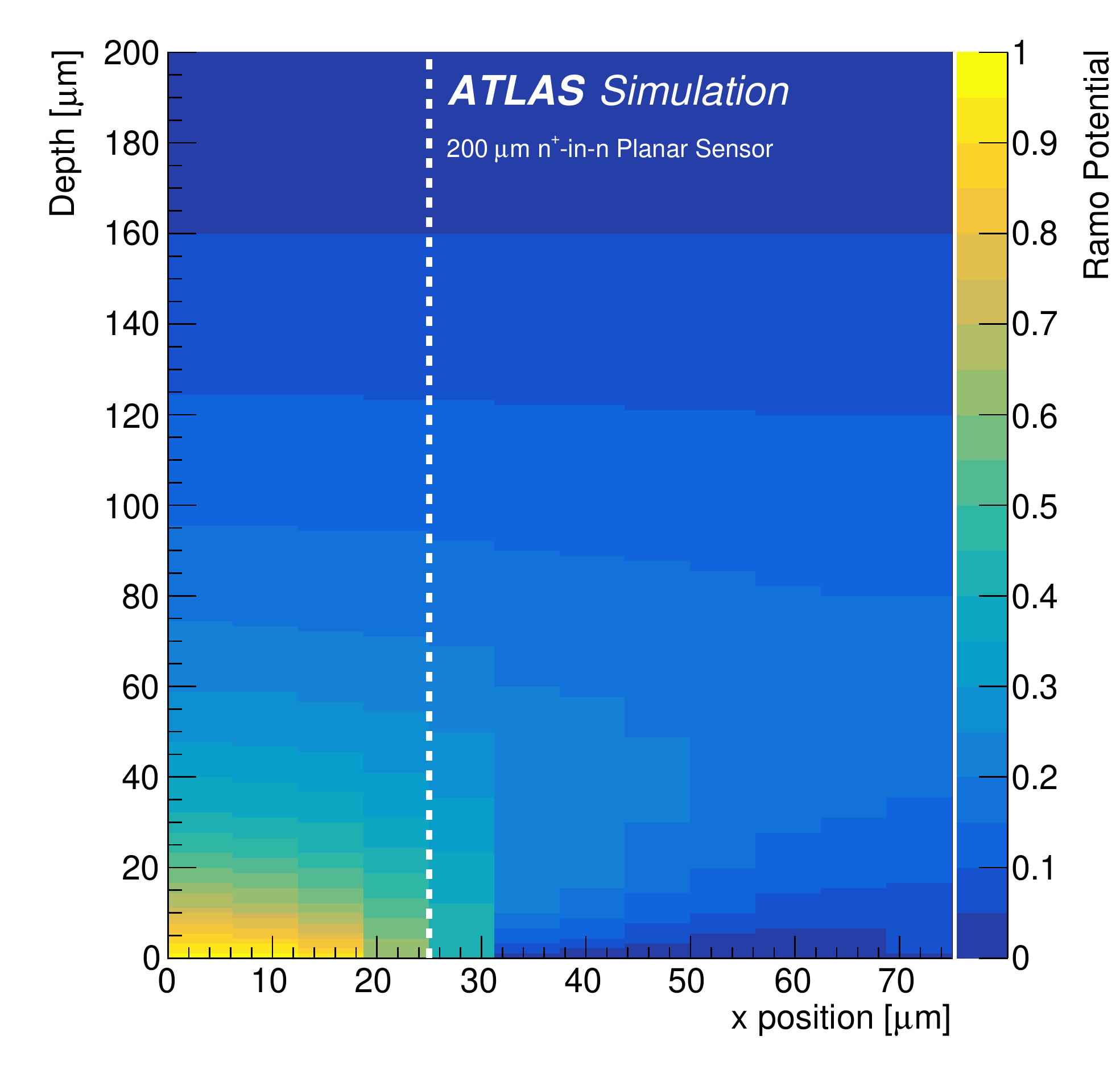} %
  \caption{A slice of the full three-dimensional ATLAS IBL planar sensor Ramo potential as computed with TCAD at $y=0$.  The dashed vertical line (at 25 \textmu m) indicates the edge of the primary pixel. }
\label{fig:ramo:planar}
\end{figure}

The Ramo potential is calculated using TCAD to solve the Poisson equation;  for planar sensors, most of the variation in the Ramo potential is in the $z$ direction, but the $x$ and $y$ dependence must also be included in order to account for charge induced on the neighbouring pixels.  Figure~\ref{fig:ramo:planar} shows a slice of the three-dimensional Ramo potential in the centre of the pixel electrode ($y=0$).   The vertical line indicates the edge of the pixels: the Ramo potential has sizeable contributions in the neighbouring pixels.

\begin{figure}[h!]
\centering
  \subfloat[]{\includegraphics[width=0.33\textwidth]{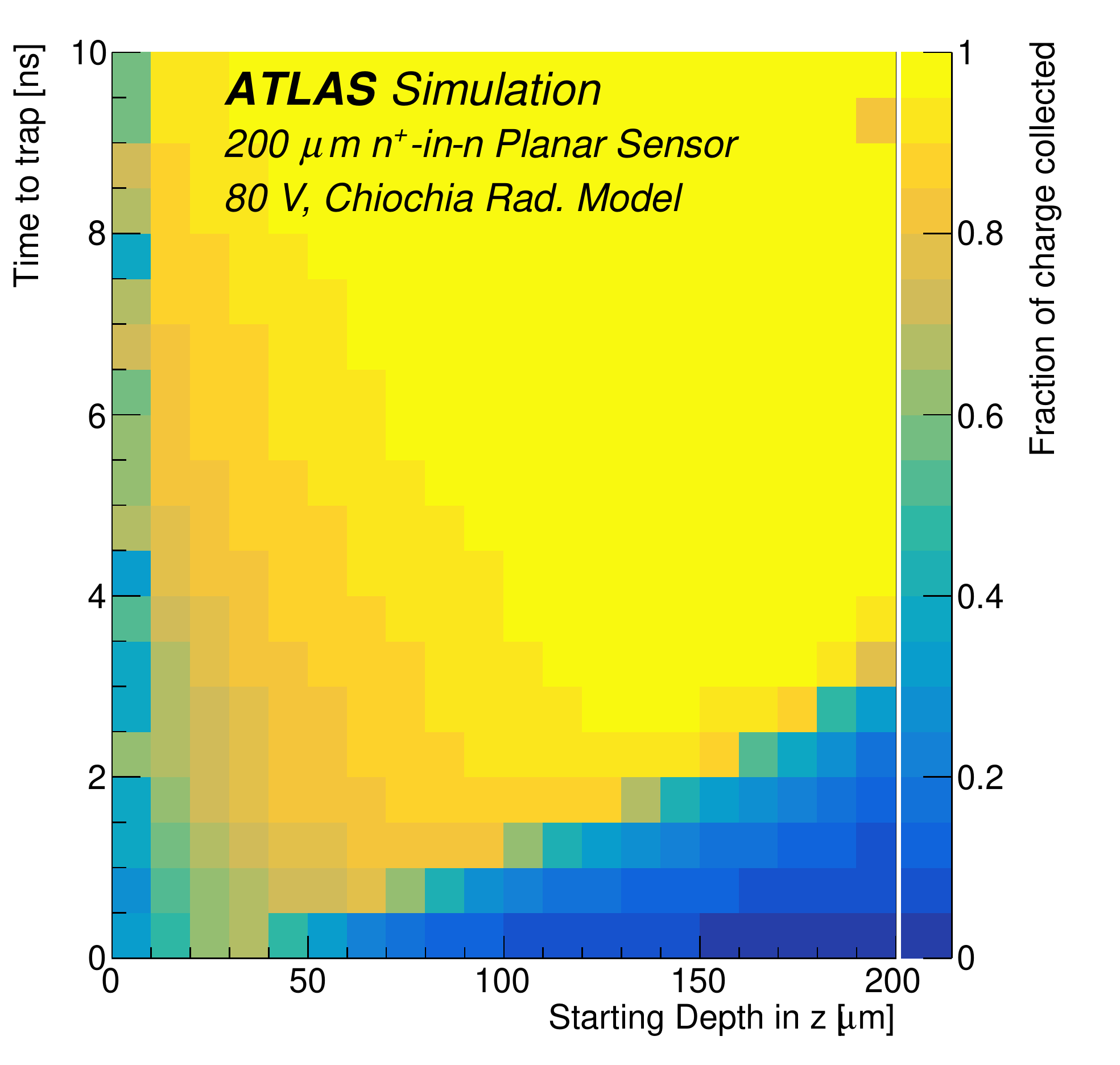}}
  \subfloat[]{\includegraphics[width=0.33\textwidth]{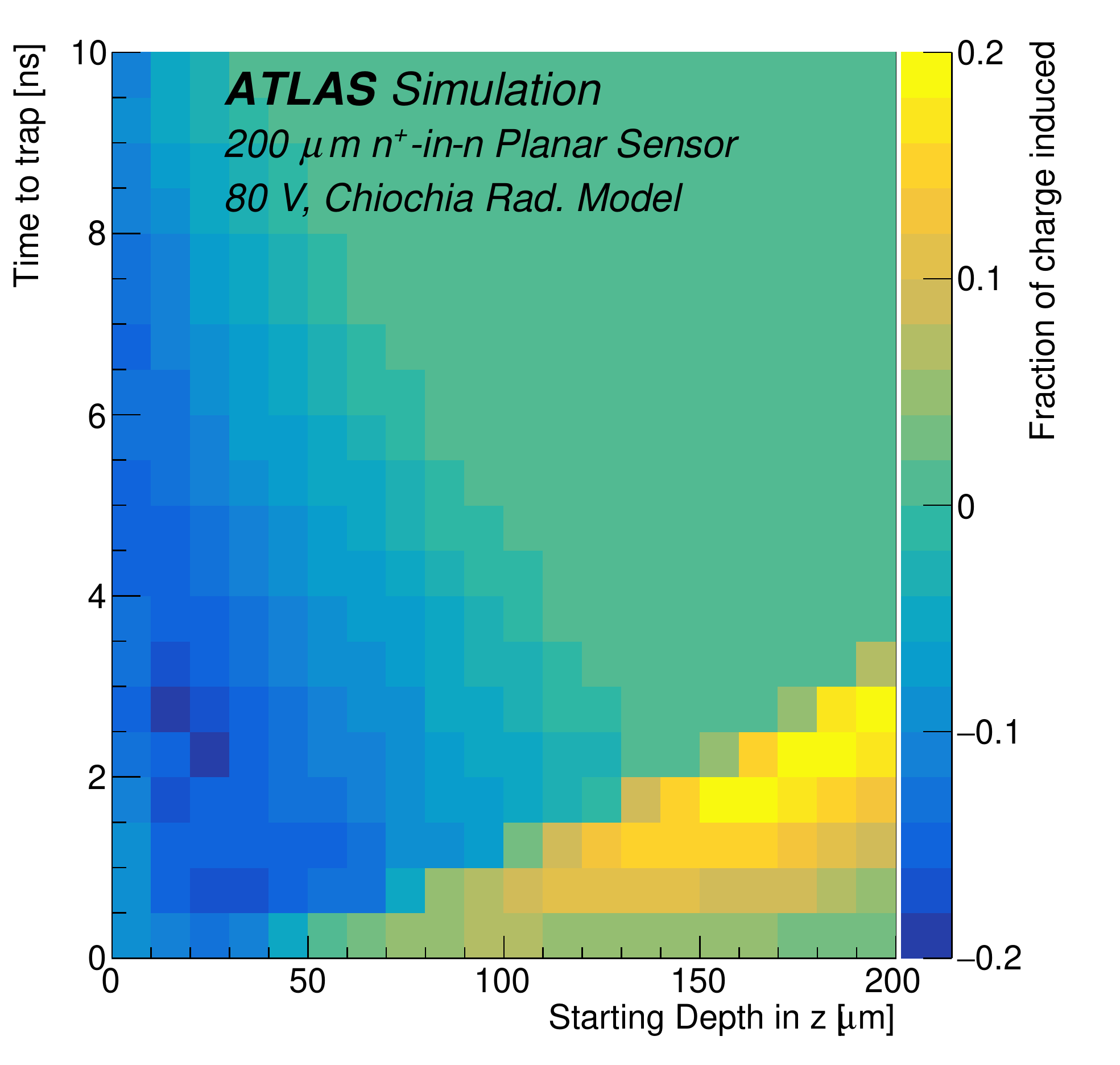}}
  \subfloat[]{\includegraphics[width=0.33\textwidth]{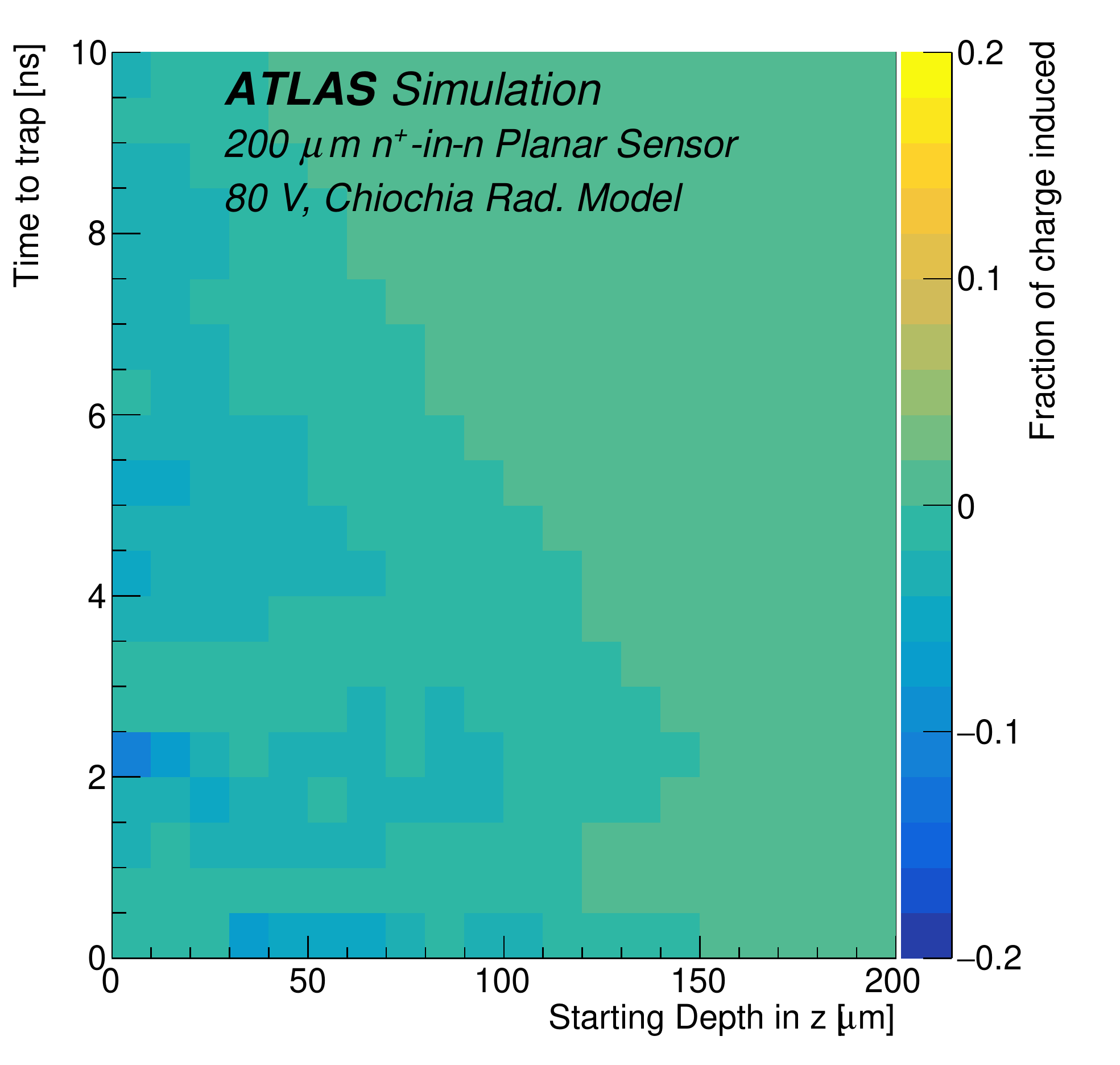}}
\caption{The average fraction of charge collected as a function of the starting location and the time to be trapped for (a) the same pixel as the electron--hole pair generation, (b) the neighbour pixel in the short-pitch ($50$ \textmu m) direction and (c) the neighbour pixel in the long-pitch ($250$ \textmu m) direction.  For illustration, in this figure only, the electric field is simulated without radiation damage and the vertical axis is the time to trap.  The induced charge includes the contribution from electrons and holes, while the electron time to trap is used for the vertical axis.}
\label{fig:ramo:check}
\end{figure}

The combination of the Ramo potential and charge trapping is illustrated in Figure~\ref{fig:ramo:check} for planar sensors.  On the electrode of the same pixel in which the electrons and holes originate, the induced charge equals the electron charge if the time to be trapped exceeds the time to drift toward the electrode (Figure~\ref{fig:ramo:check}(a)).   The average collected charge is an asymmetric function of the depth inside the sensor because the drift and trapping times are different for electrons and holes and the Ramo potential is very asymmetric: the average fraction is lower far away from the collecting electrode.  The charge induced on neighbouring pixels is shown in Figures~\ref{fig:ramo:check}(b) and \ref{fig:ramo:check}(c).  As the trapping time exceeds the drift time, the integrated induced current amounts to the full electron charge in the primary pixel and the charge in the neighbours is zero.  For some combinations of starting location and time to trap, the induced charge can have the opposite sign.  This happens when holes are trapped very close to the pixel implants. 

\clearpage

\subsection{3D sensor simulations}
\label{sec:3D}

In contrast to planar sensors where electrons and holes drift along the depth toward the pixel implant or the back plane, respectively, in 3D sensors charges drift laterally.   Columns are etched through the $p$-type silicon bulk (with an initial effective doping concentration of $6\times 10^{11}/\text{cm}^3$), and are subsequently either $p^+$ or $n^+$ doped.\footnote{In the ATLAS IBL, some sensors have columns that extend through the entire bulk while in others they only partially pass through.  This section only considers the fully passing through case.}  In the 3D sensors, two $n^+$ columns are shorted so that one pixel corresponds to two $n^+$ electrodes for collecting electrons.  The $p^+$ electrodes are connected to the bias voltage.   As the distance between electrodes can be much smaller than the sensor depth, 3D sensors are designed to be more radiation hard than planar sensors due to the reduced drift length (see e.g.\ Ref.~\cite{Lange:2015acq} and references therein).  This section describes how the digitizer presented in the previous sections can be modified to also accommodate the 3D sensor geometry.

\begin{figure}[h!]
\centering
\includegraphics[width=0.5\textwidth]{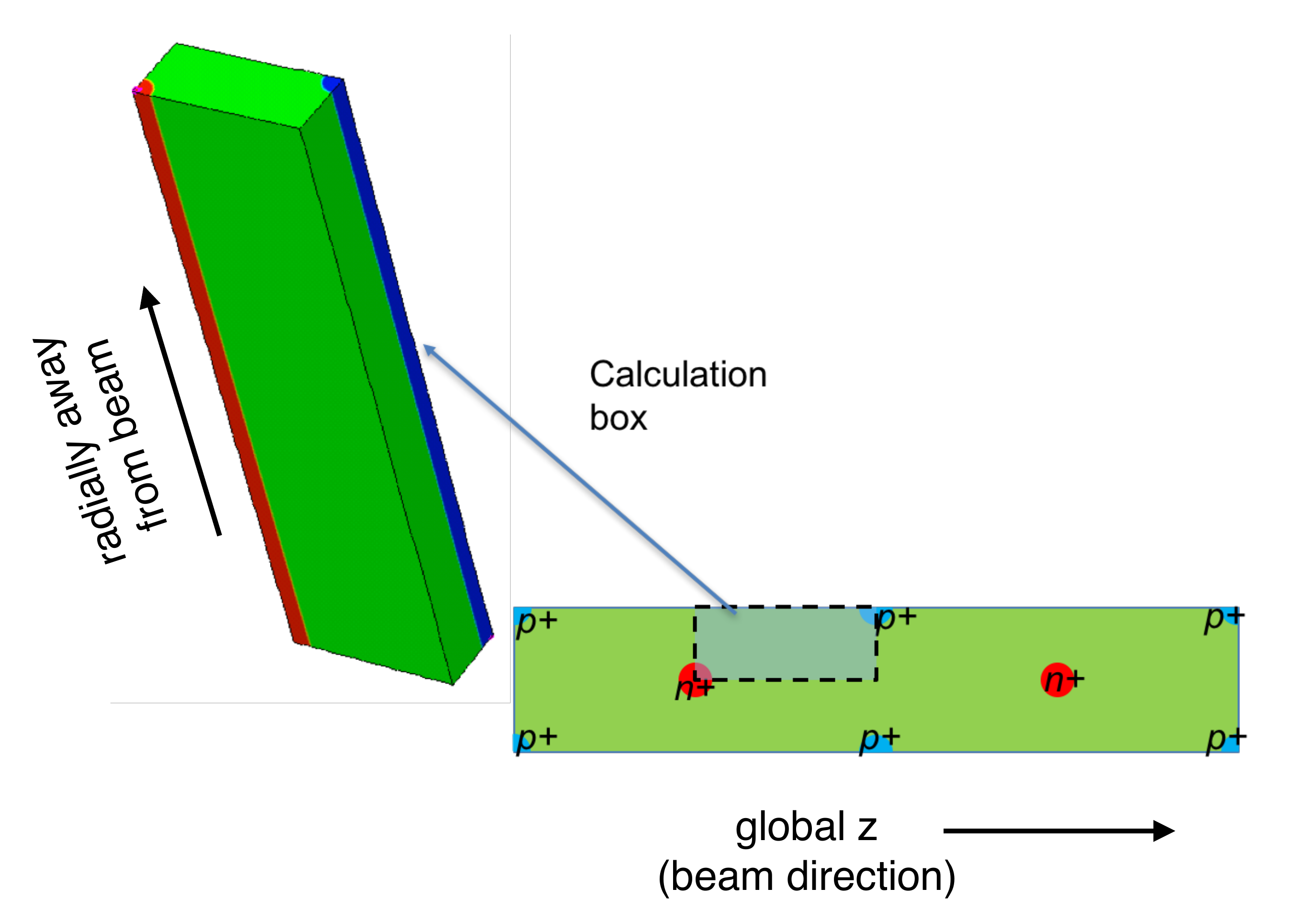}
\caption{A schematic set-up of the geometry for the TCAD simulation of 3D sensors.  The dashed box shows the unit cell that is simulated and tessellated to produce an entire pixel.}
\label{fig:tredisetup}
\end{figure}

Radiation damage effects for the 3D sensor are implemented in the Perugia model~\cite{7542192} (the Chiochia model cannot be used, as it is designed for $n$-type bulk) with the Synopsys TCAD package~\cite{synopsys}.  As shown in Figure~\ref{fig:tredisetup} one-eighth of the sensor is simulated to take advantage of the symmetry within the pixel.  In the Perugia model, there are two acceptor traps and one donor trap, with activation energies given by $E_\text{c}-0.42$~\eV, $E_\text{c}-0.46$~\eV, and $E_\text{v}+0.36$~\eV, respectively. The density of traps is predicted to 
increase linearly with fluence $\Phi$, so for each trap an introduction rate
($g_\text{int}$) is defined as: $g_\text{int}=N/\Phi$, where $N$ is the trap concentration. 
In Table~\ref{tab:Perugia2016} the values used for the simulation of 3D sensors
 reported in this paper are summarized.

\begin{table}[!htpb]
\centering
  \begin{tabular}{|c|c|c|c|c|}
    \hline
    Type & Energy~[\eV] &  $g_\text{int}~[\text{cm}^{-1}]$ & $\sigma_{e}~[{\text{cm}^{2}}]  $ & $\sigma_{h}~[{\text{cm}^{2}}]$  \\
    \hline
    Acceptor & $E_\text{c}-0.42$ & 1.613& $1\times10^{-15}$ & $1\times10^{-14}$    \\
    Acceptor & $E_\text{c}-0.46$ & 0.9  & $3\times10^{-15}$ & $3\times10^{-14}$    \\
    Donor    & $E_\text{v}+0.36$ & 0.9  & $3.23\times10^{-13}$ & $3.23\times10^{-14}$    \\
    \hline
  \end{tabular}
  \caption{\label{tab:Perugia2016}Values used in TCAD simulations for 
  deep defect energies, introduction rates ($g_\text{int}$) and for their 
  electron and hole capture cross sections ($\sigma_{e,h}$). Values are taken 
  from the Perugia model~\cite{7542192}.}
\end{table}

\begin{figure}[h!]
\centering
  \subfloat[]{\includegraphics[width=0.95\textwidth]{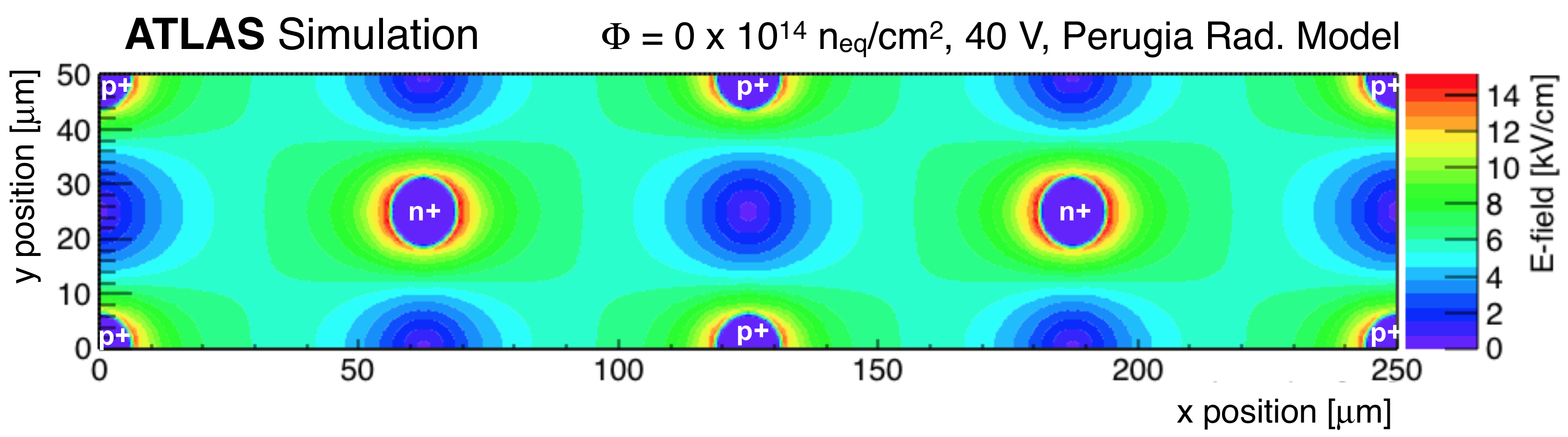}}\\
  \subfloat[]{\includegraphics[width=0.95\textwidth]{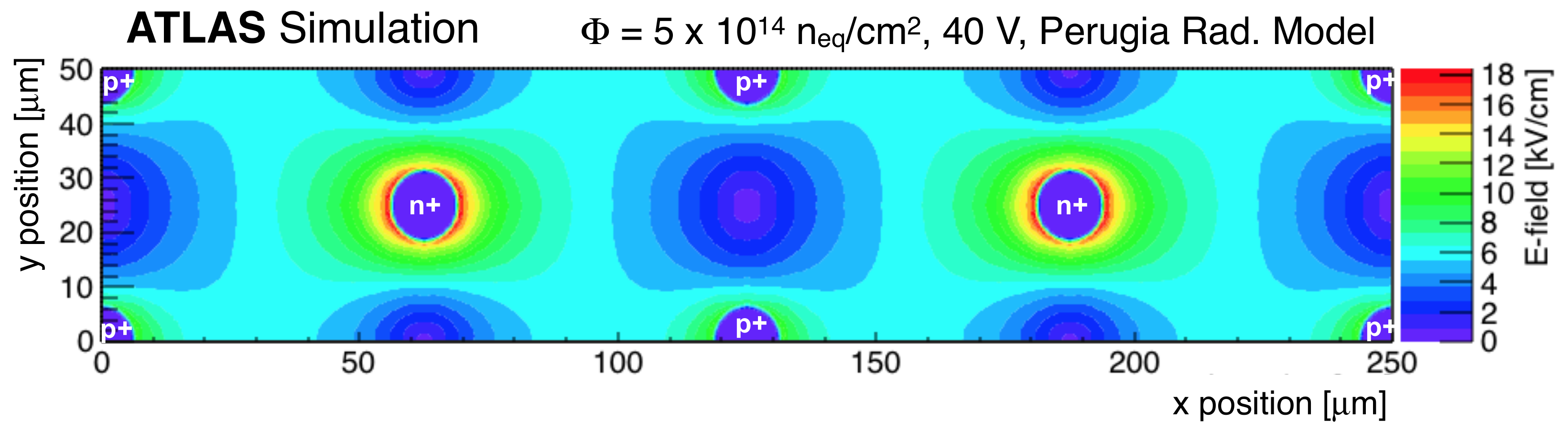}}
  \caption{Simulated magnitude of the electric field as a function of local $x$ and $y$ in an ATLAS IBL 3D sensor with a bias voltage at $V_\text{bias} = -40$~V for (a) an unirradiated sensor and (b) for a fluence of $5\times10^{14}$~$\text{n}_\text{eq}/\text{cm}^2$. The doping type of the columns is indicated.}
\label{fig:electricfield:profilesRun23_3D}
\end{figure}

Electric field profiles simulated with TCAD are shown in Figure~\ref{fig:electricfield:profilesRun23_3D}.  In contrast to planar sensors (Section~\ref{sec:Efieldprofile}), the field is nearly independent of $z$ and depends strongly on $x$ and $y$.  Therefore, the electric field magnitude is shown as a two-dimensional map for both an unirradiated sensor and a highly irradiated sensor.  The $n^+$ and $p^+$ implants are regions of no field due to their large doping and are modelled as having $0\%$ charge collection efficiency.  To illustrate the entire pixel, the one-eighth map that is simulated is tessellated.

\begin{figure}[h!]
\centering
  \subfloat[]{ \includegraphics[width=0.85\textwidth]{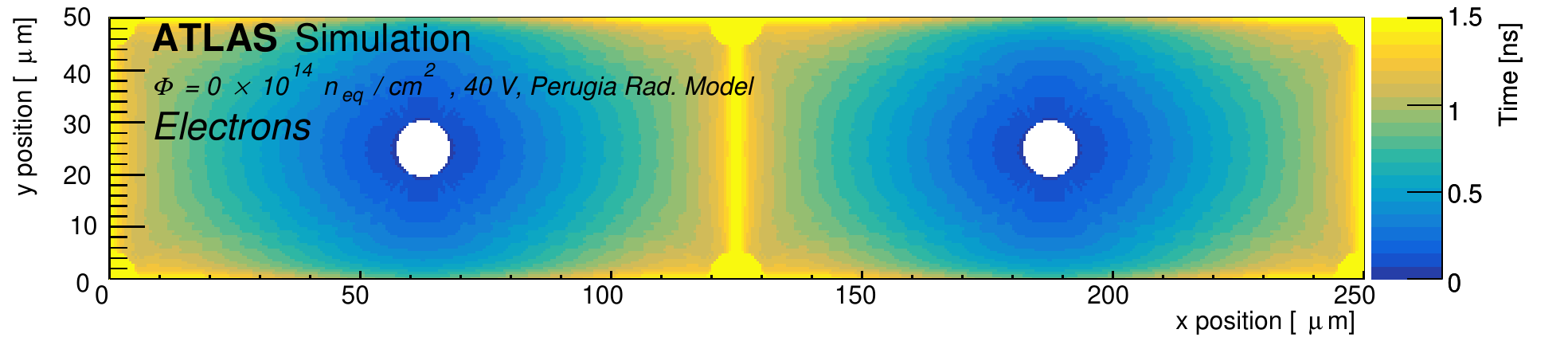}}\\
    \subfloat[]{ \includegraphics[width=0.85\textwidth]{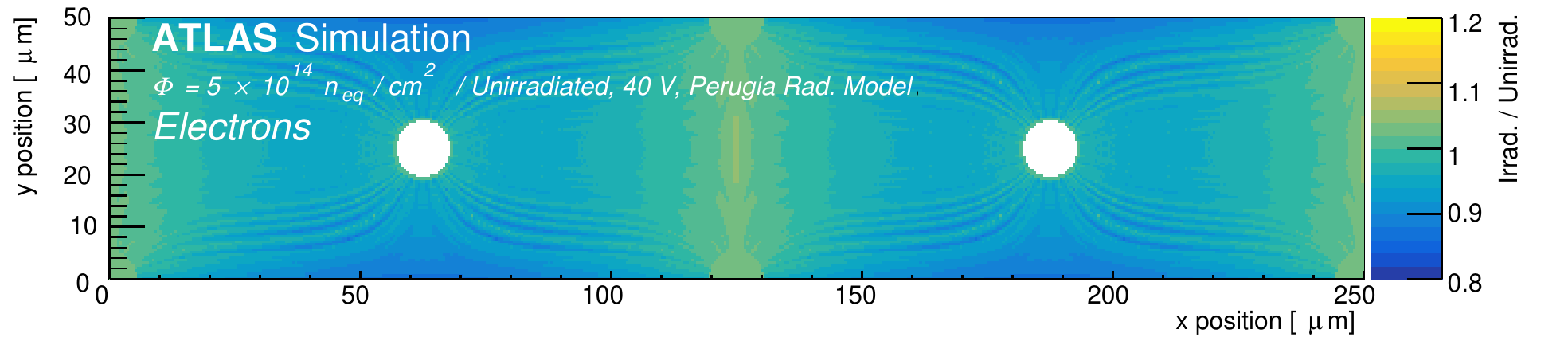}}\\
      \subfloat[]{ \includegraphics[width=0.85\textwidth]{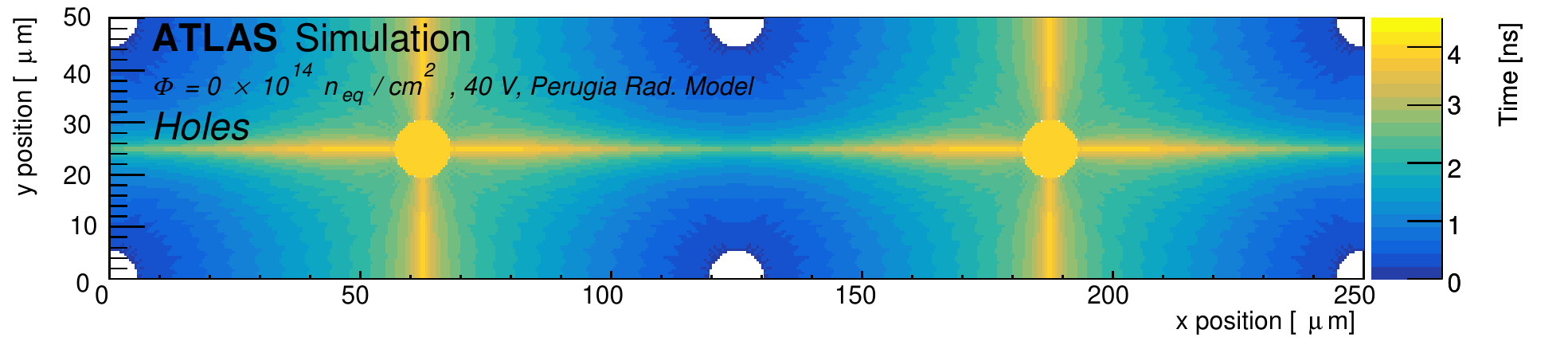}}\\
    \subfloat[]{ \includegraphics[width=0.85\textwidth]{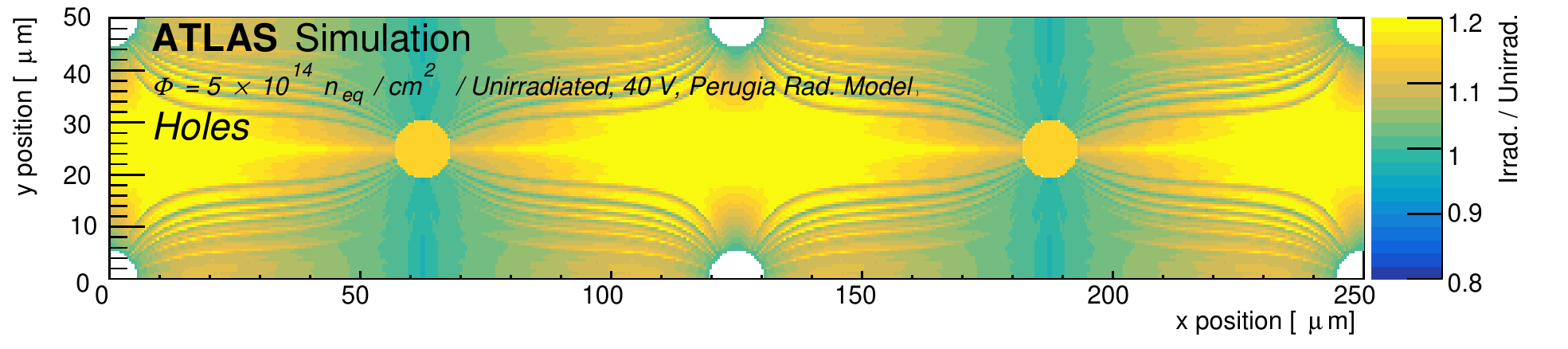}}
  \caption{The projected (in the absence of trapping) time for electrons and holes to drift from their point of generation to the corresponding collecting electrode for an ATLAS IBL 3D sensor.  (a) and (c) are computed  for unirradiated sensors, respectively, for electrons and holes; (b) and (d) show the ratio of the time for a fluence of $5\times 10^{14}$~$\text{n}_\text{eq}/\text{cm}^2$ to the time for no radiation damage for electrons and holes, respectively.}
\label{fig:timetoelectrode3D}
\end{figure}

The projected time to reach the electrode can be computed analogously to planar sensors (Section~\ref{sec:maps}) and is shown in Figure~\ref{fig:timetoelectrode3D}.  The main difference relative to planar sensors is that electrons and holes follow a non-trivial trajectory due to the more complex electric field.  However, a simplification relative to planar sensors is that the electric field is nearly parallel to the magnetic field so the Lorentz angle is negligibly small.

\begin{figure}[h!]
\centering
\includegraphics[width=0.7\textwidth]{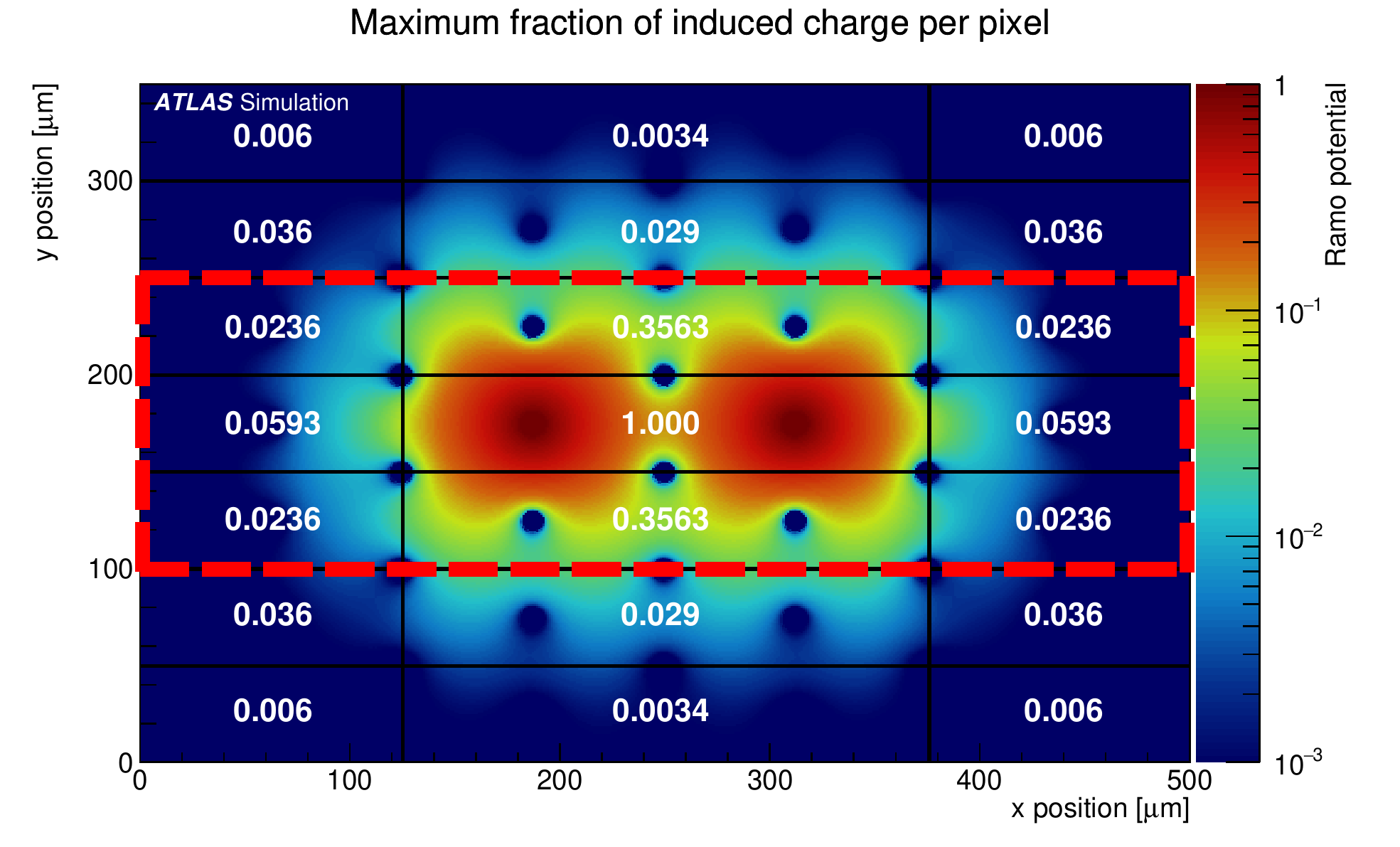}
\caption{The Ramo potential for an ATLAS 3D sensor as computed with TCAD.  The two $n^+$ electrodes of the centre pixel are electrically connected and therefore both are held at unit potential in the calculation of the Ramo potential.  The circular holes are due to the $p^+$ electrodes.  White numbers indicate the maximum induced charge (normalized to one electron charge) in that pixel considering all starting positions and trapping times in the central pixel.  A red dashed rectangle shows which pixels are included in the simulation.}
\label{fig:ramo:3D}
\end{figure}

The Ramo potential for 3D sensors is slightly more complex than for planar sensors.
In particular, the two $n^+$ columns in one pixel are electrically connected and so in the calculation of the Ramo potential, both are held at unit potential while all other electrodes are grounded.  Therefore, the calculation requires a relatively large simulation area.  This is illustrated in Figure~\ref{fig:ramo:3D}. As with the planar sensors, only the immediate neighbours are included in the calculation.  The numbers overlaid on Figure~\ref{fig:ramo:3D} show that this is a good approximation. 

\begin{figure}[h!]
\centering
\includegraphics[width=0.9\textwidth]{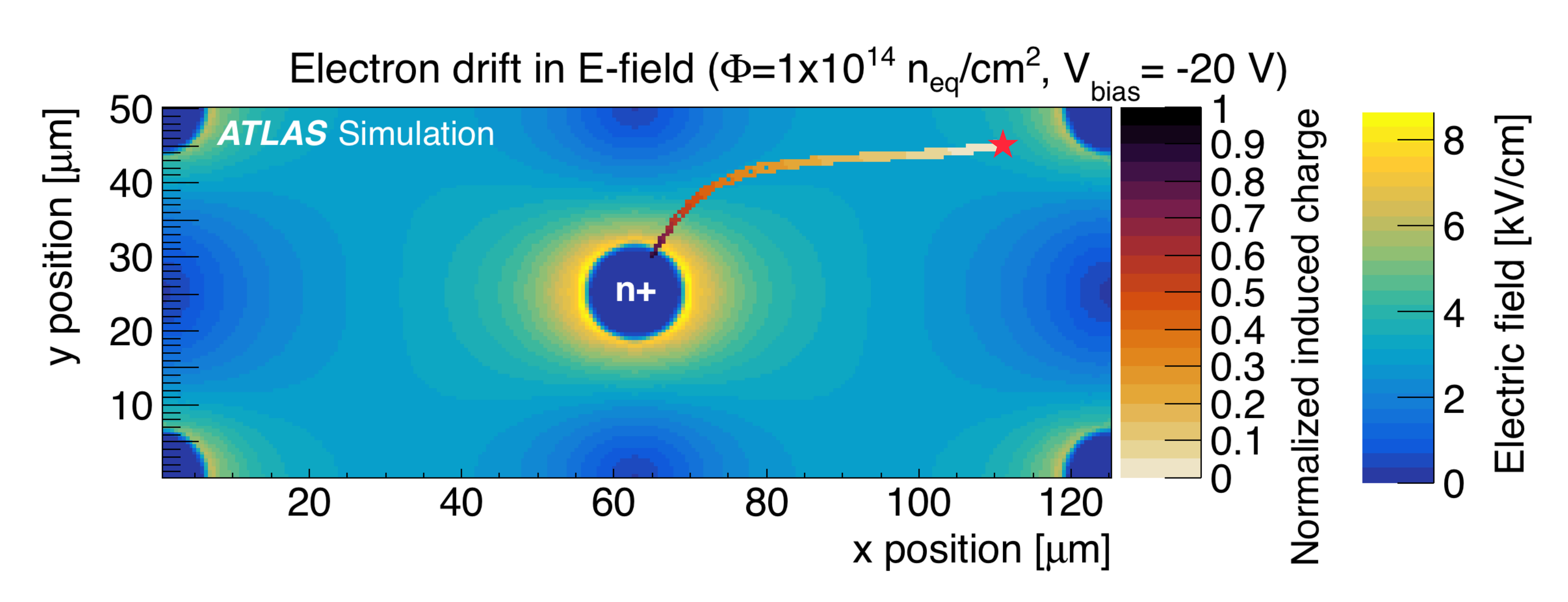}
\caption{An illustration of charge trapping and the Ramo potential in one-half of a 3D sensor.  The initial electrons originate from the top right corner of the plot (indicated by a star).  Under the influence of the electric field, they drift toward the $n^+$ electrode in the centre.  As they drift, there is a chance that they get trapped.  Markers indicate the location of trapped charges and the colour shows the induced charge.  The process is repeated many times, with diffusion. The simulated bias voltage was $-20$~V.}
\label{fig:ramo:schematic}
\end{figure}

The various digitizer components are combined in Figure~\ref{fig:ramo:schematic} to show the charge induced by electrons as they drift toward the $n^+$ electrode.  The stochastic path from the same initial electron position near the upper right part of the plot is re-simulated many times.  Markers indicate the final location of the electrons (they all have the same initial position).  The electrons that travel closer to the electrode before being trapped induce a larger charge (darker markers) than those that are trapped right away.   

Since the 3D sensors in the IBL are outside of the acceptance to reconstruct tracks ($|\eta| < 2.5$), they have not been studied as thoroughly as their planar counterparts.  However, future studies with these sensors will provide an important opportunity to test the 3D digitization model presented here, which has already been used to make projections for the inner tracking detector (ITk) at the HL-LHC~\cite{Collaboration:2285585}.  

\clearpage

\section{Model predictions and validation}
\label{sec:validation}

\subsection{Data and simulation}

The models presented in the previous sections are validated by comparing the simulations with data, both in terms of physics predictions.  This section presents two key observables for studying radiation damage: the charge collection efficiency and the Lorentz angle.  These two quantities are measured as a function of time in Run 2 for the IBL planar sensors.  The IBL is well-suited for this test because at the start of Run 2, it was unirradiated.  The data were collected in the fall of 2015 and throughout 2016 and 2017. Charged-particle tracks are reconstructed from hits in the pixel detector, silicon strip detector, and transition radiation tracker.  Clusters on the innermost pixel layer associated with tracks are considered for further analysis.  The IBL is operated with a bias voltage of 80~V, 150~V, or 350~V at a temperature ranging from $-15^{\circ}$C to $+20^{\circ}$C.  The analogue threshold is 2550~$e$ with 4~bits of ToT for the digital charge read-out~\cite{Abbott:2018ikt}.  The ToT is calibrated so that a ToT of 8 corresponds to 16~k$e^-$ and a digital threshold of ToT~$>1$ is applied.  The analogue-to-digital conversion is modelled using the same charge-to-ToT conversion.  In practice, the charge-to-ToT conversion is non-linear, especially for low charge.  This is most prominent at low charge and therefore required care in interpreting data-to-simulation comparisons in this regime. %

Simulated datasets are based on Geant4~\cite{Agostinelli:2002hh} with digitization implemented in Allpix~\cite{benoit:20xx}, which is a lightweight wrapper of Geant4 that is optimized for test-beam analysis and is a powerful test-bench for digitizer development.  The figures in this section refer to this set-up as the `Stand-alone Simulation.'

\subsection{Charge collection efficiency}
\label{sec:CCE}

\begin{figure}[h!]
\centering
\includegraphics[width=0.7\textwidth]{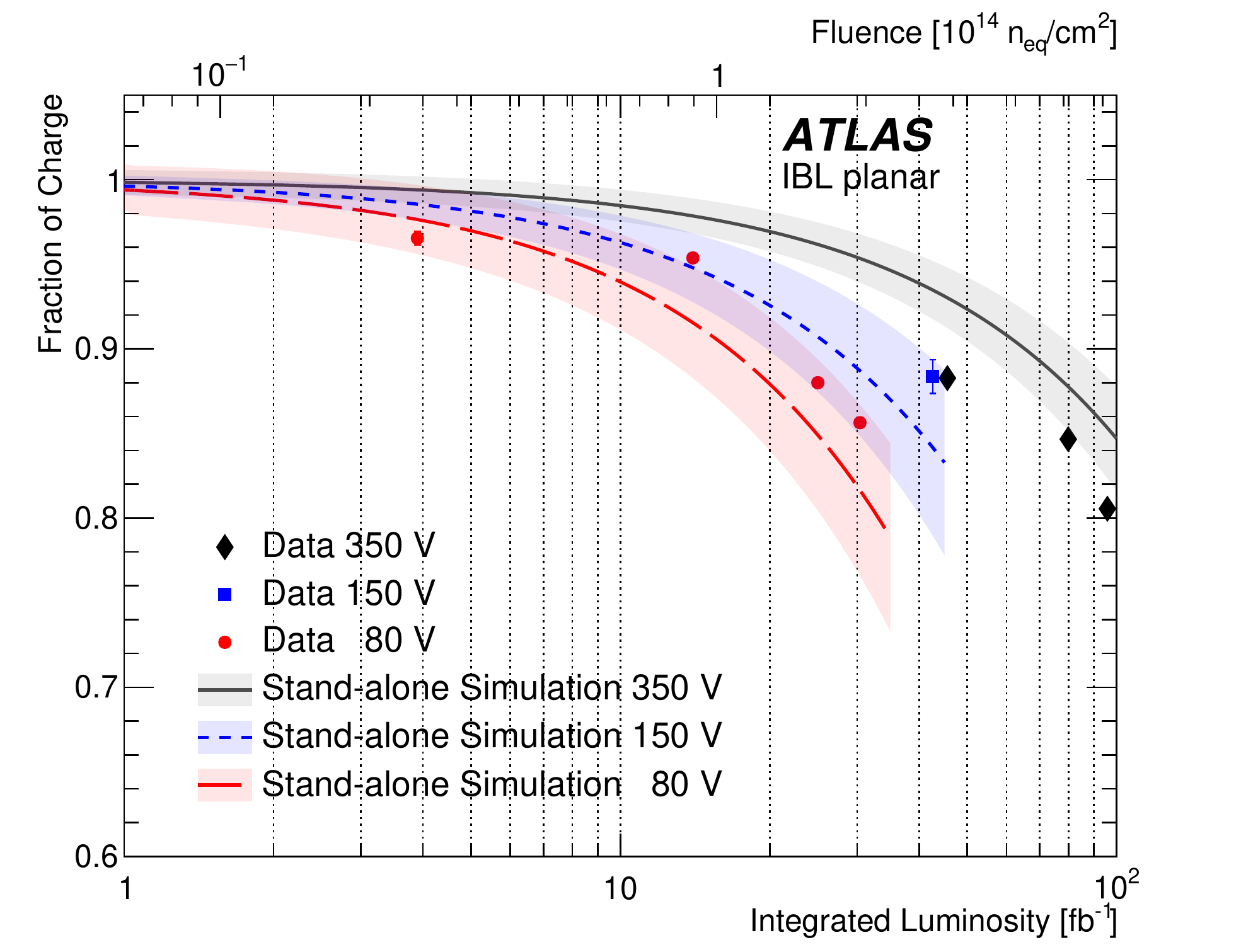}
\caption{The charge collection efficiency as a function of integrated luminosity for 80~V, 150~V, and 350~V bias voltage.  A linear trendline is added to the simulation to guide the eye.  The bias voltage was increased during data-taking, so the data points are only available at increasing high-voltage values.  The points are normalized to unity for a run near the beginning of Run 2.  The uncertainty on the simulation includes variations in the radiation damage model parameters as well as the uncertainty in the luminosity-to-fluence conversion.  Vertical uncertainty bars on the data are due to the charge calibration drift.  Horizontal error bars on the data points due to the luminosity uncertainty are smaller than the markers.} %
\label{fig:CCE:Run2}
\end{figure}

The collected charge is represented by the most probable value of the charge distribution, which is approximately Landau-distributed~\cite{Landau:1944if}.  The charge collection efficiency (CCE) is defined here to be the collected charge at one fluence divided by the charge for unirradiated sensors in over-depletion.  Figure~\ref{fig:CCE:Run2} shows the measured and predicted charge collection efficiencies as a function of integrated luminosity in Run 2. Data points are corrected in order to account for the drift in the ToT calibration.  As radiation effects cause the measured ToT to drift with integrated luminosity, regular re-tunings were performed to bring the mean ToT back to the tuning point. This is a feature of the electronics and is not due to a physical change in the charge collection.  The drift is approximated as linear and the correction is evaluated in the middle of the run (one period of stable beam) considered. An uncertainty of 30\% is assigned to this correction, which is then propagated to the charge collection efficiency value.  The size of the correction varies from run to run, but is in general below 5\% with a final uncertainty in the CCE of $2$--$3 $\%. The total uncertainty in the predicted value of the charge collection efficiency is evaluated by taking the squared sum of the differences between the nominal value and the one obtained with the variation of the radiation damage parameters, as explained in detail in Section~\ref{sec:Efieldmodelcomparisons}. An additional uncertainty is due to the trapping constant, as explained in Section~\ref{sec:chargetrapping}.  An uncertainty of 3\%~\cite{Aaboud:2016hhf} is also assigned to the luminosity value of the data points (the horizontal error bars in Figure~\ref{fig:CCE:Run2}). For the Allpix (`stand-alone') simulation points, the integrated luminosity is converted to a fluence using the information presented in Figure~\ref{fig:fluenceoverview2}, and an uncertainty of 15\% (see Section~\ref{sec:fluence}) is assigned to the conversion. As expected, the efficiency drops with luminosity ($\propto$ fluence) due to charge trapping and under-depletion.  Part of this loss was recovered after switching the bias voltage in the IBL first from 80~V to 150~V and then later to 350~V, and further increases will be necessary to recover future losses in the charge collection efficiency.  A breakdown of the impact of the variations performed to assess systematic uncertainties is reported in Table~\ref{tab:SystError}. 

\begin{table}
\begin{center}
\setlength{\tabcolsep}{0.0pc}
\begin{tabular*}{\textwidth}{@{\extracolsep{\fill}}lS[table-format=3.2]S[table-format=3.2]S[table-format=3.2]S[table-format=3.2]S[table-format=3.2]S[table-format=3.2]}
\noalign{\smallskip}\hline\noalign{\smallskip}
Bias voltage [V]		& 80 & 80 & 80& 150 & 150 &150\\
  Fluence $[10^{14} \text{n}_\text{eq}/\text{cm}^{2}]$ 					& 1 & 2 & 5 & 1 & 2 & 5\\
\noalign{\smallskip}\hline\noalign{\smallskip}

Variation 				& Impact~[\%] & Impact~[\%] & Impact~[\%] & Impact~[\%]& Impact~[\%] & Impact~[\%]\\
\noalign{\smallskip}\hline\noalign{\smallskip}

Energy acceptor $+10\%$   	& 0.4  	& 2.2  & 0.7 & 0.2 & 1.6 & 1.5  \\
Energy donor $+10\%$  			& 0.5  	&      & 4.5 & 0.2 & 0.3 & 5.7 \\
Energy acceptor $-10\%$  		&  		&  1.7 & 3.8 & 0.1 & 0.3 & 1.6 \\
Energy donor $-10\%$			& 0.5  	& 0.1  & 2.9 & 0.1 & 0.9 & \\
\noalign{\smallskip}\hline\noalign{\smallskip}

$g_\text{int}$  acceptor $+10\%$  		& 0.4   & 0.3   & 3.8   & 0.1  & 0.9   &1.3\\
$g_\text{int}$  donor $+10\%$  			& 0.2   & 1.0   & 4.2   & 0.1  & 0.4  & 5.7\\
$g_\text{int}$  acceptor $-10\%$  		& 0.3   & 1.7   & 14. & 0.1  & 0.3  & 1.5\\
$g_\text{int}$  donor $-10\%$ 			& 0.03 & 0.4   & 6.8   & 0.1  & 0.7  & 6.9\\
\noalign{\smallskip}\hline\noalign{\smallskip}

$\sigma_e$ acceptor $+10\%$  	& 0.3	& 1.4   & 0.9 & 0.06 & 0.4  & 2.3\\
$\sigma_e$  donor $+10\%$  		& 0.2  	& 1.0   & 1.8 & 0.01 & 0.4  & 0.6\\
$\sigma_e$  acceptor $-10\%$  	& 0.4  	& 0.3   & 1.9 & 0.1   & 0.8  & 4.7\\
$\sigma_e$  donor $-10\%$ 			& 0.1  	& 0.6   & 0.1 & 0.1   & 0.6  & 5.5\\

\noalign{\smallskip}\hline\noalign{\smallskip}

$\sigma_h$ acceptor $+10\%$  	& 0.3  	& 1.5  & 1.3 & 0.1  & 0.8  & 2.6\\
$\sigma_h$  donor $+10\%$  		& 0.3  	& 0.1  & 1.4 & 0.1  & 0.6  & 5.0\\
$\sigma_h$  acceptor $-10\%$  	& 0.3  	& 2.0  & 1.0 & 0.1  & 0.3  & 2.2\\
$\sigma_h$  donor $-10\%$ 			& 0.3  	& 0.5  & 0.8 & 0.01& 0.3  & 0.8\\
\noalign{\smallskip}\hline\noalign{\smallskip}
electron trapping constant $+ 33\%$ 			& 2.3  	& 1.1  & 1.1 &  1.3 & 2.5 & 1.0\\
hole trapping constant $+ 23\%$  				& 0.5  	& 2.4  & 11.8 & 1.2 & 2.1 & 6.0\\
electron trapping constant $- 33\%$ 			& 1.1  	& 2.0  & 5.1 &  0.5 & 0.1 & 6.1\\
hole trapping constant $- 23\%$  				& 2.2  	& 2.5  & 0.4 & 0.6 & 0.2 & 2.9\\
\noalign{\smallskip}\hline\noalign{\smallskip}
Total Uncertainty								& 3.4& 5.1 	& 21.3 & 1.9 & 4.2  &15.6\\
\noalign{\smallskip}\hline\noalign{\smallskip}
\end{tabular*}
   \caption{List of systematic uncertainties considered in the simulation and their relative impact on the predicted charge collection efficiency.  Blank spaces correspond to uncertainties that are below $0.005$.  The total uncertainty is the sum in quadrature of the maximum of the up and down variations.}
   \label{tab:SystError}
\end{center}

\end{table}

\subsection{Lorentz angle}
\label{sec:lorentzangle}

\begin{figure}[h!]
\centering
  \subfloat[]{\includegraphics[width=0.42\textwidth]{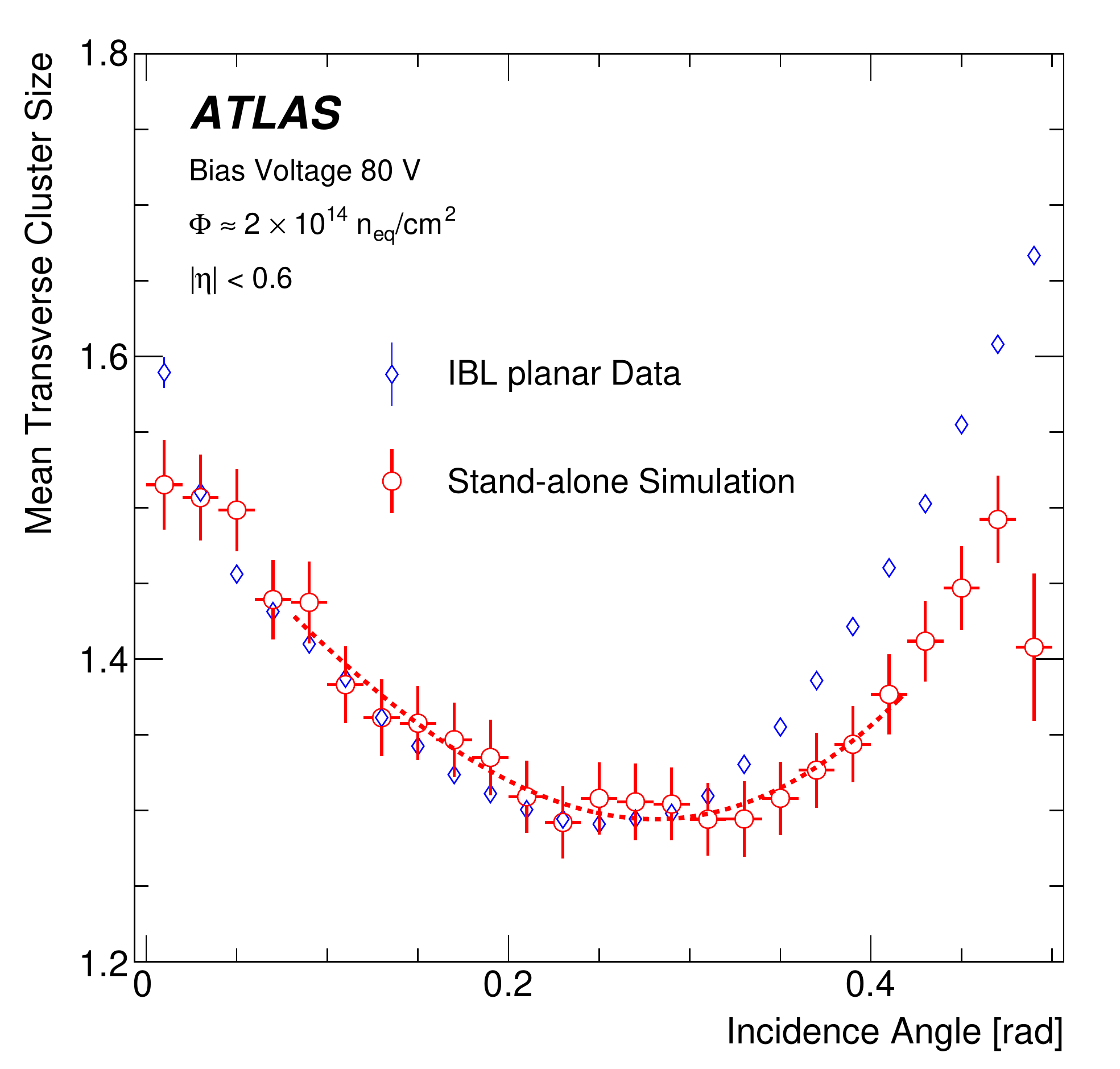}}
  \subfloat[]{\includegraphics[width=0.58\textwidth]{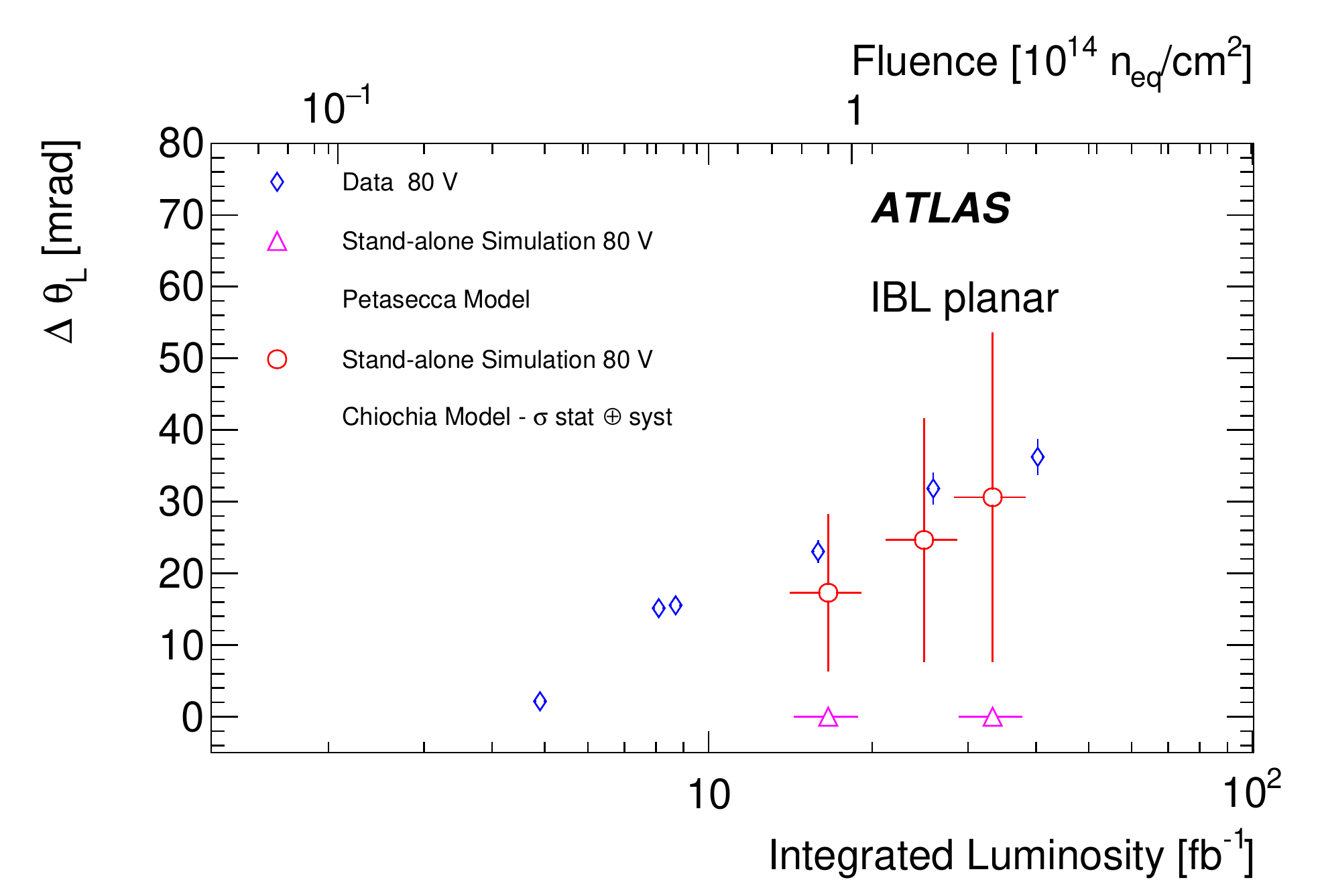}}
  \caption{(a) The mean transverse cluster size versus transverse incidence angle near the end of the 2016 run ($\sim2\times 10^{14}$~$\text{n}_\text{eq}/\text{cm}^2$) with a bias voltage of 80~V.  (b) The change in the Lorentz angle ($\theta_\text{L}$) from the unirradiated case as a function of the integrated luminosity in 2015-2016.  Two TCAD radiation damage models are considered, Chiochia and Petasecca.}
\label{fig:LA2:Run2}
\end{figure}

The Lorentz angle is determined by performing a fit to the transverse cluster size $F$ as function of the incidence angle of the associated track using the following functional form:

\begin{align*}
F(\alpha)=[a\times |\tan\alpha-\tan\theta_\text{L}|+b/\sqrt{\cos\alpha}]\otimes G(\alpha|\mu=0,\sigma),
\end{align*}

where $\alpha$ is the incidence angle,\footnote{The angle the tangent vector of the track makes with the vector normal to the sensor surface.} $\theta_\text{L}$ is the fitted Lorentz angle, $G$ is a Gaussian probability distribution evaluated at $\alpha$ with mean $0$ and standard deviation $\sigma$, and $a$ and $b$ are two additional fit parameters related to the depletion depth and the minimum cluster size, respectively.  An example input to the fit is shown in Figure~\ref{fig:LA2:Run2}(a).  In general, the simulation does not match the data at very low and high incidence angles, since the simulated points depend on many features of the simulation, but the position of the minimum should depend only on the Lorentz angle.  For example, the geometry used for this simulation is simplified and the extreme incidence angles are likely more impacted in the actual geometry.  The simulation in Figure~\ref{fig:LA2:Run2}(a) matches the low incidence angles well, but this is not seen for all fluences; it could be due in part to the uncertainty in the fluence.  

The fitted Lorentz angle as a function of integrated luminosity is shown in Figure~\ref{fig:LA2:Run2}(b).  Due to the degradation in the electric field, the mobility and thus the Lorentz angle increase with fluence.  This is not true for the Petasecca model, which does not predict regions of low electric field.
Charge trapping does not play a significant role in the Lorentz angle prediction.  The overall normalisation of the simulation prediction is highly sensitive to the radiation damage model parameters, but the increasing trend is robust.  An overall offset (not shown) is consistent with previous studies and appears even without radiation damage (zero fluence)~\cite{ATL-INDET-PUB-2018-001}, which is why only the difference in the angle is presented.

\clearpage

\section{Conclusions and future outlook}
\label{sec:conclusions}

This paper presents a digitization model for ATLAS planar and 3D sensors that includes radiation damage effects.  Predictions for the fluence at a given integrated luminosity are validated using leakage current data and stand-alone Hamburg model-based calculations.  TCAD simulations with effective traps in the silicon bulk are used to model distortions in the electric field caused by exposure to radiation.  The impact of annealing is studied by using predictions of the effective doping concentration to adjust the concentration of defect levels in the TCAD simulation.  Systematic uncertainties in all aspects of the luminosity-to-fluence conversion and radiation damage model are estimated.  

Comparisons between simulations using the radiation damage model and collision data indicate that within the current precision, the fluence-dependence is well-reproduced.  The charge collection efficiency gradually decreases with integrated luminosity until there are significant regions of the sensor with a small electric field, at which point there are significant losses.  After switching the bias voltage in the IBL from 80~V to 150~V, and then from 150~V to 350~V, the charge collection efficiency significantly increased.  The bias voltage will need to be increased further in order to recover the losses in the charge collection efficiency.  Both the simulations and the data indicate that the Lorentz angle increases with fluence.  The prediction for the Lorentz angle is quite sensitive to variations in the radiation damage parameters, but an increasing trend is a robust prediction of the simulation.  With more fluence and annealing, collision data may even be used to further constrain the radiation damage models.  

However, the current radiation damage models have known limitations.   As an example, in addition to predicting the wrong fluence for space-charge sign inversion, an alternative model (Petasecca) predicts a linear electric field profile that does not qualitatively describe the Lorentz angle dependence on fluence.   At the same time, the Hamburg model used to describe annealing does not incorporate effects from a non-uniform space-charge density distribution.  Thus far, the Hamburg model provides an excellent description of leakage current data, but this may not hold in the extreme irradiation regime when the electric field profile is sufficiently far from linear.   After the next long shutdown currently scheduled to end in 2021, annealing may produce a large enough effect so that the approximate integration of annealing into TCAD presented earlier may no longer be accurate. Another related challenge is that TCAD models such as the Chiochia one need to be improved in order to accommodate an accurate temperature dependence.
 While future collision data may be used to tune the radiation damage models, additional work may be required to combine the best of the Hamburg and TCAD models.  The flexibility of the digitizer model will allow collision data to be used to validate and test new ideas in the future.  A version of the digitizer model is publicly available on GitHub~\cite{github} and the model is also implemented in the ATLAS software framework (ATHENA) in releases designed for simulating the data in 2016 and beyond.  The ATLAS code can be found on GitLab~\cite{gitlab}.

Even though the innermost pixel layers have only been exposed to $\Phi\approx\times 10^{15}$~$\text{n}_\text{eq}/\text{cm}^{2}$, radiation damage effects are already measurable.  The projected fluence on the IBL at the end of LHC operation (300~fb$^{-1}$) is about $\Phi=2\times 10^{15}$~$\text{n}_\text{eq}/\text{cm}^{2}$; the sensors on the innermost layer of the upgraded ATLAS tracker (ITk) will have to withstand $\Phi\sim 1\times 10^{16}$~$\text{n}_\text{eq}/\text{cm}^{2}$~\cite{ITkStripsTDR}.  For such high fluences, pixel sensor radiation damage will be an important aspect of operations (setting the high voltage, deciding the time spent warm, etc.) and track reconstruction.  The simulation framework presented here can be used to inform both online and offline performance for the current pixel detector as well as for making important design decisions for the upgraded ATLAS detector that must survive the harsh HL-LHC radiation environment.

\section*{Acknowledgements}


We thank CERN for the very successful operation of the LHC, as well as the
support staff from our institutions without whom ATLAS could not be
operated efficiently.

We acknowledge the support of ANPCyT, Argentina; YerPhI, Armenia; ARC, Australia; BMWFW and FWF, Austria; ANAS, Azerbaijan; SSTC, Belarus; CNPq and FAPESP, Brazil; NSERC, NRC and CFI, Canada; CERN; CONICYT, Chile; CAS, MOST and NSFC, China; COLCIENCIAS, Colombia; MSMT CR, MPO CR and VSC CR, Czech Republic; DNRF and DNSRC, Denmark; IN2P3-CNRS, CEA-DRF/IRFU, France; SRNSFG, Georgia; BMBF, HGF, and MPG, Germany; GSRT, Greece; RGC, Hong Kong SAR, China; ISF and Benoziyo Center, Israel; INFN, Italy; MEXT and JSPS, Japan; CNRST, Morocco; NWO, Netherlands; RCN, Norway; MNiSW and NCN, Poland; FCT, Portugal; MNE/IFA, Romania; MES of Russia and NRC KI, Russian Federation; JINR; MESTD, Serbia; MSSR, Slovakia; ARRS and MIZ\v{S}, Slovenia; DST/NRF, South Africa; MINECO, Spain; SRC and Wallenberg Foundation, Sweden; SERI, SNSF and Cantons of Bern and Geneva, Switzerland; MOST, Taiwan; TAEK, Turkey; STFC, United Kingdom; DOE and NSF, United States of America. In addition, individual groups and members have received support from BCKDF, CANARIE, CRC and Compute Canada, Canada; COST, ERC, ERDF, Horizon 2020, and Marie Sk{\l}odowska-Curie Actions, European Union; Investissements d' Avenir Labex and Idex, ANR, France; DFG and AvH Foundation, Germany; Herakleitos, Thales and Aristeia programmes co-financed by EU-ESF and the Greek NSRF, Greece; BSF-NSF and GIF, Israel; CERCA Programme Generalitat de Catalunya, Spain; The Royal Society and Leverhulme Trust, United Kingdom. 

The crucial computing support from all WLCG partners is acknowledged gratefully, in particular from CERN, the ATLAS Tier-1 facilities at TRIUMF (Canada), NDGF (Denmark, Norway, Sweden), CC-IN2P3 (France), KIT/GridKA (Germany), INFN-CNAF (Italy), NL-T1 (Netherlands), PIC (Spain), ASGC (Taiwan), RAL (UK) and BNL (USA), the Tier-2 facilities worldwide and large non-WLCG resource providers. Major contributors of computing resources are listed in Ref.~\cite{ATL-GEN-PUB-2016-002}.

\clearpage

\printbibliography

\clearpage

 
\begin{flushleft}
{\Large The ATLAS Collaboration}

\bigskip

M.~Aaboud$^\textrm{\scriptsize 35d}$,    
G.~Aad$^\textrm{\scriptsize 101}$,    
B.~Abbott$^\textrm{\scriptsize 128}$,    
D.C.~Abbott$^\textrm{\scriptsize 102}$,    
O.~Abdinov$^\textrm{\scriptsize 13,*}$,    
D.K.~Abhayasinghe$^\textrm{\scriptsize 93}$,    
S.H.~Abidi$^\textrm{\scriptsize 167}$,    
O.S.~AbouZeid$^\textrm{\scriptsize 40}$,    
N.L.~Abraham$^\textrm{\scriptsize 156}$,    
H.~Abramowicz$^\textrm{\scriptsize 161}$
% ...
\bigskip
\\

% ...
$^{13}$Institute of Physics, Azerbaijan Academy of Sciences, Baku; Azerbaijan.\\
% ...
$^{35}$$^{(a)}$Facult\'e des Sciences Ain Chock, R\'eseau Universitaire de Physique des Hautes Energies - Universit\'e Hassan II, Casablanca;$^{(b)}$Facult\'{e} des Sciences, Universit\'{e} Ibn-Tofail, K\'{e}nitra;$^{(c)}$Facult\'e des Sciences Semlalia, Universit\'e Cadi Ayyad, LPHEA-Marrakech;$^{(d)}$Facult\'e des Sciences, Universit\'e Mohamed Premier and LPTPM, Oujda;$^{(e)}$Facult\'e des sciences, Universit\'e Mohammed V, Rabat; Morocco.\\
% ...
$^{40}$Niels Bohr Institute, University of Copenhagen, Copenhagen; Denmark.\\
% ...
$^{93}$Department of Physics, Royal Holloway University of London, Egham; United Kingdom.\\
% ...
$^{101}$CPPM, Aix-Marseille Universit\'e, CNRS/IN2P3, Marseille; France.\\
$^{102}$Department of Physics, University of Massachusetts, Amherst MA; United States of America.\\
% ...
$^{128}$Homer L. Dodge Department of Physics and Astronomy, University of Oklahoma, Norman OK; United States of America.\\
% ...
$^{156}$Department of Physics and Astronomy, University of Sussex, Brighton; United Kingdom.\\
% ...
$^{161}$Raymond and Beverly Sackler School of Physics and Astronomy, Tel Aviv University, Tel Aviv; Israel.\\
% ...
$^{167}$Department of Physics, University of Toronto, Toronto ON; Canada.\\
% ...
$^{a}$ Also at Borough of Manhattan Community College, City University of New York, New York NY; United States of America.\\
$^{b}$ Also at Centre for High Performance Computing, CSIR Campus, Rosebank, Cape Town; South Africa.\\
$^{c}$ Also at CERN, Geneva; Switzerland.\\
$^{d}$ Also at CPPM, Aix-Marseille Universit\'e, CNRS/IN2P3, Marseille; France.\\
$^{e}$ Also at D\'epartement de Physique Nucl\'eaire et Corpusculaire, Universit\'e de Gen\`eve, Gen\`eve; Switzerland.\\
% ...
$^{*}$ Deceased

\end{flushleft}


\end{document}